\documentclass[aps,prx,amsmath,amssymb]{revtex4-2}

\usepackage{graphicx}
\usepackage{dcolumn}
\usepackage{bm}
\usepackage{array,booktabs}
\usepackage{mathptmx}
\DeclareSymbolFont{epsilon}{OML}{ntxmi}{m}{it}
\DeclareMathSymbol{\epsilon}{\mathord}{epsilon}{"0F}

\usepackage{multirow}
\usepackage{caption}
\usepackage{subcaption}
\usepackage[referable]{threeparttablex}

\DeclareMathAlphabet{\mathcal}{OMS}{cmsy}{m}{n}

\usepackage[utf8]{inputenc}
\usepackage[T1]{fontenc}
\usepackage{ulem}
\usepackage{mathptmx}

\usepackage[breaklinks=true,colorlinks=true,linkcolor=blue,urlcolor=blue,citecolor=blue,bookmarks=false]{hyperref}
\begin{document}

\title[]{Relativistic relationship between nuclear-spin-dependent parity-violating NMR shielding and nuclear spin-rotation tensors}

\author{I.~Agust{\'{\i}}n Aucar}
\email[Author to whom correspondence should be addresed. Electronic mail: ]{agustin.aucar@conicet.gov.ar}
\affiliation{Instituto de Modelado e Innovaci{\'o}n Tecnol{\'o}gica (UNNE-CONICET), Facultad de Ciencias Exactas y Naturales y Agrimensura, Universidad Nacional del Nordeste, Avda.~Libertad 5460, Corrientes, Argentina}

\author{Mariano T.~Colombo Jofr{\'e}}
\affiliation{Instituto de Modelado e Innovaci{\'o}n Tecnol{\'o}gica (UNNE-CONICET), Facultad de Ciencias Exactas y Naturales y Agrimensura, Universidad Nacional del Nordeste, Avda.~Libertad 5460, Corrientes, Argentina}

\author{Gustavo A.~Aucar}
%\email{gaa@unne.edu.ar}
\affiliation{Instituto de Modelado e Innovaci{\'o}n Tecnol{\'o}gica (UNNE-CONICET), Facultad de Ciencias Exactas y Naturales y Agrimensura, Universidad Nacional del Nordeste, Avda.~Libertad 5460, Corrientes, Argentina}

\date{\today}

\begin{abstract}

It is known that the nuclear-spin-dependent parity-violating contributions to the NMR shielding and the nuclear spin-rotation tensors ($\bm{\sigma}^{PV}$ and $\bm{M}^{PV}$, respectively) are formally related each other within the non-relativistic (NR) regime. Such a formal relationship is not any longer valid within the relativistic domain. A new more general formal relationship, that is valid within the relativistic framework is shown here, being developed through the use of the LRESC model. The formalism of polarization propagators is applied to write the different contributions to both properties within both regimes, relativistic and NR. In the relativistic regime the Dirac-Coulomb Hamiltonian was selected as the unperturbed Hamiltonian. Theoretical developments together with results of calculations performed on the H$_2X_2$ series of molecules ($X =$ $^{17}$O, $^{33}$S, $^{77}$Se, $^{125}$Te and $^{209}$Po) show that also within the relativistic regime there is a close relationship between the parity-violation contributions to both properties. In particular, spin-dependent contributions are the most important in the four-component calculations of electroweak effects on the isotropic values of both tensors, $\bm{\sigma}^{PV}$ and $\bm{M}^{PV}$, being also responsible for the breakdown of the previously mentioned NR relationship among them. This last relationship is still fulfilled when the scalar-relativistic effects are considered.

\end{abstract}

\maketitle

\section{\label{sec:intro}Introduction}

The pioneering work by Lee and Yang postulated the existence of parity non conservation in weak interactions\cite{LeeYang56}. Since then many attempts were made to demonstrate experimentally that processes involving weak interactions are asymmetric under spatial inversion. The first experimental verification of these hypothesis was provided by Wu~\textit{et~al.}, who observed an asymmetry in the $\beta$-decay process of polarized $^{60}$Co nuclei\cite{Wu57}. Furthermore, the parity non-conserving effects were observed not only at nuclear scale, but also at the atomic level\cite{Bouchiat12}.

Because of parity violation (PV), the two enantiomers of an isolated chiral molecule in gas phase have different energies\cite{Zeldovich77,Quack05}. This phenomenon should be observable and the experimental detection of the frequency shift due to PV effects have been, and still is, considered a huge challenge\cite{Jones12}. This is the reason why several research programs were proposed during the last decades to detect PV effects in molecules. Among them we can mention the frequency shifts in the rotational, vibrational and electronic spectroscopy\cite{Letokhov75,Berger04-bookchapter,DarStoShe10,Hobi2013,CouManPie19}, as well as studies on nuclear magnetic resonance (NMR) parameters, like the shielding, ${\bm \sigma}$, and the indirect spin-spin coupling, ${\bm J}$, tensors\cite{Gorshkov82,Barra1986,Barra1988,Barra1996,Robert2001,Soncini2003,Laubender2003,Weijo2005,LauBer06,Bast2006,Nahrwold2009,Eills2017}. In spite of the ever-improving precision of such experiments, they are still unable to detect the so elusive PV effects in molecules.

In order to introduce another clue for getting deeper understandings of the physics that is behind the PV effects on molecules, we give here a novel analysis of the relativistic electronic mechanisms that are involved in the PV contributions to two molecular response properties, the NMR shielding and the nuclear spin-rotation (NSR), ${\bm M}$, tensors. Furthermore, using the linear response within the elimination of small components (LRESC) formalism we have uncovered a new generalized relationship among both properties that is valid within the relativistic framework. We show then the grounds for this new theoretical developments and the results of some numerical calculations that endorse it.

In 1986 Barra, Robert, and Wiesenfeld published a non-relativistic (NR) study of the role of nuclear-spin-dependent (NSD) PV contributions to ${\bm \sigma}$, ${\bm J}$ and ${\bm M}$ tensors ($\bm{\sigma}^{PV}$, $\bm{J}^{PV}$ and $\bm{M}^{PV}$, respectively)\cite{Barra1986}. Two years later, a four-component (4c) formalism for the PV effects in NMR shielding and indirect spin-spin coupling tensors was proposed by the same authors, where some numerical estimations were made within a semi-empirical relativistically parameterised extended Hückel approach\cite{Barra1988}. Laubender and Berger afterwards published the first NR {\it ab initio} calculations of PV frequency shifts in NMR spectra of chiral compounds\cite{Laubender2003}. The effects of special relativity were included in a two-component model by Weijo and co-workers, who applied a perturbational analysis and performed first-principles calculations of relativistic leading-order PV contributions based on the Breit-Pauli Hamiltonian, retaining only one-electron effects\cite{Weijo2005}. Finally, Bast and co-workers performed the first 4c {\it ab initio} calculations of PV-NMR-shielding tensors\cite{Bast2006}.

Bouchiat and Bouchiat pointed out that a considerable enhancement of sensitivity to the PV effects should be expected for heavy atoms\cite{Bouchiat74}. From their findings they suggested that searching for these effects in experimental measurements of both NMR spectroscopic parameters and the ${\bm M}$ tensor in molecules would likely result in the first observations of PV interactions in static systems. They argued that atomic PV experiments involve transition processes\cite{Bouchiat97}. Concerning the contributions $\bm{M}^{PV}$, Aucar and Borschevsky have recently developed a 4c relativistic formalism for including them. They calculated those contributions applying the polarization propagator approach at Dirac-Hartree-Fock (DHF) and density functional theory (DFT) levels of theory to a series of model molecules\cite{Aucar-PVSR2021}.

On the other hand, in usual measurements --considering only parity-conserving (PC) contributions-- it is expected to detect equal ${\bm M}$ and chemical shift tensors for a nucleus in both enantiomers of a chiral molecule. It is well known that the chemical shift is the NMR shielding of a given nucleus in a molecule relative to some reference system. Nevertheless, if the NSD-PV contributions to both properties are taken into account, the chemical shift and the tensor ${\bm M}$ of that nucleus are no longer the same in both enantiomers. The ${\bm \sigma}$ and the ${\bm M}$ tensors for a given nucleus in a chiral molecule should then be written as the sum of two contributions: a PC term with equal value for identical nuclei in both enantiomers, and a PV contribution with equal absolute value but opposite sign for each enantiomer.

As mentioned above we shall consider here the isotropic values of the ${\bm \sigma}$ and ${\bm M}$ tensors ($\sigma_{iso}$ and $M_{iso}$, respectively), and also analyze the theoretical relationship between them within the 4c relativistic formalism. Besides, we report a study of the relativistic effects on both $\sigma^{PV}_{iso}$ and $M^{PV}_{iso}$ in the $P$ enantiomers of the chiral series of molecules H$_2X_2$ (with $X =$ $^{17}$O, $^{33}$S, $^{77}$Se, $^{125}$Te and $^{209}$Po), within the random phase approximation (RPA) to calculate linear response functions. We use the Dirac-Coulomb (DC) Hamiltonian to study the relation between both properties in the relativistic domain, as well as the Lévy-Leblond (LL) Hamiltonian for the NR limit. The analysis of spin-dependent (SD) and scalar-relativistic effects is made employing the spin-free (SF) Hamiltonian.

\medskip

This work is organized as follows: in Sec.~\ref{sec:theory}, we give an introduction to the 4c theories of the PV-NSR and PV-NMR-shielding tensors within the relativistic polarization propagator formalism. Then, we use the LRESC model to propose two-component expansions for both properties, and the relationship between them. In Sec.~\ref{sec:comp-det} computational details are given. In Sec.~\ref{sec:res-disc} we show the results from computations of isotropic PV-NSR and PV-NMR-shielding constants for the $X$ nuclei ($X =$ $^{17}$O, $^{33}$S, $^{77}$Se, $^{125}$Te and $^{209}$Po) of the H$_2X_2$ molecules, focusing on the relation between both properties in the relativistic domain.

\section{\label{sec:theory}Theory}

Molecular properties can be calculated at different levels of approximation within wave-function, DFT, or polarization propagator theories\cite{Oddershede1978,Helgaker2012,Rusakov2013}.
In the latter, static second-order molecular response properties are written as\cite{Aucar2014}
\begin{equation}\label{eq:E2-PolProp}
E^{(2)}_{PQ}= \textnormal{Re}\left[  \langle\langle \, \hat{H}^P \, ; \, \hat{H}^Q \, \rangle\rangle_{\omega=0}  \right]
\end{equation}
\noindent where the linear response function at zero frequency
\begin{equation}\label{eq:LR-PQ-PolProp}
    \langle\langle \, \hat{H}^P \, ; \, \hat{H}^Q \, \rangle\rangle_{\omega=0} =
\textbf{b}^P \, \textbf{M}^{-1} \, \textbf{b}^Q
\end{equation}
\noindent arises as a product of the perturbators $\textbf{b}^P$, $\textbf{b}^Q$ (i.e., the property matrix elements) and the principal propagator $\textbf{M}^{-1}$ (i.e., the inverse of the electronic Hessian)\cite{Aucar2010}. As the explicit calculation of the principal propagator is computationally too expensive, the linear response function is usually obtained by solving the following response equation
\begin{equation}\label{eq:resp-eq}
%E_0^{[2]} \textbf{X}_Q(\omega) = - \textbf{E}_Q^{[1]},
\textbf{M} \; \textbf{X}^Q(\omega) = \textbf{b}^Q,
\end{equation}
\noindent where the solution vector $\textbf{X}^Q(\omega) = \textbf{M}^{-1} \, \textbf{b}^Q$, is first expanded in a linear combination of trial vectors. Then, it is contracted with the property matrix $\textbf{b}^P$ as shown in Eq.~\eqref{eq:LR-PQ-PolProp}\cite{Saue2003}. Since we only consider static properties in this work, we omit the subscript $\omega=0$ henceforth.

\subsection{\label{subsec:PV-prop} Parity-violating contributions to response properties}

For a given nucleus $N$ in a molecule, the NMR shielding tensor ($\bm{\sigma}_N$) and the NSR tensor ($\bm{M}_N$) are defined, respectively, from the following energy derivatives at zero frequency:
\begin{eqnarray}
 \bm{\sigma}_N &=& \frac{\partial^2 E}{\partial \bm{\mu}_N \partial \bm{B}_0}\bigg|_{\bm{\mu}_N=\bm{B}_0=0}, \label{eq:sigma-energy} \\
    \bm{M}_N &=& - \hslash \frac{\partial^2 E}{\partial \bm{I}_N \partial \bm{J}} \bigg|_{\bm{I}_N=\bm{J}=0}, \label{eq:M-energy}
\end{eqnarray}
\noindent where $\bm{\mu}_N = \frac{e \hslash}{2m_p} g_N \bm{I}_N$ [in Système International (SI) units] is the dipole magnetic moment of nucleus $N$, with $e$, $\hslash=\frac{h}{2\pi}$, $m_p$, $g_N$ and $\bm{I}_N$ being the elementary charge, the reduced Planck constant, the proton mass, the g-factor of nucleus $N$ and the dimensionless spin angular momentum of nucleus $N$, respectively. $\bm{B}_0$ is a uniform external magnetic field, and $\bm{J}$ is the molecular rotational angular momentum around the molecular center of mass (CM). In Eq.~\eqref{eq:M-energy} the NSR tensor is given in units of energy, and can therefore be converted to units of frequency by dividing it by the Planck constant $h$. SI units are adopted in the present work.

At the 4c level of theory, the perturbative Hamiltonians coupling the electronic dynamics with an external magnetic field ($\hat{H}^{\bm{B}}$), and with the molecular rotation ($\hat{H}^{\bm{J}}$) are given by\cite{Agus2012,Agus2013-1,aucar-aucar2019}
\begin{eqnarray}
    \hat{H}^{\bm{B}} &=& \frac{e}{2} \; \bm{B}_0 \cdot \sum_i \bm{r}_{GO} \times c\, \bm{\alpha}_i, \label{eq:H-B-SI}
\\
    \hat{H}^{\bm{J}} &=& -\bm{\omega}\cdot\bm{J}_e + \hat{H}^{\bm{J}-Breit}, \label{eq:H_J}
\end{eqnarray}
\noindent with $\bm{r}_{GO} = \bm{r}_i - \bm{R}_{GO}$ being the electron position operator relative to a fixed gauge origin for the external magnetic potential. Besides, $\bm{\omega}=\bm{I}^{-1} \cdot \bm{L}_N$ is the molecular angular velocity, $\bm{I}^{-1}$ the inverse molecular inertia tensor with respect to the CM in the equilibrium geometry, and $\bm{J}_e=\bm{L}_e+\bm{S}_e$ the $4 \times 4$ total electronic angular momentum operator (with $\bm{L}_e = \sum_i (\bm{r}_i - \bm{R}_{CM}) \times \bm{p}_i$). Breit interactions between moving nuclei and electrons, given by the perturbative term $\hat{H}^{\bm{J}-Breit}$, will not be considered in the present work because of their negligible influence with respect to the first term in the right-hand-side of Eq.~\eqref{eq:H_J}\cite{Agus2013-2}.

It is also known that the 4c electron-nucleus effective interaction operator, to the lowest order $Z^0$-exchange between electrons and nuclei, is given by the perturbative Hamiltonian\cite{Berger04-bookchapter,Flambaum1980,Blundell1992}
\begin{eqnarray}\label{eq:H-PV-complete}
 \hat{H}^{PV} &=& \frac{G_F}{2\sqrt{2}} \sum_{i,N} \left\{ Q_{w,N} \gamma_i^5 \rho_N(\bm{r}_i) + \kappa_N \; \bm{\alpha}_i \cdot \bm{I}_N \; \rho_N(\bm{r}_i) \right\}.
\end{eqnarray}

\noindent The sums with indices $i$ and $N$ run over all electrons and nuclei, respectively. $G_F$ is the Fermi coupling constant, whose most recent value is $G_F/(\hslash \, c_0)^3=1.1663787 \times 10^{-5}$~GeV$^{-2}$, or equivalently $G_F\simeq 2.222516 \times10^{-14} \, E_h \, a_0^3$\cite{codata2018}. $Q_{w,N}=Z_N \left( 1-4\textnormal{sin}^2\theta_W \right)- N_N$ is the weak nuclear charge with $Z_N$ and $N_N$ the number of protons and neutrons of nucleus $N$, respectively. We have defined $\kappa_N = -2\lambda_N \left(1-4\, \textnormal{sin}^2\theta_W\right)$ for the sake of simplicity. While the most recent value of the sine-squared weak mixing angle $\theta_W$ is 0.23857(5)\cite{Zyla2020}, we use $\textnormal{sin}^2\theta_W=0.2319$\cite{sintheta} as the Weinberg parameter throughout this work for ease of comparison with earlier investigations\cite{Laubender2003,Bast2006,Nahrwold2009,Aucar-PVSR2021,Aucar-PVSR2022}. $\gamma_i^5$ and $\bm{\alpha}_i$ are the $4 \times 4$ Dirac pseudo-scalar (chirality operator) and the vector comprised by the Dirac matrices in standard representation, respectively, given as
\begin{equation}
 \gamma_i^5 = \begin{pmatrix} \bm{0} & \bm{1} \\ \bm{1} & \bm{0} \end{pmatrix},\qquad
 \bm{\alpha}_i = \begin{pmatrix} \bm{0} & \bm{\sigma}_i \\ \bm{\sigma}_i & \bm{0} \end{pmatrix},
\end{equation}
\noindent where both operate on the electron $i$'s spinors and with $\bm{0}$, $\bm{1}$ and $\bm{\sigma}_i$ being the $2 \times 2$ zero, identity and Pauli spin matrices, respectively. In addition, $\bm{r}_i$ is the position of electron $i$ with respect to the coordinate origin, $\rho_N(\bm{r}_i)$ is the normalized nuclear electric charge density of nucleus $N$ at the position of electron $i$ (given in units of the inverse of cube distances), and $\lambda_N$ is a nuclear state dependent parameter. We have set $\lambda_N=1$ in our calculations to facilitate the comparison with previous works\cite{Laubender2003,Bast2006,Aucar-PVSR2021,Aucar-PVSR2022}.

We do not address the nuclear-spin-independent terms of Eq.~\eqref{eq:H-PV-complete} but retain only the NSD one-electron contributions to the perturbative PV Hamiltonian, given in the second term on the right-hand-side of Eq.~\eqref{eq:H-PV-complete},
\begin{equation}\label{eq:H-PV-SD}
 \hat{H}^{PV} = \hat{H}_{SD}^{PV} = \; \frac{G_F}{2\sqrt{2}} \sum_{i,N} \kappa_N \; \bm{\alpha}_i \cdot \bm{I}_N \; \rho_N(\bm{r}_i).
\end{equation}
\noindent Then, the 4c PV contributions to NMR shielding and NSR tensors are given by the following linear response functions\cite{Laubender2003,Bast2006,Nahrwold2009,Aucar-PVSR2021,Aucar-PVSR2022}:
\begin{eqnarray}
 \bm{\sigma}_N^{PV} &=& \frac{G_F\,m_p}{2\sqrt{2}\,\hslash\,c_0} \; \frac{\kappa_N}{g_N}
 \langle\langle \; \rho_N(\bm{r}) \; c \bm{\alpha} \; ; \;  \bm{r}_{GO} \times c \bm{\alpha} \; \rangle\rangle, \label{eq:sigma-PV}
\\
 \bm{M}_N^{PV} &=& \frac{G_F\,\hslash}{2\sqrt{2}\,c_0} \; \kappa_N
 \langle\langle \; \rho_N(\bm{r}) \; c \bm{\alpha} \; ; \;  \bm{J}_e \; \rangle\rangle \; \cdot \bm{I}^{-1}, \label{eq:M-PV}
\end{eqnarray}
\noindent where $c$ is the speed of light in vacuum, and the factor $\frac{1}{c_0}$ is linearly proportional to the fine structure constant (which in SI units is given as $\frac{1}{4\pi \varepsilon_0}\frac{e^2}{\hslash c_0}$). It must be noted that the NR limit is recovered when $c \to \infty$.

\subsection{\label{subsec:LRESC} Relativistic contributions to parity-violating properties within the LRESC approach}

The LRESC approach is a useful model to write theoretical expressions of 4c second-order molecular properties as a Taylor's series expansion in terms of the fine structure constant. The zeroth order term of this series expansion yields the NR contribution of a given property, while the relativistic corrections are obtained by applying the elimination of small components (ESC) approach\cite{melo-lresc,Review-LRESC}. The LRESC model unveil the physical mechanisms behind the relativistic effects of molecular properties.

In this Section, the LRESC model is applied to the 4c expressions of the PV-NMR-shielding and PV-NSR tensors given in Eqs.~\eqref{eq:sigma-PV} and \eqref{eq:M-PV}, respectively. We only focus on the main steps to obtain the series expansions for these properties. The detailed derivations involved in this approach are extensively discussed elsewhere for second-order molecular response properties\cite{melo-lresc,Agus2012,Agus_g_2014,Review-LRESC}.

\bigskip

Within the relativistic polarization propagator theory, 4c second-order response molecular properties may be written as the sum of two terms:
\begin{equation}\label{eq:LR-PQ-PolProp-ee-pp}
    \langle\langle \, \hat{H}^P \, ; \, \hat{H}^Q \, \rangle\rangle \approx \langle\langle \, \hat{H}^P \, ; \, \hat{H}^Q \, \rangle\rangle^{(e-e)} + \langle\langle \, \hat{H}^P \, ; \, \hat{H}^Q \, \rangle\rangle^{(p-p)},
\end{equation}
\noindent with $\hat{H}^P$ and $\hat{H}^Q$ being perturbative Hamiltonians. The first term in the right-hand-side of Eq.~\eqref{eq:LR-PQ-PolProp-ee-pp} involves the positive energy spectrum of electronic states (namely ($e$-$e$) excitations), whereas in the second one only excitations to negative energy electronic orbitals are allowed, i.e., virtual electron-positron pairs in the QED picture ($e$-$p$ excitations)\cite{Review-LRESC}.

It is assumed that the off-diagonal contributions to the principal propagator of Eq.~\eqref{eq:LR-PQ-PolProp} are smaller than the diagonal ones. Then, in this representation their contributions are neglected in the right-hand side of Eq.~\eqref{eq:LR-PQ-PolProp-ee-pp}\cite{Aucar1999}. In the special case of magnetic properties, the NR limits of $\langle\langle \, \hat{H}^P \, ; \, \hat{H}^Q \, \rangle\rangle^{(e-e)}$ and $\langle\langle \, \hat{H}^P \, ; \, \hat{H}^Q \, \rangle\rangle^{(p-p)}$ are their paramagnetic and diamagnetic contributions, respectively.

In order to have a power series expansion in $c^{-1}$ for the ($e$-$e$) contribution we apply the ESC approach and write the 4c matrix elements of $\hat{H}^P$ and $\hat{H}^Q$ in terms of the ``normalized'' Pauli spinors $|\tilde{\phi}\rangle$, as\cite{melo-lresc}
\begin{equation}\label{eq:matrixelementsH}
\langle \phi_i^{(4)} | \hat{H}^{P(Q)} | \phi_j^{(4)} \rangle \simeq \langle \tilde{\phi}_i | \hat{O}(\hat{H}^{P(Q)}) | \tilde{\phi}_j \rangle.
\end{equation}

Another source of relativistic effects for the ($e$-$e$) contributions can be traced back to the unperturbed molecular Breit-Pauli Hamiltonian
\begin{equation}\label{eq:HBP-corrections}
\hat{H}^{BP} = \hat{H}^{S} + \hat{D}_1 + \hat{D}_2,
\end{equation}
\noindent where $\hat{H}^{S}$ stands for the $N$-electron Schr\"odinger Hamiltonian and $\hat{D}_1$ and $\hat{D}_2$ are generalized $N$-particle space operators arising from the one-body ($\hat{D}_1$) and two-body ($\hat{D}_2$) leading order expansions of the Dirac Hamiltonian\cite{Review-LRESC}. In particular,
\begin{equation}
 \hat{D}_1=\hat{H}^{Mv} + \hat{H}^{Dw} + \hat{H}^{SO},
\end{equation}
\noindent being $\hat{H}^{Mv}$, $\hat{H}^{Dw}$ and $\hat{H}^{SO}$ the well-known mass-velocity, one-body Darwin and one-body spin-orbit perturbative relativistic corrections, respectively. On the other hand, $\hat{D}_2$ corresponds to the sum of the orbit–orbit interaction $\hat{H}^{OO}$, the two-body Darwin term $\hat{H}^{Dw(2)}$, the two-body spin-orbit interaction $\hat{H}^{SO(2)}$, the spin–other orbit interaction $\hat{H}^{SOO}$, and two spin–spin interactions: the Fermi-contact $\hat{H}^{FC-SS}$ and the dipole–dipole terms $\hat{H}^{SD-SS}$\cite{melo-lresc}.

\medskip

Retaining contributions up to the leading order (in an expansion in terms of $c^{-2}$) in the right-hand-side of Eq.~\eqref{eq:matrixelementsH}, we can write operator $\hat{O}(\hat{H}^{P(Q)})$ as
\begin{equation}\label{eq:Onr2}
\hat{O}(\hat{H}^{P(Q)}) = \hat{O}^{NR}(\hat{H}^{P(Q)}) + \hat{O}^{(2)}(\hat{H}^{P(Q)}) + \mathcal{O}(c^{-4}).
\end{equation}

By neglecting two-body relativistic effects in Eq.~\eqref{eq:HBP-corrections} (i.e., those contained in the $\hat{D}_2$ operator), we obtain
\begin{eqnarray}\label{eq:ee-expansion}
\langle\langle \, \hat{H}^P \, ; \, \hat{H}^Q \, \rangle\rangle^{(e-e)} &=& \;\;\;\; \langle\langle \, \hat{O}^{NR}(\hat{H}^P) \, ; \, \hat{O}^{NR}(\hat{H}^Q) \, \rangle\rangle  \nonumber \\ &&
+ \; \langle\langle \, \hat{O}^{NR}(\hat{H}^P) \, ; \, \hat{O}^{(2)}(\hat{H}^Q) \, \rangle\rangle  \nonumber \\ &&
+ \; \langle\langle \, \hat{O}^{(2)}(\hat{H}^P) \, ; \, \hat{O}^{NR}(\hat{H}^Q) \, \rangle\rangle  \nonumber \\
&& + \; \langle\langle \, \hat{O}^{NR}(\hat{H}^P) \, ; \, \hat{D}_1 \, ; \, \hat{O}^{NR}(\hat{H}^Q) \, \rangle\rangle \nonumber \\ &&
+ \; \mathcal{O}(c^{-4}),
\end{eqnarray}
\noindent where $\langle\langle ; ; \rangle\rangle$ stands for a quadratic response. The first term in the right-hand-side of Eq.~\eqref{eq:ee-expansion}, $\langle\langle \, \hat{O}^{NR}(\hat{H}^P) \, ; \, \hat{O}^{NR}(\hat{H}^Q) \, \rangle\rangle$ corresponds to the NR contribution to $\langle\langle \, \hat{H}^P \, ; \, \hat{H}^Q \, \rangle\rangle^{(e-e)}$, while the second, third and fourth terms correspond to the leading order relativistic contributions.

On the other hand, neglect of two-body contributions from the unperturbed Dirac-Coulomb-Breit molecular Hamiltonian\cite{Review-LRESC}, yields the LRESC series expansion for the ($p$-$p$) contribution
\begin{eqnarray}\label{eq:pp-expansion}
\langle\langle \, \hat{H}^P \, ; \, \hat{H}^Q \, \rangle\rangle^{(p-p)} &=& \frac{1}{2mc^2} \left[ \, \langle \Psi_0 | \hat{H}^P \hat{P}_p \hat{X}(\hat{H}^Q) | \Psi_0 \rangle  \right. \nonumber \\
&& \qquad
+
\left.
\langle \Psi_0 | \hat{H}^Q \hat{P}_p \hat{X}(\hat{H}^P) | \Psi_0 \rangle \, \right],
\end{eqnarray}
\noindent where $m$ is the electron mass, $\Psi_0$ is the 4c wave function corresponding to the ground state solution of the DHF approximation, and $\hat{P}_p$ is the projector onto the positronic states. Besides, $\hat{X}(\hat{H}^{P(Q)})$ is defined as
\begin{equation}\label{eq:X(HP)}
\hat{X}(\hat{H}^{P(Q)}) = 2 \; \hat{H}^{P(Q)} + \frac{1}{2mc^2} [\hat{H}^D,\hat{H}^{P(Q)}]  + \; \mathcal{O}(c^{-4}),
\end{equation}
where the commutator of the one-electron part of the Dirac-Coulomb-Breit Hamiltonian $\hat{H}^D$ and the $\hat{H}^{P(Q)}$ perturbative correction appears in the second term at the right-hand-side of Eq.~\eqref{eq:X(HP)}\cite{melo-lresc,Review-LRESC}.

\subsubsection{Parity-violation Hamiltonian in the LRESC framework}\label{sec:LRESC-PV}

In the particular cases of the PV-NMR-shielding and PV-NSR tensors, the involved perturbative Hamiltonians are $\hat{H}^{PV}$, $\hat{H}^{\bm{J}}$ and $\hat{H}^{\bm{B}}$.

According to Eq.~\eqref{eq:matrixelementsH}, it can be seen that the 4c matrix elements of operator $\hat{H}^{PV}$ can be written as a series in terms of $c^{-2}$. Since we only analyze leading order relativistic contributions, terms of order $c^{-4}$ and higher are not fully considered here. Therefore,
\begin{widetext}
\begin{eqnarray}\label{eq:O(H-PV)}
  \langle \phi_i ^{(4)} | \bm{\alpha} \cdot \bm{I}_N \; \rho_N(\bm{r}) | \phi_j ^{(4)} \rangle  \simeq  % \nonumber \\
  \langle \tilde{\phi}_i |  \left[    N \left( \frac{\bm{\sigma}\cdot\bm{p}}{2mc} \right)
                                       \left( 1 + \frac{V_C-E_i}{2mc^2}\right)  \bm{\sigma} \cdot \bm{I}_N \; \rho_N(\bm{r}) \; N  %\right. \nonumber \\
  %\left.
  + \; N \;  \bm{\sigma} \cdot \bm{I}_N \; \rho_N(\bm{r})
                                       \left( 1 + \frac{V_C-E_j}{2mc^2}\right)
                                       \left( \frac{\bm{\sigma}\cdot\bm{p}}{2mc} \right) N
  \right]
  | \tilde{\phi}_j \rangle, \nonumber \\
\end{eqnarray}
\end{widetext}
\noindent where $V_C$ and $E_{i(j)}$ stand for the electron-nucleus Coulomb potential and the energy eigenvalue of the one-electron Dirac wave function $| \phi_{i(j)}^{(4)} \rangle$, respectively.

In Eq.~\eqref{eq:O(H-PV)} it was further assumed that, at first order of approximation, the large and small components of the wave function $| \phi_i^{(4)} \rangle$ are related by\cite{melo-lresc}
\begin{equation}\label{eq:psiS-psiL}
| \phi_i^{S} \rangle \approx  \frac{1}{2mc} \left[ 1 + \frac{V_C-E_i}{2mc^2}\right] \bm{\sigma} \cdot \bm{p} \; | \phi_i^{L} \rangle,
\end{equation}
\noindent and that $N$ is a constant defined through normalization of $| \phi^L_i \rangle$\cite{melo-lresc},
\begin{equation}\label{eq:norm-phiL}
| \phi_i^{L} \rangle \simeq N | \tilde{\phi}_i \rangle = \left( 1 - \frac{p^2}{8 m^2 c^2} \right) | \tilde{\phi}_i \rangle.
\end{equation}

Then, we can use the zeroth order relation\cite{melo-lresc,Review-LRESC}
\begin{equation}
 (V_C-E_i) | \tilde{\phi}_i \rangle = - \frac{p^2}{2m}| \tilde{\phi}_i \rangle
\end{equation}
and retain only terms up to order $c^{-3}$ in Eq.~\eqref{eq:O(H-PV)} to obtain
\begin{widetext}
\begin{eqnarray}\label{eq:O(H-PV)-exp}
  \langle \phi_i ^{(4)} | \bm{\alpha} \cdot \bm{I}_N \; \rho_N(\bm{r}) | \phi_j ^{(4)} \rangle  &\simeq& \;\;\;\;
\frac{1}{2mc} \langle \tilde{\phi}_i | \left\{ \bm{\sigma}\cdot\bm{p} \; , \; \bm{\sigma} \cdot \bm{I}_N \, \rho_N(\bm{r}) \right\} | \tilde{\phi}_j \rangle + \; \frac{1}{16m^3c^3} \times \nonumber \\
&& \hspace{-0.8cm} % + \; \frac{1}{16m^3c^3}
\langle \tilde{\phi}_i | \Big( 2 \left\{ p^2 \; , \; \left\{ \bm{\sigma}\cdot\bm{p} \; , \; \bm{\sigma} \cdot \bm{I}_N \, \rho_N(\bm{r}) \right\} \right\}
+ \left[ p^2 \; , \; \left[ \bm{\sigma}\cdot\bm{p} \; , \; \bm{\sigma} \cdot \bm{I}_N \, \rho_N(\bm{r}) \right] \right]
- 8 m \bm{\sigma}\cdot\bm{\nabla} V_C \times \bm{I}_N \, \rho_N(\bm{r}) \Big)
  | \tilde{\phi}_j \rangle .% \nonumber \\
%
%&& + \; \mathcal{O}(c^{-5}) .
\end{eqnarray}
\end{widetext}

Further use of the following identity for the Pauli matrices:
\begin{equation}
\left(\bm{\sigma} \cdot \bm{A}\right) \; \left(\bm{\sigma} \cdot \bm{B}\right) = \bm{A} \cdot \bm{B} + i \bm{\sigma} \cdot \left( \bm{A} \times \bm{B}\right),
\end{equation}

\noindent and of Eqs.~\eqref{eq:matrixelementsH} and \eqref{eq:Onr2}, yields the result

\begin{widetext}
\begin{eqnarray}
\label{eq:Onr(PV)-exp}
 \hat{O}^{NR}(\hat{H}^{PV})  &=& \; \frac{G_F}{2\sqrt{2}} \sum_{i,N} \frac{\kappa_N}{2mc} \left(
 \bm{I}_N \cdot \left\{ \bm{p}_i \; , \; \rho_N(\bm{r}_i ) \right\}
 + i \left[ \bm{p}_i \; , \; \rho_N(\bm{r}_i ) \right] \cdot \bm{I}_N \times \bm{\sigma}_i
 \right) \\
\label{eq:O2(PV)}
 \hat{O}^{(2)}(\hat{H}^{PV})  &=& \; \frac{G_F}{2\sqrt{2}} \sum_{i,N} \frac{\kappa_N}{16m^3c^3} \left(
 \; 2 \left\{ p^2_i , \left\{ \bm{\sigma}_i \cdot \bm{p}_i , \bm{\sigma}_i \cdot \bm{I}_N \, \rho_N (\bm{r}_i) \right\} \right\}
 + \left[ p^2_i , \left[ \bm{\sigma}_i \cdot \bm{p}_i , \bm{\sigma}_i \cdot \bm{I}_N \, \rho_N (\bm{r}_i)\right] \right]
 \; - \, 8m \, \bm{\sigma}_i \cdot \bm{\nabla} V_C \times \bm{I}_N \, \rho_N (\bm{r}_i)
 \; \right). \nonumber \\
\end{eqnarray}
\end{widetext}

\bigskip

Finally, we employ Eq.~\eqref{eq:pp-expansion} in order to get an expansion for the ($p$-$p$) contributions of PV second-order response properties according to the LRESC model. For this, the perturbative 4c Hamiltonian $\hat{H}^{PV}$ should be replaced in Eq.~\eqref{eq:X(HP)}. After retaining leading order terms, $\hat{X}(\hat{H}^{PV})$ can be written as
%
%\begin{widetext}
\begin{eqnarray}\label{eq:XHPV}
\hat{X}(\hat{H}^{PV})
% &=&
% \; \frac{G_F}{2\sqrt{2}} \;
% \sum_{i,N} \kappa_N \left( 2 \, \bm{\alpha}_i \cdot \bm{I}_N \; \rho_N(\bm{r}_i) + \frac{1}{2mc^2} [\hat{H}^D,\bm{\alpha}_i \cdot \bm{I}_N \; \rho_N(\bm{r}_i)]  \right) \nonumber \\
&=& \; \frac{G_F}{2\sqrt{2}} \;
\sum_{i,N} \kappa_N \Big( (2\,I+\beta) \, \bm{\alpha}_i \cdot \bm{I}_N \; \rho_N(\bm{r}_i)
\nonumber \\
& & \quad \quad \quad \quad
+ \frac{1}{2mc} [\bm{\alpha}_i \cdot \bm{p}_i,\bm{\alpha}_i \cdot \bm{I}_N \; \rho_N(\bm{r}_i) ]  \Big),
\end{eqnarray}
%\end{widetext}
%
\noindent with $I$ and $\beta$ being the 4$\times$4 identity and Dirac matrices, respectively.

\subsubsection{LRESC applied to the PV-NMR-shielding tensor}\label{sec:LRESC-PV-shi}

According to LRESC model, the application of Eqs.~\eqref{eq:matrixelementsH} and \eqref{eq:Onr2} to the perturbative Hamiltonian $\hat{H}^{\bm{B}}$ of Eq.~\eqref{eq:H-B-SI}, yields (see Appendix \ref{sec:appendix})\cite{melo-lresc,Review-LRESC}
%that describes the interaction of the electronic system with an external uniform magnetic field
%
\begin{eqnarray}
 \hat{O}^{NR}(\hat{H}^{\bm{B}})  &=& \hat{H}^{OZ} + \hat{H}^{SZ}                  \label{eq:Onr(AB)}, \\
 \hat{O}^{(2)}(\hat{H}^{\bm{B}}) &=& \hat{H}^{OZ-K} + \hat{H}^{SZ-K} + \hat{H}^{B-SO}.   \label{eq:O2(AB)}
\end{eqnarray}

In Eq.~\eqref{eq:Onr(AB)}, the \textit{orbital Zeeman} ($\hat{H}^{OZ}$) and the \textit{spin Zeeman} ($\hat{H}^{SZ}$) terms are due to the interaction of the electrons with an external uniform magnetic field $\bm{B}_0$ in the NR limit. The \textit{Kinetic orbital-Zeeman} ($\hat{H}^{OZ-K}$), the \textit{Kinetic spin-Zeeman} ($\hat{H}^{SZ-K}$) and the so-called \textit{magnetic field induced spin-orbit} ($\hat{H}^{B-SO}$) contributions given in Eq.~\eqref{eq:O2(AB)} are the leading order relativistic corrections to $\hat{H}^{\bm{B}}$, whose derivation and explicit expressions are extensively discussed elsewhere (see also Appendix \ref{sec:appendix})\cite{Review-LRESC}.

\medskip

Making the substitutions $\hat{H}^{P}\to\hat{H}^{PV}$ and $\hat{H}^{Q}\to\hat{H}^{\bm{B}}$ in Eq.~\eqref{eq:ee-expansion}, retaining only the zeroth-order term in the resulting expression, and then using Eq.~\eqref{eq:sigma-energy} yields the NR limit of $\bm{\sigma}_N^{PV(e-e)}$, which is the paramagnetic contribution to $\bm{\sigma}_N^{PV-NR}$. Using Eqs.~\eqref{eq:Onr(PV)-exp} and~\eqref{eq:Onr(AB)} we obtain
%
%\begin{widetext}
\begin{eqnarray}\label{eq:sigma-PV-NR-complete}
\bm{\sigma}_N^{PV-NR-para}
% &=& \frac{\partial^2}{\partial \bm{\mu}_N \partial \bm{B}_0} \langle \langle \hat{O}^{NR}(\hat{H}^{PV}) \; ; \; \hat{O}^{NR}(\hat{H}^{\bm{B}}) \rangle \rangle \nonumber \\
&=&
\; \frac{G_F \, m_p}{4\sqrt{2}\,\hslash\,m^2\,c} \frac{\kappa_N}{g_N} \bigg( \langle\langle \left\{ \bm{p} \, , \, \rho_N(\bm{r}) \right\} \, ; \, \bm{r}_{GO} \times \bm{p} \rangle\rangle
\nonumber \\
&& \quad\quad\quad + \; \hslash^2 \langle\langle
  \bm{\sigma} \times \left[ \bm{\nabla} \, , \, \rho_N(\bm{r}) \right] \, ; \, \bm{\sigma} \rangle \rangle \bigg).
\end{eqnarray}
%\end{widetext}

For molecular systems with closed-shell electronic configurations, the last term of Eq.~\eqref{eq:sigma-PV-NR-complete} is exactly zero because the eigenvalue of the total spin operator in the NR electronic ground state is zero. Then, for this particular case we have
\begin{eqnarray}\label{eq:sigma-PV-NR}
\bm{\sigma}_N^{PV-NR-para} &=&
\; \frac{G_F\,m_p}{4\sqrt{2}\, \hslash \, m^2 c} \frac{\kappa_N}{g_N}
\langle\langle \left\{ \bm{p} \, , \, \rho_N(\bm{r}) \right\} \, ; \, \bm{r}_{GO} \times \bm{p} \rangle\rangle. \nonumber \\
\end{eqnarray}

The leading order relativistic corrections to $\bm{\sigma}_N^{PV(e-e)}$ are obtained following the same procedure, meaning calculating the second, third and fourth terms in the right-hand-side of Eq.~\eqref{eq:ee-expansion}, and using the operators given in Eqs.~\eqref{eq:Onr(PV)-exp}, \eqref{eq:O2(PV)}, \eqref{eq:Onr(AB)} and \eqref{eq:O2(AB)}. A complete expression of these contributions will be given in Sec.~\ref{sec:app-MV-PV}.

\bigskip

Besides, the LRESC model permits the calculation of the NR limit and leading order relativistic corrections for the ($p$-$p$) contributions to $\bm{\sigma}_N^{PV}$. To get these expressions, Eq.~\eqref{eq:pp-expansion} must be used with $\hat{H}^{PV}$ and $\hat{H}^{\bm{B}}$ as perturbative Hamiltonians. In addition, $\hat{X}(\hat{H}^{PV})$ will be taken from Eq.~\eqref{eq:XHPV}, and $\hat{X}(\hat{H}^{\bm{B}})$ can be obtained from previous works, where its expression is shown to be\cite{melo-lresc,Review-LRESC}
\begin{eqnarray}
\hat{X}(\hat{H}^{\bm{B}}) %&=&  2 \; \bm{\alpha} \cdot \bm{A}_B + \frac{1}{2mc^2}\left[ \hat{H}^D , \bm{\alpha}\cdot \bm{A}_B \right] \nonumber \\
&=& ( 2 \, I + \beta ) \; \bm{\alpha}\cdot \bm{A}_B
+ \frac{1}{2mc} \left[ \bm{\alpha} \cdot \bm{p} \, , \, \bm{\alpha} \cdot \bm{A}_B \right].
\end{eqnarray}

The projection operator over the negative-energy electronic states $\hat{P}_p=1-\hat{P}_e$ is expressed retaining terms up to order $c^{-1}$ as the $4 \times 4$ matrix of spinor components
\begin{equation}
 \hat{P}_p =
\begin{pmatrix}
                0                           &     -\frac{\bm{\sigma}\cdot\bm{p}}{2mc}   \\
 -\frac{\bm{\sigma}\cdot\bm{p}}{2mc}        &                       1
\end{pmatrix}.
\end{equation}

\noindent Therefore, the ($p$-$p$) contributions to $\bm{\sigma}_N^{PV}$ can be written as
\begin{eqnarray}\label{eq:sigma-pp-lresc-sum}
\bm{\sigma}_N^{PV(p-p)} &\simeq& \frac{\partial^2}{\partial \bm{\mu}_N \partial \bm{B}_0} \left[ \frac{G_F}{2\sqrt{2}} \frac{e}{2mc} \kappa_N \,
\langle \tilde{\Psi}_0 | K | \tilde{\Psi}_0 \rangle
\right] ,
\end{eqnarray}
\noindent where $\tilde{\Psi}_0$ stands for the solution to the Breit-Pauli Hamiltonian of Eq.~\eqref{eq:HBP-corrections}, and the $K$ operator is given by
\begin{widetext}
\begin{eqnarray}\label{eq:sigma-pp-lresc-sum-k}
K &=&
\left\{ \bm{\sigma} \cdot \bm{A}_{\bm{B}} \; , \; \bm{\sigma} \cdot \bm{I}_N \rho_N(\bm{r}) \right\}
- \frac{1}{4m^2c^2} \bigg[
\frac{1}{2} \left\{ p^2 \; , \; \left\{ \bm{\sigma} \cdot \bm{A}_{\bm{B}} \; , \; \bm{\sigma} \cdot \bm{I}_N \rho_N(\bm{r}) \right\} \right\}
+3 \bm{\sigma} \cdot \bm{A}_{\bm{B}} \bm{\sigma} \cdot \bm{p} \bm{\sigma} \cdot \bm{I}_N \rho_N(\bm{r}) \bm{\sigma} \cdot \bm{p}
 \nonumber \\
&&
+ \, \bm{\sigma} \cdot \bm{p} \bm{\sigma} \cdot \bm{A}_{\bm{B}} \bm{\sigma} \cdot \bm{p} \bm{\sigma} \cdot \bm{I}_N \rho_N(\bm{r})
+ 3 \bm{\sigma} \cdot \bm{I}_N \rho_N(\bm{r}) \bm{\sigma} \cdot \bm{p} \bm{\sigma} \cdot \bm{A}_{\bm{B}} \bm{\sigma} \cdot \bm{p}
+ \bm{\sigma} \cdot \bm{p} \bm{\sigma} \cdot \bm{I}_N \rho_N(\bm{r}) \bm{\sigma} \cdot \bm{p} \bm{\sigma} \cdot \bm{A}_{\bm{B}}
\bigg] . %+ \mathcal{O}(c^{-4}).
\end{eqnarray}
\end{widetext}

The NR limit of $\bm{\sigma}_N^{PV(p-p)}$ gives the diamagnetic contribution to $\bm{\sigma}_N^{PV-NR}$ in the form of a tensor, whose $ij$-th element can be written:
\begin{eqnarray}\label{eq:sigma-pp-lresc-NR}
\bm{\sigma}_{N,ij}^{PV-NR-dia} &=& \frac{G_F\,m_p}{2\sqrt{2}\,\hslash m c} \, \frac{\kappa_N}{g_N}
\, \epsilon_{ijk} \, \langle \tilde{\Psi}_0 | (\bm{r}_{GO})_k \; \rho_N(\bm{r}) | \tilde{\Psi}_0 \rangle, \nonumber \\
\end{eqnarray}
\noindent where $\epsilon_{ijk}$ stands for the Levi-Civita tensor.

From Eq.~\eqref{eq:sigma-pp-lresc-NR} it is easily seen that the diagonal tensor elements of $\bm{\sigma}_N^{PV-NR-dia}$ (and, consequently, its isotropic component) vanish identically because $\epsilon_{ijk} = 0$ when $i=j$. This statement holds for an arbitrary gauge origin position, and not only for the special case mentioned by Soncini and co-workers where it is placed at the nucleus of interest\cite{Soncini2003}.

\subsubsection{LRESC applied to the PV-NSR tensor}\label{sec:LRESC-PV-SR}

The derivation developed in Sec.~\ref{sec:LRESC-PV-shi} can be equally applied to the PV-NSR tensor. For this, the perturbative Hamiltonian $\hat{H}^{\bm{J}}$ of Eq.~\eqref{eq:H_J} has to be replaced in Eqs.~\eqref{eq:matrixelementsH} and \eqref{eq:Onr2}. This gives the known LRESC expressions (see Appendix \ref{sec:appendix})\cite{Agus2012,Review-LRESC}
\begin{eqnarray}
 \hat{O}^{NR}(\hat{H}^{\bm{J}})  &=& \hat{H}^{BO-J} = \hat{H}^{BO-L} +  \hat{H}^{BO-S} , \label{eq:Onr(J)} \\
 \hat{O}^{(2)}(\hat{H}^{\bm{J}}) &=& 0 , \label{eq:O2(J)}
\end{eqnarray}
\noindent where $\hat{H}^{BO-L}$ and $\hat{H}^{BO-S}$ are the $2 \times 2$ Born-Oppenheimer perturbation operators due to the effect of the nuclear rotation on the electronic dynamics, and are associated with the total electronic orbital (with respect to the molecular CM) and spin angular momenta\cite{Agus_PCCP_2016,aucar-aucar2019}.

The ($e$-$e$) contribution is then expanded according to Eq.~\eqref{eq:ee-expansion}, making the substitutions $\hat{H}^{P}\to\hat{H}^{PV}$ and $\hat{H}^{Q}\to\hat{H}^{\bm{J}}$. Further use of Eq.~\eqref{eq:M-energy} and retaining only the zeroth-order term yields the NR limit for $\bm{M}_N^{PV(e-e)}$,
%
%\begin{widetext}
\begin{eqnarray}\label{eq:LRESC-M-NR-2terms}
\bm{M}_N^{PV-NR}
% &=& - \hslash \frac{\partial^2}{\partial \bm{I}_N \partial \bm{J}} \langle \langle \hat{O}^{NR}(\hat{H}^{PV}) \; ; \; \hat{O}^{NR}(\hat{H}^{\bm{J}}) \rangle \rangle \nonumber \\
&=&
\frac{G_F\,\hslash}{4\sqrt{2}\,mc} \kappa_N
 \bigg[ \langle\langle \left\{ \bm{p} \, , \, \rho_N(\bm{r}) \right\} \, ; \, \bm{r}_{CM}  \times \bm{p} \rangle\rangle
\nonumber \\
&& \quad
+ \frac{\hslash^2}{2} \; \langle\langle
  \bm{\sigma} \times \left[ \bm{\nabla} \, , \, \rho_N(\bm{r}) \right]
 \, ; \, \bm{\sigma} \rangle \rangle
 \bigg] \cdot \bm{I}^{-1}.
\end{eqnarray}
%\end{widetext}
%
\noindent where $\bm{r}_{CM}=\bm{r}-\bm{R}_{CM}$ is the electronic position with respect to the molecular CM.

For molecular systems with closed-shell electronic structure, only the first term of Eq.~\eqref{eq:LRESC-M-NR-2terms} is non-vanishing. For this particular case, thus,
\begin{eqnarray}\label{eq:LRESC-M-NR}
\bm{M}_N^{PV-NR} &=&
\frac{G_F\,\hslash}{4\sqrt{2}\,mc} \kappa_N \,
 \langle\langle \left\{ \bm{p} \, , \, \rho_N(\bm{r}) \right\} \, ; \, \bm{r}_{CM}  \times \bm{p} \rangle\rangle \cdot \bm{I}^{-1}. \nonumber \\
\end{eqnarray}

As can be seen, the only difference between the NR contributions to the linear response functions for PV-NMR-shielding and PV-NSR tensors (Eqs.~\eqref{eq:sigma-PV-NR} and \eqref{eq:LRESC-M-NR}, respectively) stems in the electronic orbital angular momentum operator, which in the case of PV-NSR, has to be evaluated with respect to the molecular CM and not with respect to the gauge origin, as was previously reported\cite{Barra1986,Hobi2013}. Then, both properties are related by the well-known Ramsey-Flygare relation between their PC homologous $\bm{\sigma}_N^{NR-para}$ and $\bm{M}_N^{NR-elec}$\cite{Flygare64,Flygare74,Agus2012}.

\medskip

Besides, to get the zeroth- and first-order terms of the series expansion of the ($p$-$p$) contribution to $\bm{M}_N^{PV}$, Eqs.~\eqref{eq:pp-expansion} and~\eqref{eq:psiS-psiL} must be used in conjunction with the perturbative Hamiltonians $\hat{H}^{PV}$ and $\hat{H}^{\bm{J}}$, and also taking $\hat{X}(\hat{H}^{PV})$ from Eq.~\eqref{eq:XHPV}, and $\hat{X}(\hat{H}^{\bm{J}})$ from previous works\cite{Agus2012,Review-LRESC}, being
\begin{eqnarray}
\hat{X}(\hat{H}^{\bm{J}}) %&=& -2 \; \bm{\omega} \cdot \bm{J}_e - \frac{1}{2mc^2} \left[ \hat{H}^D , \bm{\omega} \cdot \bm{J}_e \right] \nonumber \\
&=& -2 \; \bm{\omega} \cdot \bm{J}_e - \frac{1}{2mc} \left[ \bm{\alpha} \cdot \bm{p} \, , \, \bm{\omega} \cdot \bm{J}_e \right].
\end{eqnarray}

Thus,
% %
% \begin{widetext}
% \begin{eqnarray}\label{eq:pp-expansion-SR-dev}
% \langle\langle \, \hat{H}^{PV} \, ; \, \hat{H}^{\bm{J}} \, \rangle\rangle^{(p-p)} &\simeq&
% \frac{G_F}{2\sqrt{2}} \frac{1}{2mc^2} \kappa_N
% \bigg[ \langle \Psi_0^L |
% \bm{\omega} \cdot \bm{J}_e \frac{\bm{\sigma} \cdot \bm{p}}{2mc} \bm{\sigma} \cdot \bm{I}_N \rho_N(\bm{r})
% | \Psi_0^L \rangle
% + \langle \Psi_0^L |
% \bm{\sigma} \cdot \bm{I}_N \rho_N(\bm{r}) \frac{\bm{\sigma} \cdot \bm{p}}{2mc} \bm{\omega} \cdot \bm{J}_e
% | \Psi_0^L \rangle  \nonumber \\
% %
% && + \langle \Psi_0^S |
% (-\bm{\omega} \cdot \bm{J}_e) \bm{\sigma} \cdot \bm{I}_N \rho_N(\bm{r})
% | \Psi_0^L \rangle
% + \langle \Psi_0^L |
% \bm{\sigma} \cdot \bm{I}_N \rho_N(\bm{r}) (-2 \bm{\omega} \cdot \bm{J}_e )
% | \Psi_0^S \rangle + \mathcal{O}(c^{-3}) \bigg].
% \end{eqnarray}
% \end{widetext}
%
% Further use of Eq.~\eqref{eq:psiS-psiL} yields the following expression for the zeroth- and first-order terms of the series expansion:
%
\begin{widetext}
\begin{eqnarray}
\langle\langle \, \hat{H}^{PV} \, ; \, \hat{H}^{\bm{J}} \, \rangle\rangle^{(p-p)} &\simeq& \frac{G_F}{2\sqrt{2}} \frac{1}{4m^2c^3} \kappa_N
\langle\tilde{\Psi}_0 | \, N
\Big(
\left[ \bm{\omega} \cdot \bm{J}_e \, , \, \bm{\sigma} \cdot \bm{p} \right] \bm{\sigma} \cdot \bm{I}_N \rho_N(\bm{r})
+ 2 \;
\bm{\sigma} \cdot \bm{I}_N \rho_N(\bm{r}) \left[ \bm{\sigma} \cdot \bm{p} \, , \, \bm{\omega} \cdot \bm{J}_e \right]
 %+ \mathcal{O}(c^{-2})
 \Big) N \,
 | \tilde{\Psi}_0 \rangle .
\end{eqnarray}
\end{widetext}
\noindent where $N$ is taken from Eq.~\eqref{eq:norm-phiL}.

Since $\left[ \bm{\sigma} \cdot \bm{p} \, , \, \bm{\omega} \cdot \bm{J}_e \right]=0$ holds\cite{Agus2012}, then
%
% \begin{equation}
% \langle\langle \, \hat{H}^{PV} \, ; \, \hat{H}^{\bm{J}} \, \rangle\rangle^{(p-p)}=0+\mathcal{O}(c^{-5})
% \end{equation}
% %
% \noindent and, finally, we arrive at
%
\begin{eqnarray}\label{eq:SR-pp-lresc-final}
\bm{M}_N^{PV(p-p)} &\simeq& -\hslash \frac{\partial^2}{\partial \bm{I}_N \partial \bm{J}} \left( \langle\langle \, \hat{H}^{PV} \, ; \, \hat{H}^{\bm{J}} \, \rangle\rangle^{(p-p)} \right) \nonumber \\
&=& 0 + \mathcal{O}(c^{-5}).
\end{eqnarray}

As the NR limit of $\bm{M}_N^{PV}$ is of order $c^{-1}$ [see Eq.~\eqref{eq:LRESC-M-NR-2terms}], then Eq.~\eqref{eq:SR-pp-lresc-final} implies that the NR limit and the leading order relativistic contributions to $\bm{M}_N^{PV(p-p)}$ vanish identically. The same holds for the ($p$-$p$) contribution to the PC-NSR tensor, $\bm{M}_N^{(p-p)}$\cite{Agus2012}.

\bigskip

Comparing Eqs.~\eqref{eq:sigma-PV-NR} and \eqref{eq:LRESC-M-NR} it is evident that both PV-NMR-shielding and PV-NSR tensors are closely related in the NR limit when the gauge origin of the external magnetic potential is placed at the molecular CM. This can be seen from
\begin{equation}
\bm{\sigma}^{PV-NR-para}_N = \frac{m_p}{m\hslash^2} \frac{1}{g_N} \bm{M}^{PV-NR}_N \cdot \bm{I}.
\end{equation}

In particular, their isotropic values are related according to
\begin{equation}
\sigma^{PV-NR}_{N,iso} = \frac{m_p}{m\hslash^2} \frac{1}{g_N} \bm{M}^{PV-NR}_{N,iso} \cdot \bm{I},
\end{equation}
\noindent since $\sigma^{PV-NR-dia}_{N,iso} = 0$.

\subsection{Application of the M-V model to PV properties}\label{sec:app-MV-PV}

Based on the above development, a close relationship among the PV-NSR and PV-NMR-shielding tensors are found. This relation is similar to the one previously found for their PC homologous\cite{Agus_PCCP_2016}.

For closed-shell molecules, the leading order relativistic contributions to the ($e$-$e$) part of both properties reduce to
\begin{widetext}
\begin{eqnarray}\label{eq:delta-sigma-M-ee-lresc}
\bm{\sigma}_N^{PV(e-e)}  - \bm{\sigma}_N^{PV-NR-para} &\simeq& \frac{\partial^2}{\partial \bm{\mu}_N \partial \bm{B}_0} \bigg[ \langle \langle \hat{O}^{NR}(\hat{H}^{PV}) \; ; \; \hat{O}^{(2)}(\hat{H}^{\bm{B}}) \rangle \rangle
+ \langle \langle \hat{O}^{(2)}(\hat{H}^{PV}) \; ; \; \hat{H}^{OZ} \rangle \rangle
+ \langle \langle \hat{O}^{NR}(\hat{H}^{PV}) \; ; \; \hat{D}_1 \; ; \; \hat{O}^{NR}(\hat{H}^{\bm{B}}) \rangle \rangle \bigg], \nonumber \\
\bm{M}_N^{PV(e-e)} - \bm{M}_N^{PV-NR} &\simeq& -\hslash \frac{\partial^2}{\partial \bm{I}_N \partial \bm{J}} \bigg[  \langle \langle \hat{O}^{(2)}(\hat{H}^{PV}) \; ; \; \hat{H}^{BO-L} \rangle \rangle
+ \langle \langle \hat{O}^{NR}(\hat{H}^{PV}) \; ; \; \hat{D}_1 \; ; \; \hat{O}^{NR}(\hat{H}^{\bm{J}}) \rangle \rangle \bigg] .
\end{eqnarray}
\end{widetext}

Therefore, we can extend the arguments given in Refs.~\citenum{Agus_PCCP_2016}, \citenum{Agus_Letter_2016} and \citenum{aucar-aucar2019} when analyzing the relativistic relation between the PC-NMR-shielding and PC-NSR tensors, and apply the M-V model proposed in Ref.~\citenum{aucar-aucar2019} to their PV homologous as follows
\begin{equation}\label{eq:M5-PV}
\bm{\sigma}_N^{PV,M-V} =
  \bm{\sigma}_N^{SR,PV}
+ \bm{\sigma}_N^{FA,PV}
+ \bm{\nu}^{S,PV}_N - \bm{\nu}^{FA-S,PV}_N,
\end{equation}
\noindent being
\begin{eqnarray}
 \bm{\sigma}_N^{SR,PV} &=&
\frac{m_p}{m\hslash^2} \frac{1}{g_N} \bm{M}^{PV}_N \cdot \bm{I}, \label{eq:sigmaSRPV} \\
\bm{\nu}_N^{S,PV} &=& \frac{G_F\,m_p}{2\sqrt{2}\,m \,\hslash \,c_0} \frac{\kappa_N}{g_N} \; \langle\langle \; \rho_N(\bm{r}) \; c \bm{\alpha} \; ; \;  \bm{S}_e \; \rangle\rangle . \label{eq:nuSPV}
\end{eqnarray}
In Eq.~\eqref{eq:M5-PV}, $\bm{\sigma}_N^{FA,PV}$ is the PV-NMR-shielding tensor of nucleus $N$ in a free atom [see Eq.~\eqref{eq:sigma-PV}], and $\bm{\nu}^{FA-S,PV}_N$ is obtained by employing Eq.~\eqref{eq:nuSPV} for nucleus $N$ in a free atom instead of the molecular system.

For free atoms with closed-shell electronic structures, the second and fourth terms at the right-hand-side of Eq.~\eqref{eq:M5-PV} are identically zero due to the spherical symmetry of the electronic density around the nucleus. For these systems, we can rewrite Eq.~\eqref{eq:M5-PV} as
\begin{equation}\label{eq:M5-PV-final}
\bm{\sigma}_N^{PV,M-V} =
  \bm{\sigma}_N^{SR,PV} + \bm{\nu}^{S,PV}_N.
\end{equation}

In Sec.~\ref{sec:res-disc} we analyze the validity of Eq.~\eqref{eq:M5-PV-final} for the isotropic PV-NMR shieldings and PV-NSR constants in a set of model molecular systems.

\section{\label{sec:comp-det}Computational details}

The calculations of $\bm{\sigma}^{PV}$ and $\bm{M}^{PV}$ were performed using a development version of the \textsc{Dirac} program package\cite{DIRAC21,dirac-paper}.

\begin{figure}[ht]
    \centering
    \includegraphics[width=\linewidth]{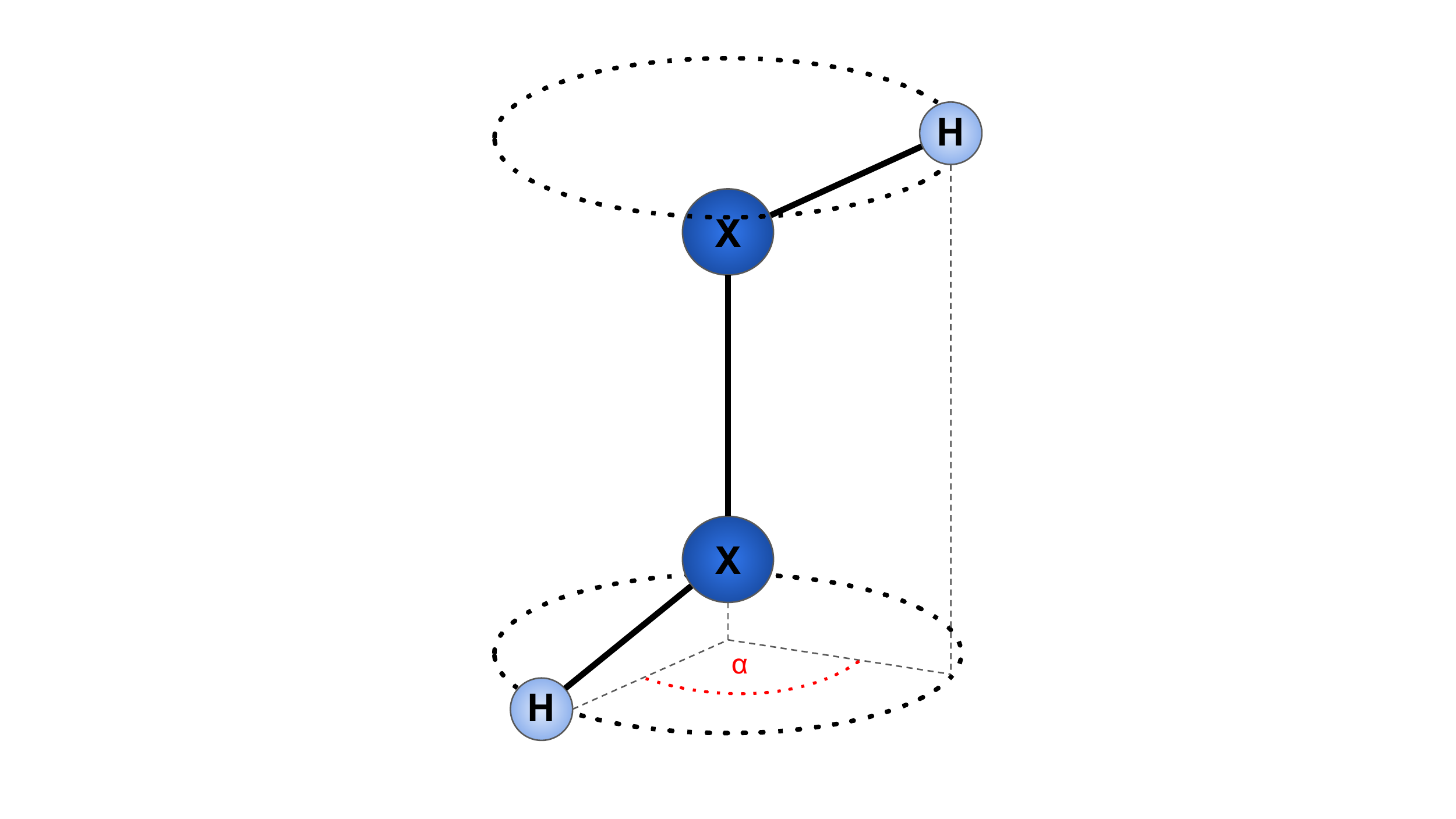}
    \caption{Schematic representation of the set of molecules studied in this work.}
    \label{fig:H2X2}
\end{figure}

The general $C_2$ symmetric $P$-conformation of the H$_2X_2$ molecular systems (with $X =$ $^{17}$O, $^{33}$S, $^{77}$Se, $^{125}$Te and $^{209}$Po) is illustrated in Fig.~\ref{fig:H2X2}. All systems have been analyzed with fixed bond lengths and H-$X$-$X$ angles, while the dihedral angle $\alpha$ varied between 0º and 180º in steps of 15º. In Table \ref{tab:geom} we list all internal structural parameters, taken from Ref.~\citenum{Laerdahl1999}, as well as the nuclear g-factors used in all calculations, extracted from Ref.~\citenum{Raghavan89}.

\begin{table}[htp]
\centering
\caption{\label{tab:geom} Structural parameters for the H$_2X_2$ molecules and nuclear g-factors ($X =$ O, S, Se, Te and Po) used in the calculations of $\bm{\sigma}_X^{PV}$ and $\bm{M}_X^{PV}$.}
\begin{tabular*}{\linewidth}{@{\extracolsep{\fill}} l ccccc}
\toprule
& $^{17}$O & $^{33}$S & $^{77}$Se & $^{125}$Te & $^{209}$Po \\
\midrule
$d (X$--$X) /$ \r{A}   & 1.49 & 2.055 & 2.48 & 2.84 & 2.91 \\[1ex]
$d (X$--H$) /$ \r{A} & 0.97 & 1.352 & 1.45 & 1.64 & 1.74 \\[1ex]
$\theta$(H$XX$)     & 100º & 92º   & 92º  & 92º  & 92º  \\[1ex]
$g_X$             & $-0.757516$ & $0.429214$ & $1.070084$ & $-1.77701$ & $1.376$\\
\bottomrule
\end{tabular*}
\end{table}

Dyall's uncontracted relativistic all-electron with extra diffuse functions quadruple-$\zeta$ quality basis sets (dyall.aae4z) were employed for all considered elements\cite{KD06,KD12,KD16}. This choice was made after a recent study on basis set analysis on PV-NSR tensors\cite{Aucar-PVSR2021,Aucar-PVSR2022}. The common-gauge-origin approach was used, and the gauge origin for the external magnetic potential was placed at the molecular CM.

The DC, SF and LL Hamiltonians were used\cite{Saue2005}. NR values of $\bm{\sigma}^{PV}$ and $\bm{M}^{PV}$ were obtained with the LL Hamiltonian. The SF contributions to PV-NMR-shielding and PV-NSR tensors were obtained by omitting all electronic spin-dependent (SD) integrals in perturbation-free calculations (by deleting in the matrix representation of the quaternion modified Dirac equation all the imaginary terms)\cite{Visscher2000}, and then keeping all 4c operators in the calculation of the linear response functions. By following these procedure, the relativistic contributions to both properties are scalar relativistic effects. On the other hand, the SD contributions to the PV-NMR-shielding and PV-NSR tensor were calculated as the difference between calculations employing the DC Hamiltonian and the SF contributions to $\bm{\sigma}^{PV}$ and $\bm{M}^{PV}$ calculated as described above. (SS$\mid$SS) integrals were neglected in both the perturbation-free (i.e., the self-consistent field procedure) and the linear response steps for DC and SF calculations and, instead, energy corrections were used to avoid their explicit calculation\cite{Visscher1997-SSSS}.

The small component basis sets were generated by means of the unrestricted kinetic balance (UKB) prescription in all the calculations,\cite{Aucar1999} but not for SF calculations, where the restricted kinetic balance (RKB) prescription was used. A spherically symmetric Gaussian-type nuclear charge density distribution model was used to calculate the unperturbed ground state energy and for describing the NSD-PV perturbed Hamiltonian\cite{Visscher1997}. It has been previously shown that the use of a finite nuclear model is important to adequately describe the PC-NSR tensors\cite{Agus_NChDE_RSC2018}.

All the linear response calculations presented in this work were carried out at the 4c RPA level of theory. DHF wave functions based on the DC, SF and LL Hamiltonians were employed, and in order to study the influence of electron correlation effects, the Dirac-Kohn-Sham-DFT wave functions were used as well. PBE0 NR exchange-correlation hybrid functional was used\cite{PBE0}, due to its good performance in the 4c calculations of PC-NSR constants, when compared with experimental values\cite{Agus_NChDE_RSC2018,Aucar_CH3X}. This functional has also been chosen to compare the present results with other previously published\cite{Aucar-PVSR2021,Aucar-PVSR2022}.

The response equations given in Eq.~\eqref{eq:resp-eq} were solved for the property gradient associated with the electronic orbital and spin angular momenta and for the external uniform magnetic field.

\section{\label{sec:res-disc}Results and discussion}

We first analyze the relation between the isotropic PV-NMR-shielding and PV-NSR constants of the $X$ nuclei in the H$_2X_2$ series of molecules (with $X =$ $^{17}$O, $^{33}$S, $^{77}$Se, $^{125}$Te and $^{209}$Po). These systems were used to test our theoretical developments as they were also the choice in several previous publications\cite{Laubender2003,LauBer06,Bast2006,Nahrwold2009}, particularly in Ref.~\citenum{Aucar-PVSR2021} where the relativistic theory of PV-NSR tensors was proposed. As the electronic correlation effects, taken as the difference between DFT- and DHF-based calculations, have been shown to be non negligible\cite{Nahrwold2009,Aucar-PVSR2021,Aucar-PVSR2022}, we present here results obtained using both methods.

\begin{figure*}[ht!]
\centering
\begin{subfigure}{.48\textwidth}
  \centering
  \includegraphics[width=\linewidth]{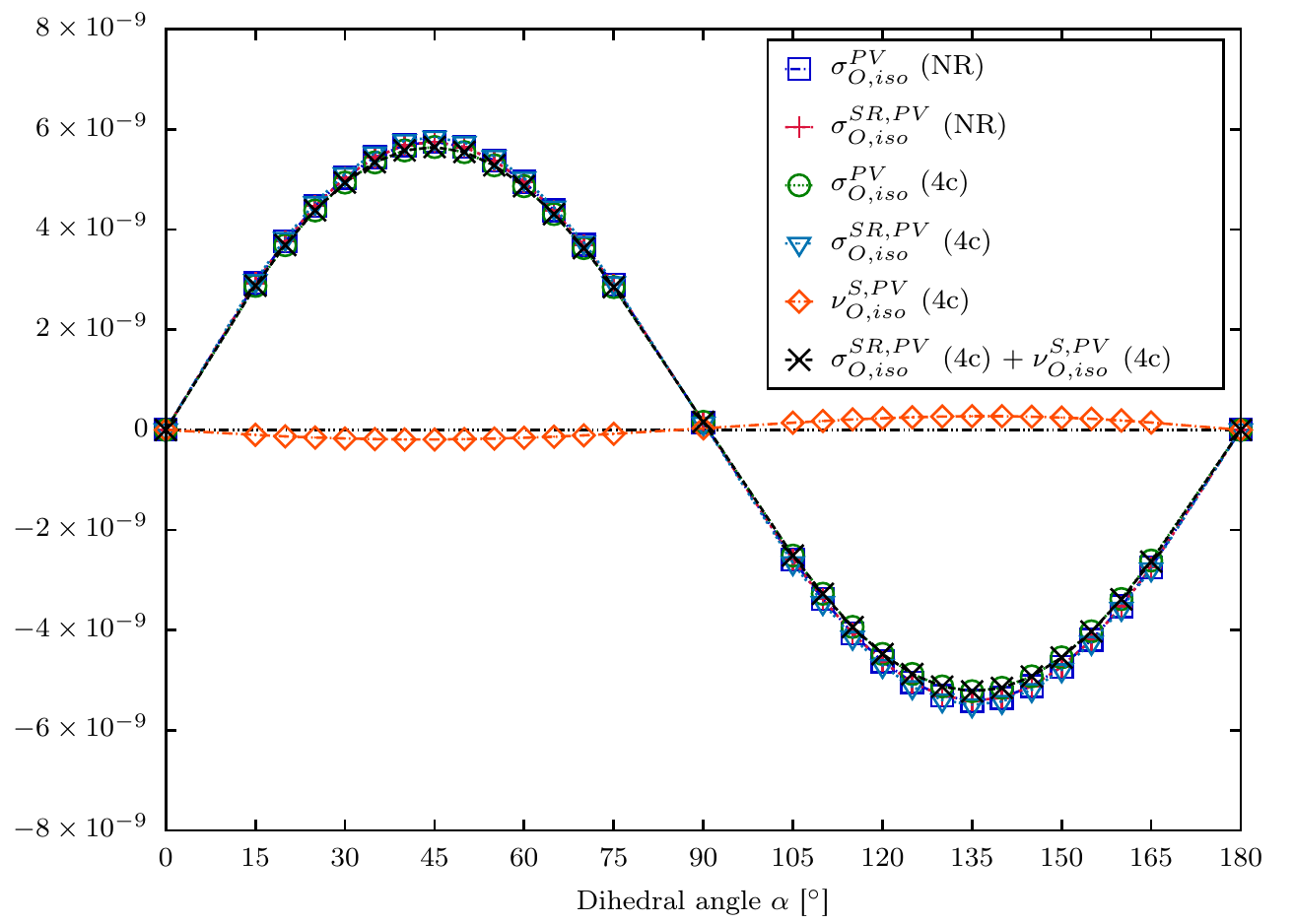}
  \caption{}
  \label{fig:O-dihedral-iso-RPA}
\end{subfigure}%
\begin{subfigure}{.48\textwidth}
  \centering
  \includegraphics[width=\linewidth]{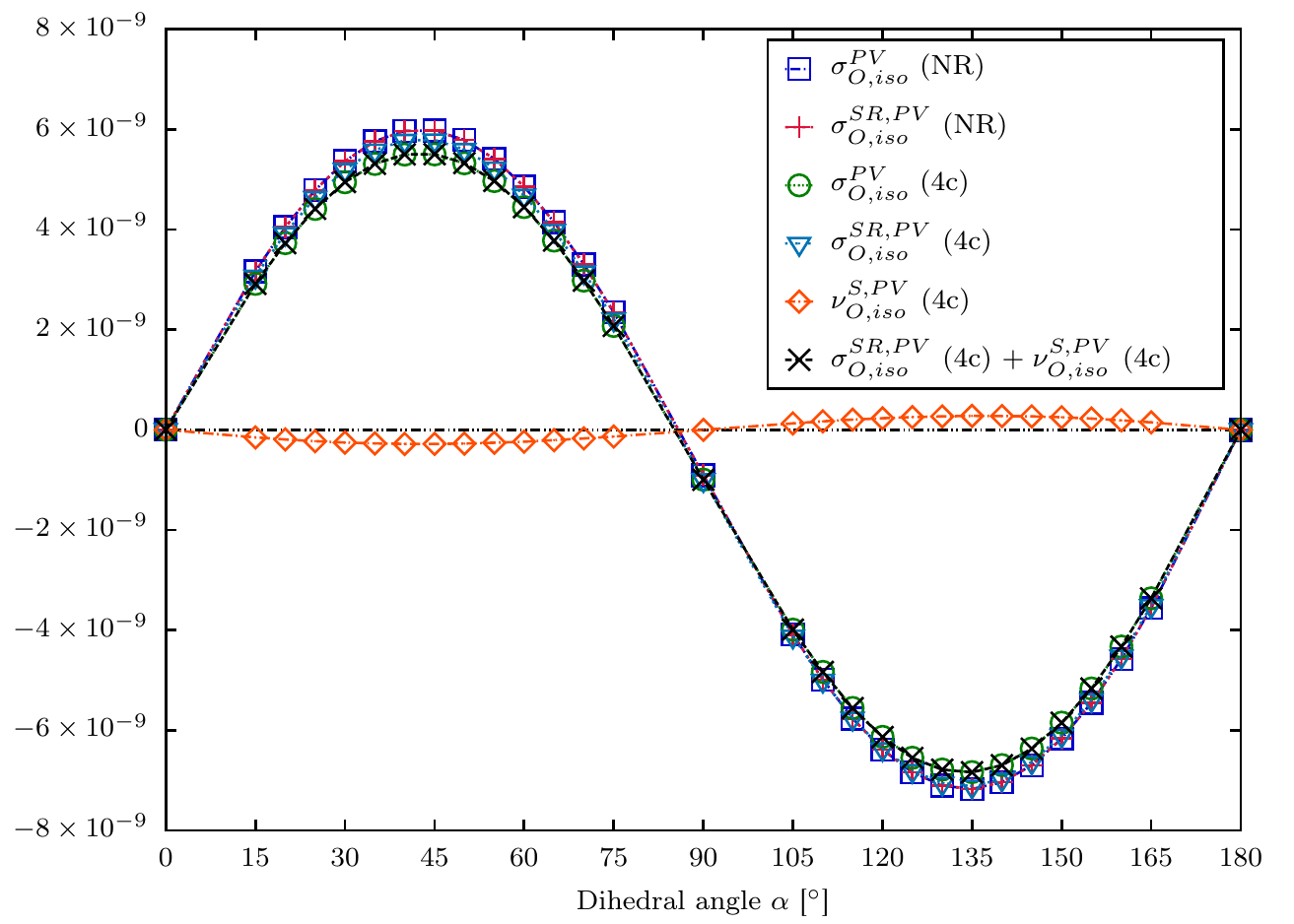}
  \caption{}
  \label{fig:O-dihedral-iso-PBE0}
\end{subfigure}
\caption{Values of $\sigma^{PV}_{iso}$, $\sigma^{SR,PV}_{iso}$ and $\nu^{S,PV}_{iso}$ for $^{17}$O in H$_2$O$_2$ (in ppm) at different dihedral angles employing the LL and DC Hamiltonians and UKB prescription at the (a) DHF and (b) DFT-PBE0 levels of approach, with dyall.aae4z basis set.}
\label{fig:O-dihedral-RPA-PBE0}
\end{figure*}

\begin{figure*}[ht!]
\centering
\begin{subfigure}{.48\textwidth}
  \centering
  \includegraphics[width=\linewidth]{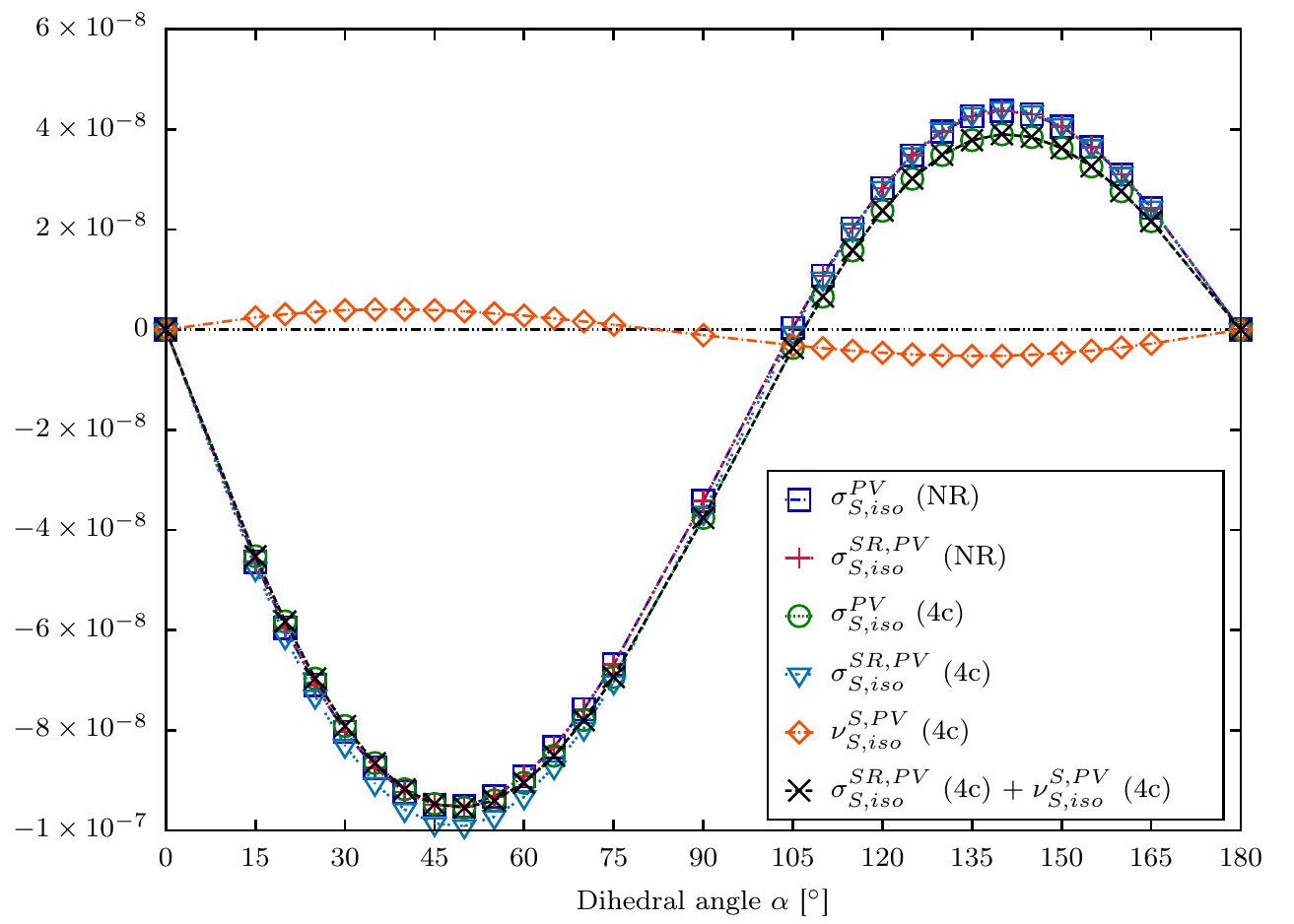}
  \caption{}
  \label{fig:S-dihedral-iso-RPA}
\end{subfigure}%
\begin{subfigure}{.48\textwidth}
  \centering
  \includegraphics[width=\linewidth]{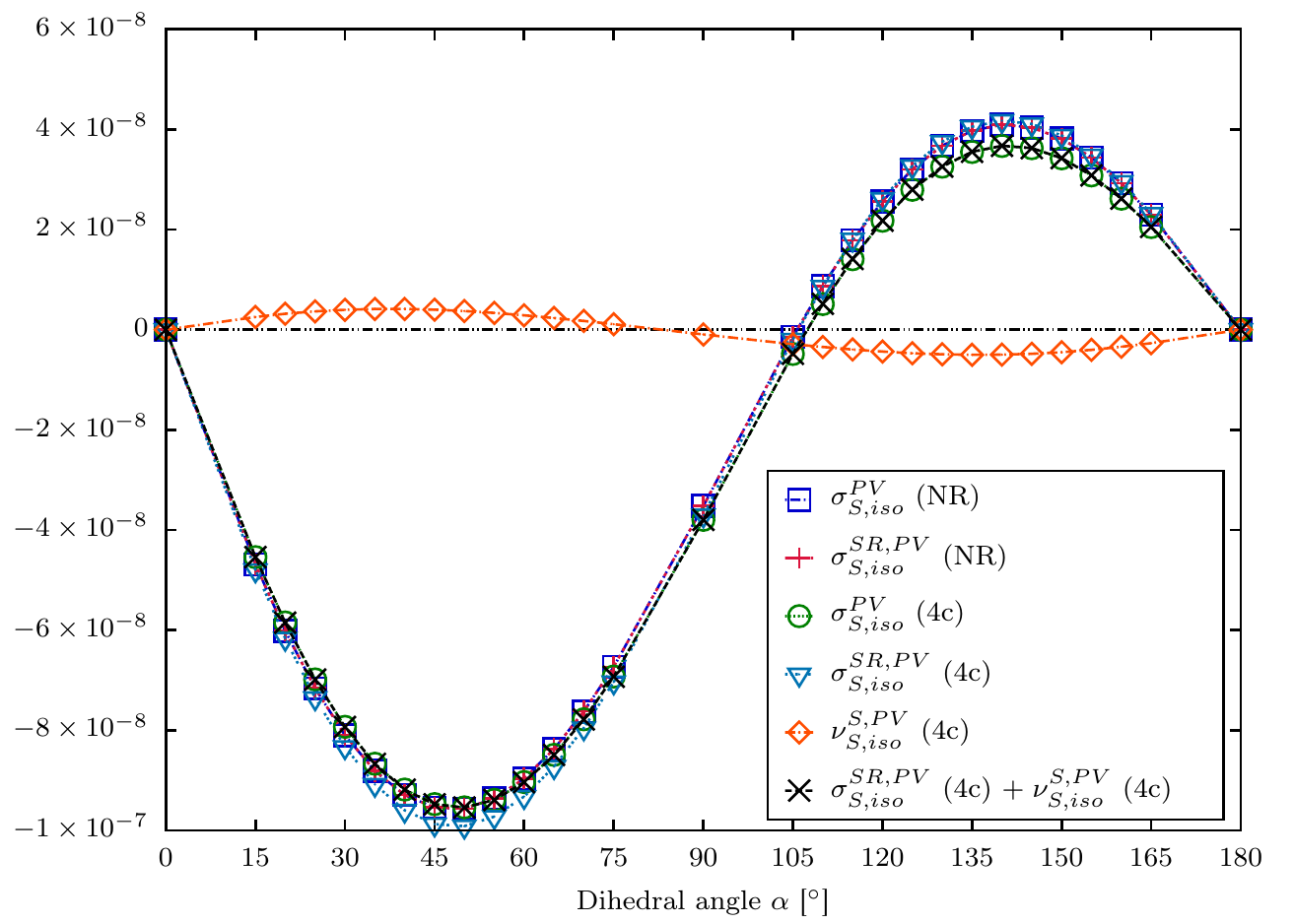}
  \caption{}
  \label{fig:S-dihedral-iso-PBE0}
\end{subfigure}
\caption{Same as Fig.~\ref{fig:O-dihedral-RPA-PBE0}, but for $^{33}$S in H$_2$S$_2$.}
\label{fig:S-dihedral-RPA-PBE0}
\end{figure*}

\begin{figure*}[ht!]
\centering
\begin{subfigure}{.48\textwidth}
  \centering
  \includegraphics[width=\linewidth]{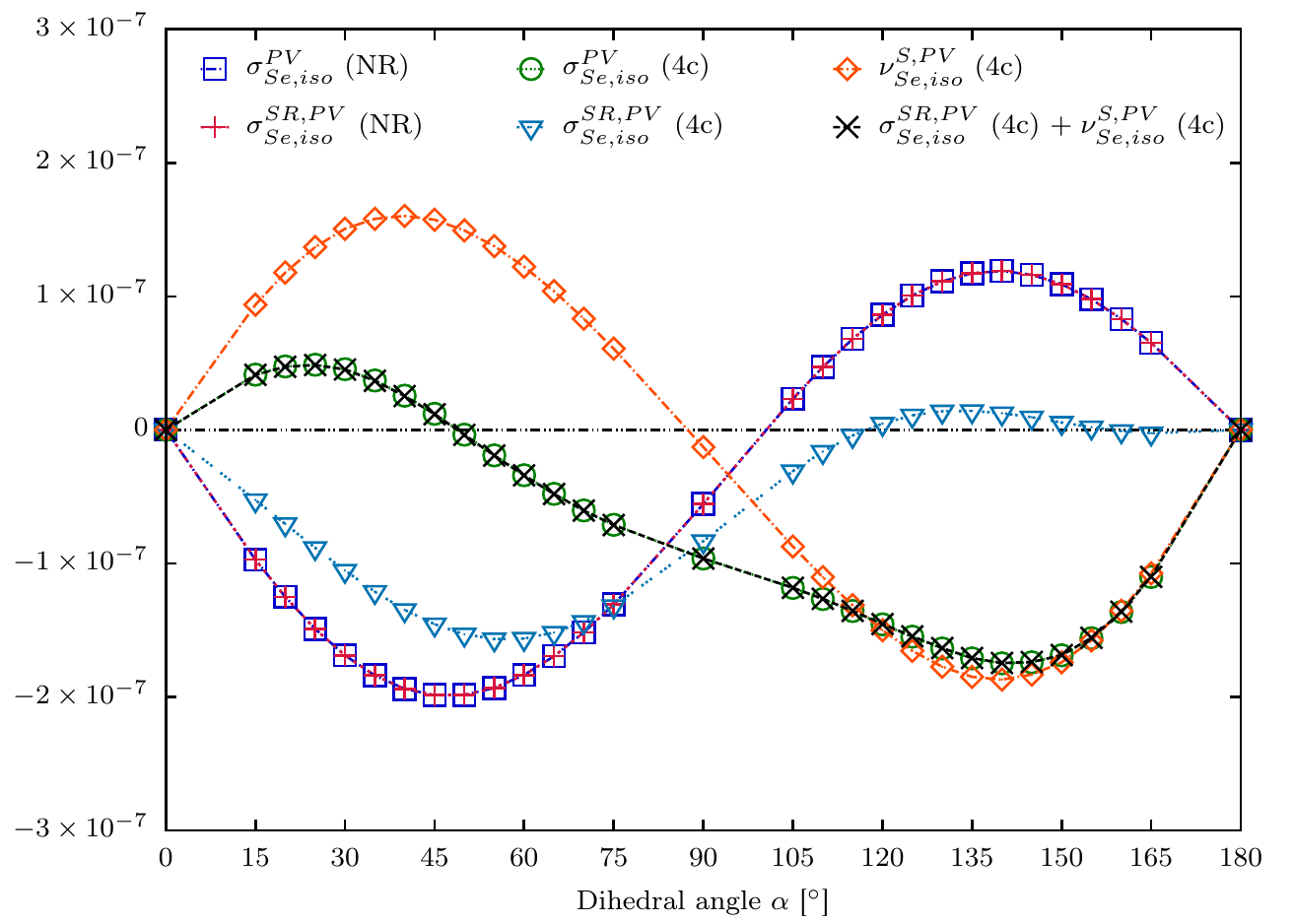}
  \caption{}
  \label{fig:Se-dihedral-iso-RPA}
\end{subfigure}%
\begin{subfigure}{.48\textwidth}
  \centering
  \includegraphics[width=\linewidth]{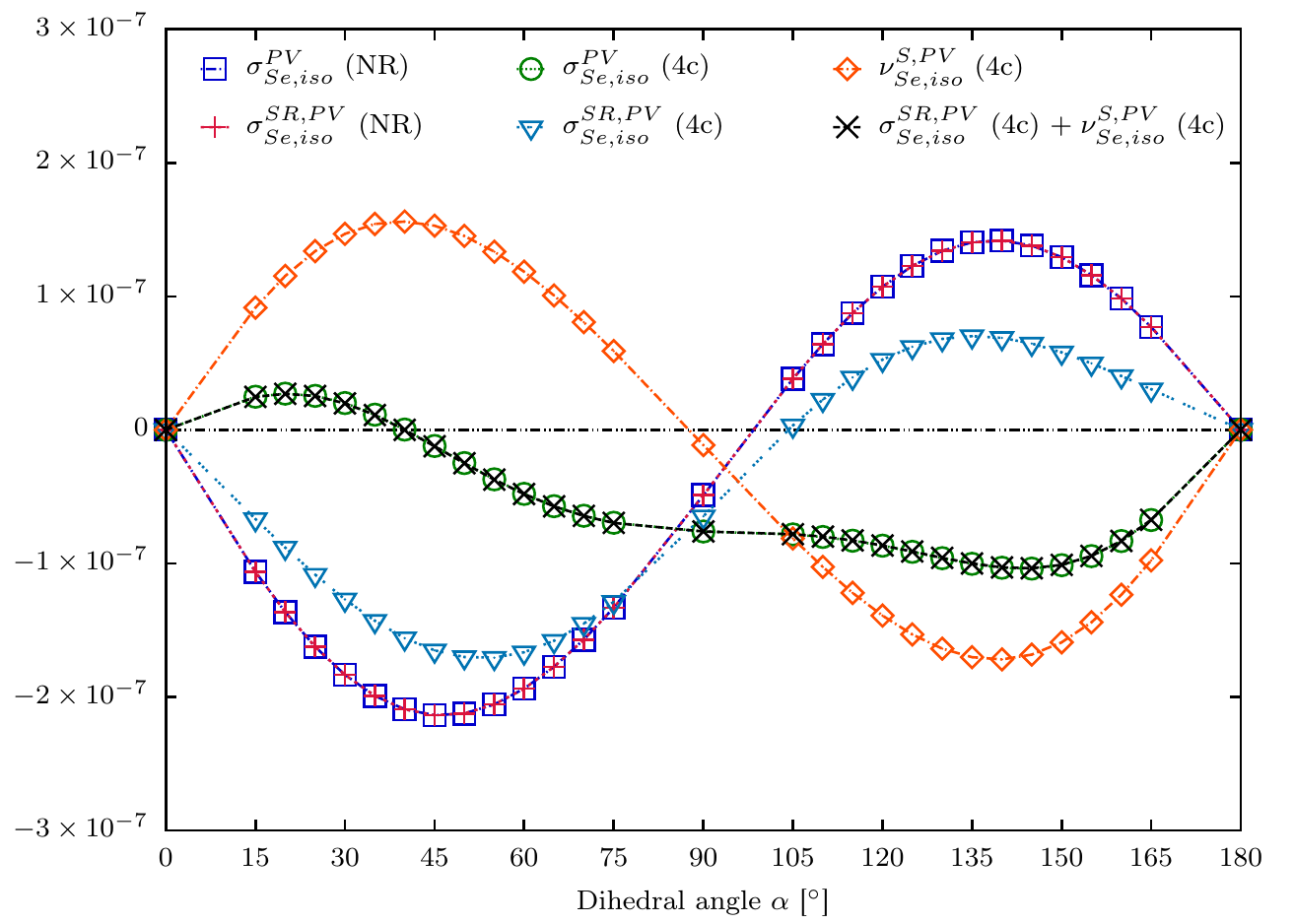}
  \caption{}
  \label{fig:Se-dihedral-iso-PBE0}
\end{subfigure}
\caption{Same as Fig.~\ref{fig:O-dihedral-RPA-PBE0}, but for $^{77}$Se in H$_2$Se$_2$.}
\label{fig:Se-dihedral-RPA-PBE0}
\end{figure*}

\begin{figure*}[ht!]
\centering
\begin{subfigure}{.48\textwidth}
  \centering
  \includegraphics[width=\linewidth]{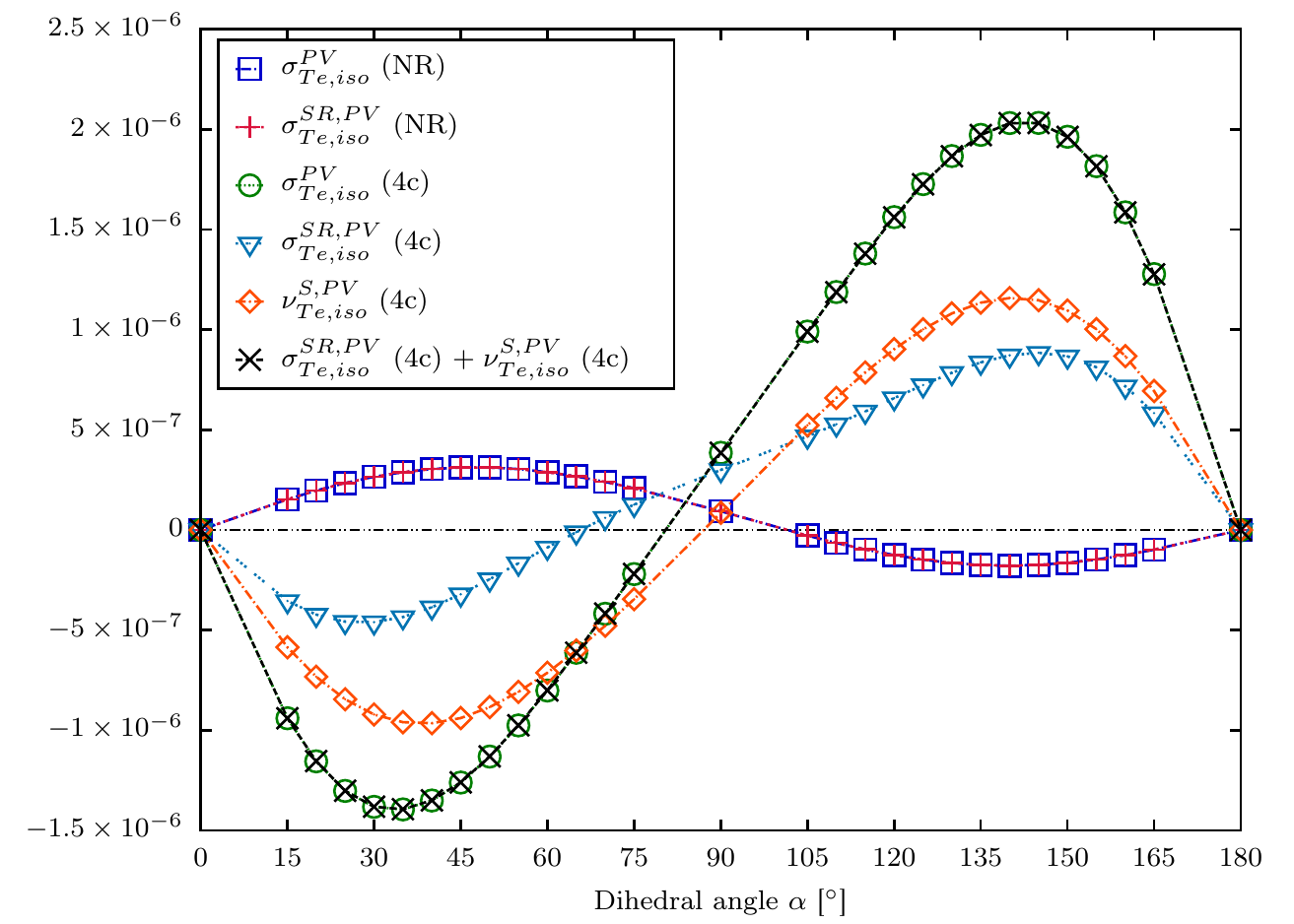}
  \caption{}
  \label{fig:Te-dihedral-iso-RPA}
\end{subfigure}%
\begin{subfigure}{.48\textwidth}
  \centering
  \includegraphics[width=\linewidth]{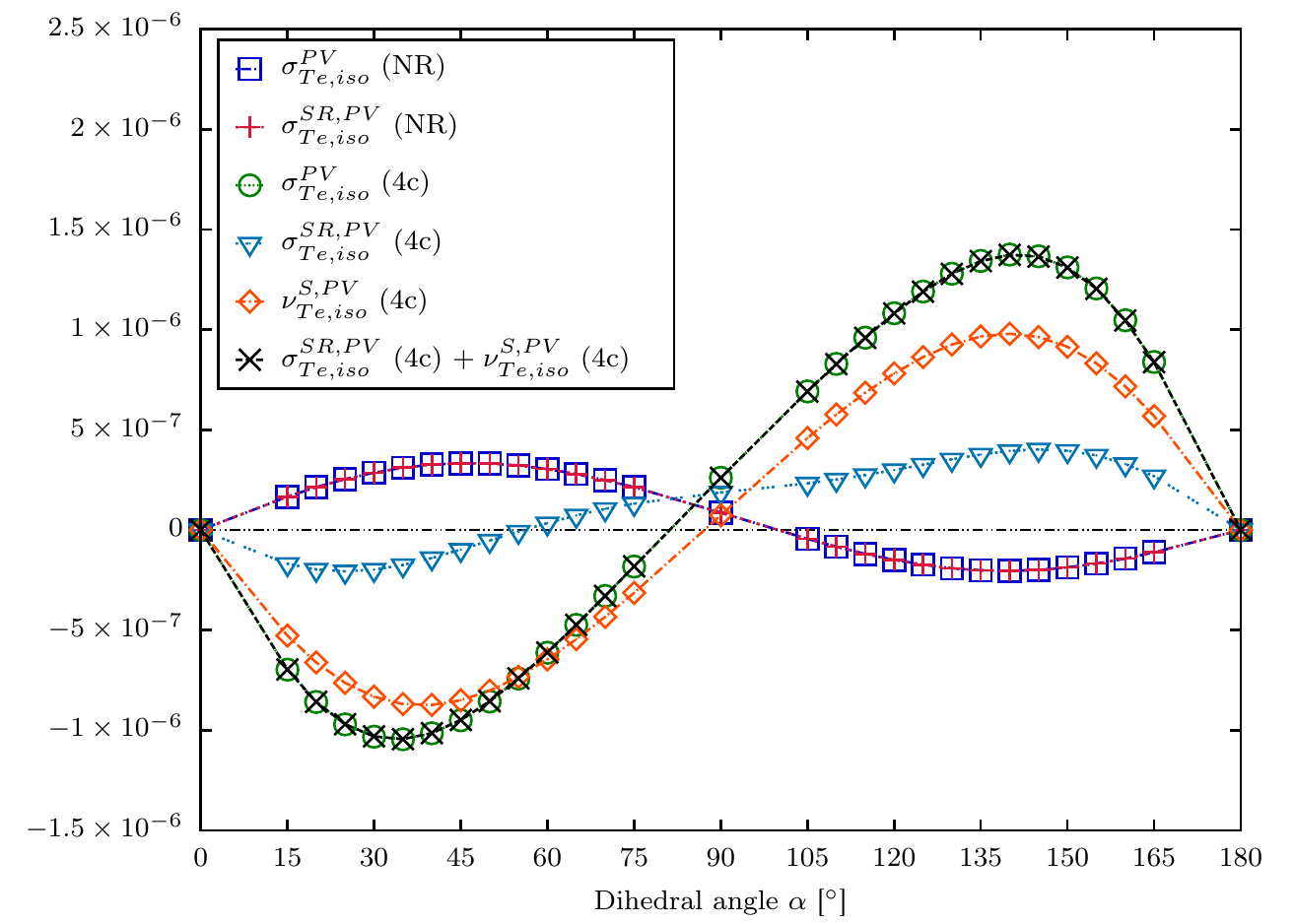}
  \caption{}
  \label{fig:Te-dihedral-iso-PBE0}
\end{subfigure}
\caption{Same as Fig.~\ref{fig:O-dihedral-RPA-PBE0}, but for $^{125}$Te in H$_2$Te$_2$.}
\label{fig:Te-dihedral-RPA-PBE0}
\end{figure*}

\begin{figure}[ht!]
\centering
  \includegraphics[width=\linewidth]{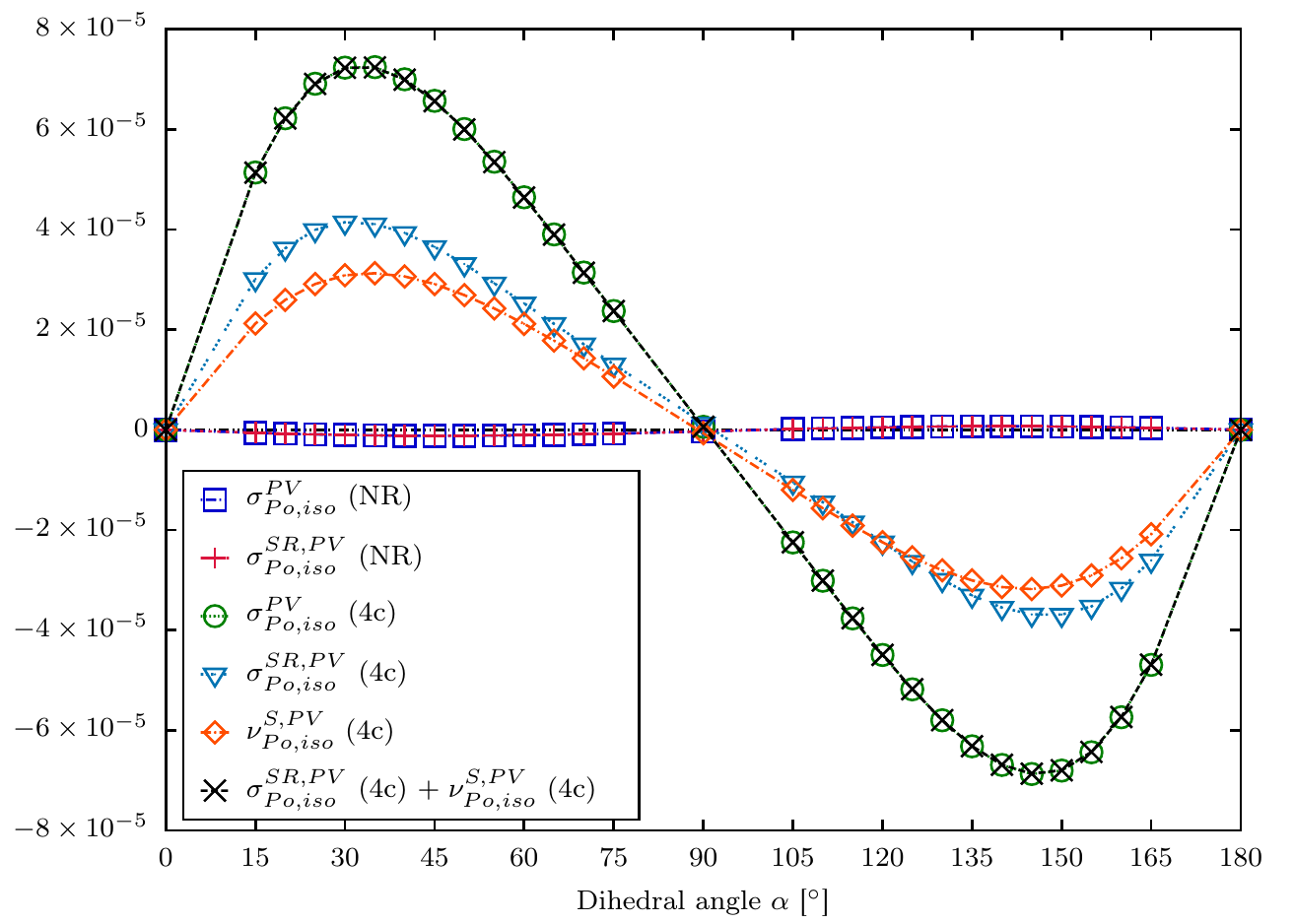}
\caption{Values of $\sigma^{PV}_{iso}$, $\sigma^{SR,PV}_{iso}$ and $\nu^{S,PV}_{iso}$ for $^{209}$Po in H$_2$Po$_2$ (in ppm) at different dihedral angles employing the LL and DC Hamiltonians and UKB prescription at the DFT-PBE0 level of approach, with dyall.aae4z basis set.}
\label{fig:Po-dihedral-PBE0}
\end{figure}

Figs.~\ref{fig:O-dihedral-RPA-PBE0} to \ref{fig:Po-dihedral-PBE0} show calculations of $\bm{\sigma}^{PV}_{iso}(X)$, $\bm{\sigma}^{SR,PV}_{iso}(X)$, $\bm{\nu}_N^{S,PV}(X)$ and $\bm{\sigma}^{SR,PV}_{iso}(X)+\bm{\nu}_N^{S,PV}(X)$ for $X =$ $^{17}$O, $^{33}$S, $^{77}$Se, $^{125}$Te and $^{209}$Po, respectively, at different dihedral angles. They were obtained using both, DHF and DFT-PBE0 wave functions. The NR limits of $\bm{\sigma}^{PV}_{iso}(X)$ and $\bm{\sigma}^{SR,PV}_{iso}(X)$, obtained employing the LL Hamiltonian, are also shown. For the 4c calculations, the DC Hamiltonian was employed instead. It was not possible to explore the whole range of dihedral angles in the case of polonium due to the presence of quasi-instabilities for the Kramer's restricted DHF wave functions with respect to the time reversal odd perturbations\cite{Aucar-PVSR2021}.

Relativistic effects do not play an important role in $\bm{\sigma}^{PV}_{iso}$ and $\bm{\sigma}^{SR,PV}_{iso}$ for the cases of oxygen and sulfur, as can be seen in Figs.~\ref{fig:O-dihedral-RPA-PBE0} and \ref{fig:S-dihedral-RPA-PBE0}, although they are not entirely negligible. It is known that they are mostly due to SD contributions\cite{Aucar-PVSR2021}. However, these effects turn relevant for a ``semi-heavy'' element as selenium, and they increase with increasing atomic number of element $X$, as expected. Figs.~\ref{fig:Se-dihedral-RPA-PBE0}-- \ref{fig:Po-dihedral-PBE0} show that the dependence of PV contributions to NMR shieldings and NSR tensors on the dihedral angle in 4c and NR calculations notably differ when going from selenium to polonium. At the same time, there is an expected equivalence in the NR values of $\sigma_{iso}^{PV}$ and $\sigma_{iso}^{SR,PV}$ for all dihedral angles. Nevertheless, this does not occur with their 4c counterparts. In fact, Fig.~\ref{fig:Se-dihedral-RPA-PBE0} evidences a different behavior in the relation of 4c values of $\sigma_{iso}^{PV}$ and $\sigma_{iso}^{SR,PV}$ with respect to the dihedral angle $\alpha$ in the case of selenium.

\begin{figure*}[ht!]
\centering
\begin{subfigure}{.48\textwidth}
  \centering
  \includegraphics[width=\linewidth]{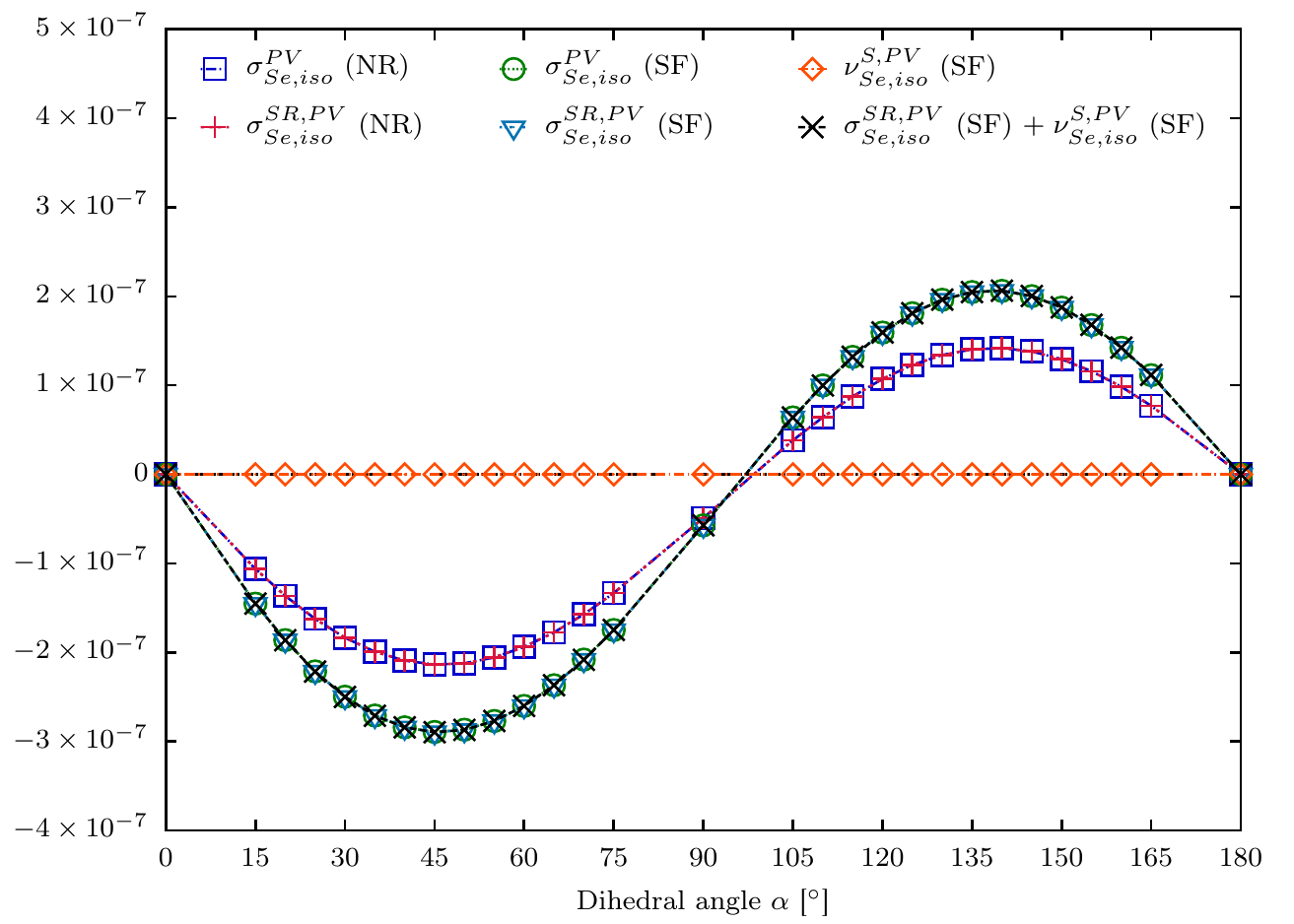}
  \caption{}
  \label{fig:Se-dihedral-SF-PBE0}
\end{subfigure}%
\begin{subfigure}{.48\textwidth}
  \centering
  \includegraphics[width=\linewidth]{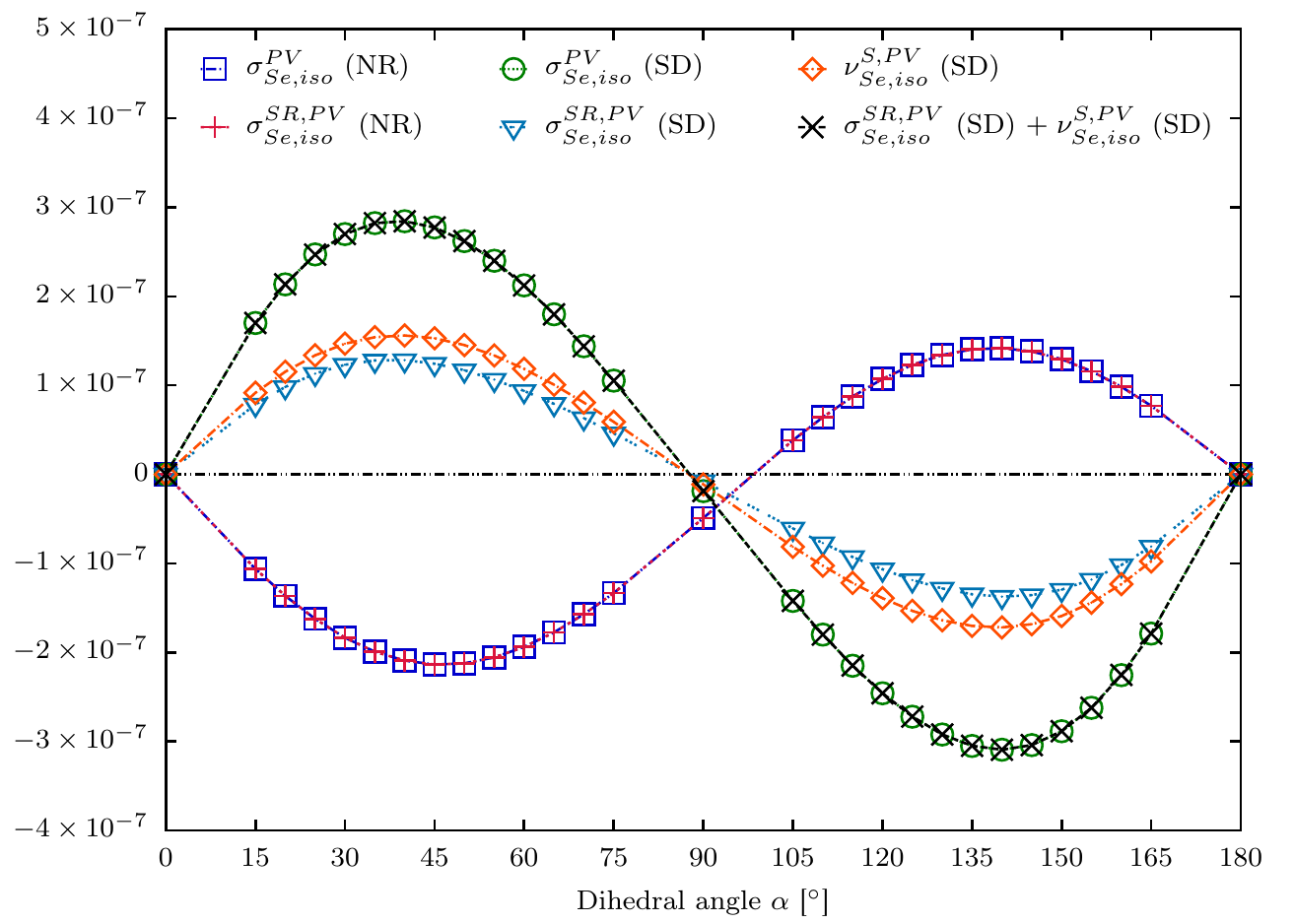}
  \caption{}
  \label{fig:Se-dihedral-SD-PBE0}
\end{subfigure}
\caption{Values of $\sigma^{PV}_{iso}$, $\sigma^{SR,PV}_{iso}$ and $\nu^{S,PV}_{iso}$ for $^{77}$Se in H$_2$Se$_2$ (in ppm) at different dihedral angles employing (a) SF and (b) SD contributions to the unperturbed Hamiltonian and RKB prescription at the DFT-PBE0 level of approach, with dyall.aae4z basis set. The SD contribution is given as the difference between DC and SF calculations.}
\label{fig:Se-dihedral-SF-SD-PBE0}
\end{figure*}

The observed trend for $\sigma^{PV}_{iso}$ shown in  Fig.~\ref{fig:Se-dihedral-RPA-PBE0} is explained by the partial cancellation of its SF and SD contributions, as displayed in Fig.~\ref{fig:Se-dihedral-SF-PBE0} (SF calculations) and \ref{fig:Se-dihedral-SD-PBE0} (SD values), where it can be seen that the dependence with the dihedral angle is closely proportional to $-\textnormal{sin}(2\alpha)$ and $\textnormal{sin}(2\alpha)$, respectively, with different roots and amplitudes.

\begin{figure*}[ht!]
\centering
\begin{subfigure}{.48\textwidth}
  \centering
  \includegraphics[width=\linewidth]{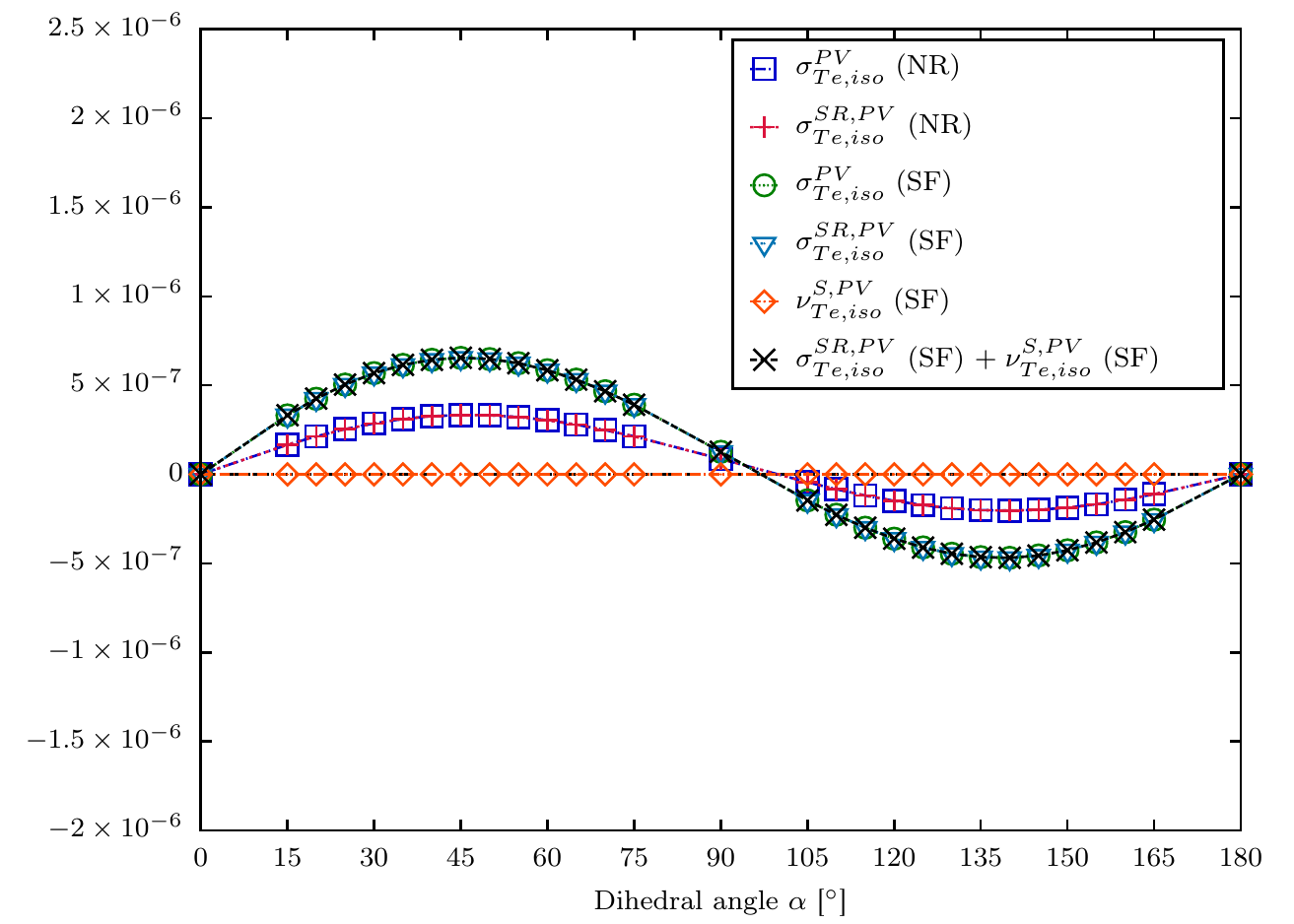}
  \caption{}
  \label{fig:Te-dihedral-SF-PBE0}
\end{subfigure}%
\begin{subfigure}{.48\textwidth}
  \centering
  \includegraphics[width=\linewidth]{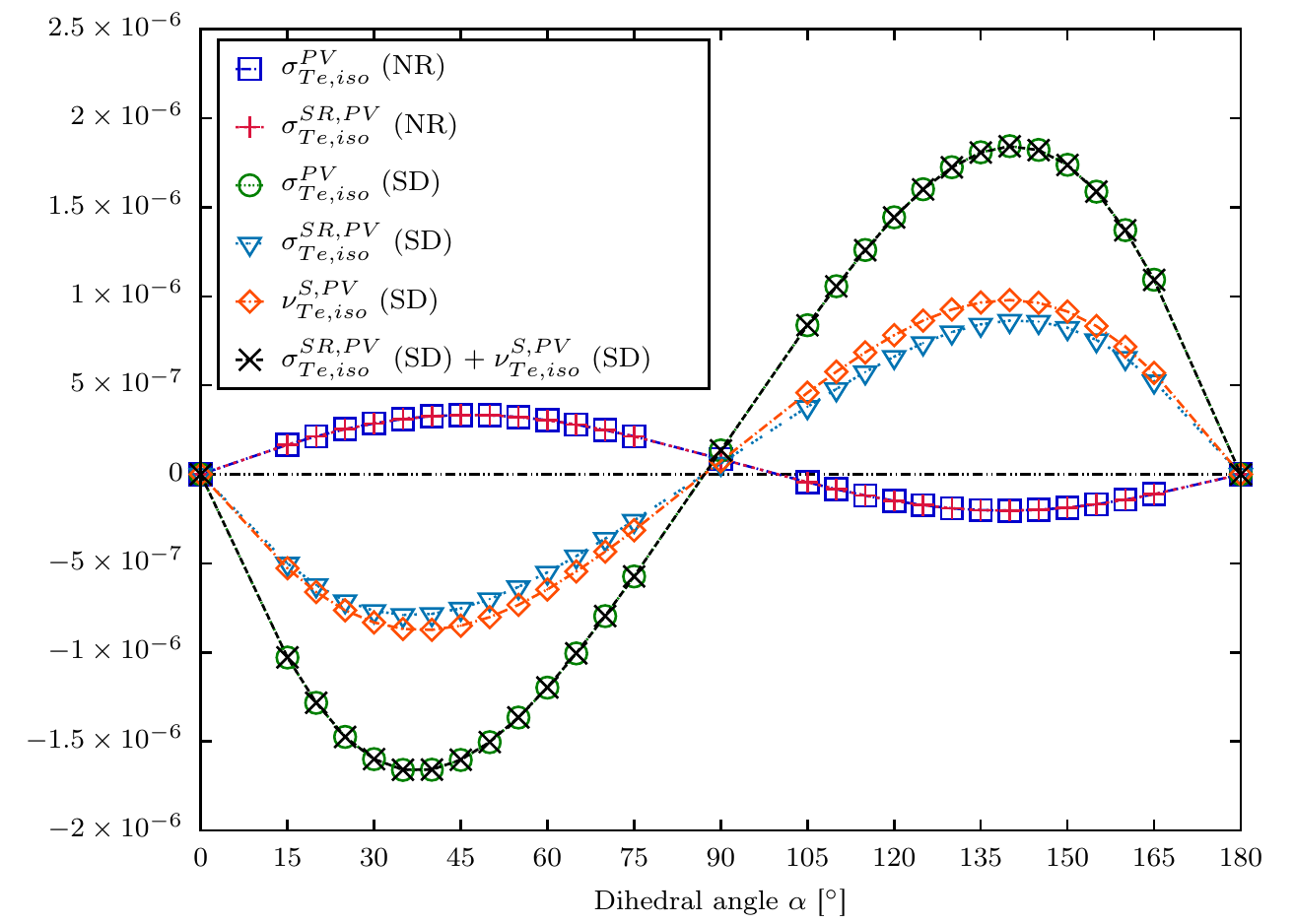}
  \caption{}
  \label{fig:Te-dihedral-SD-PBE0}
\end{subfigure}
\caption{Same as Fig.~\ref{fig:Se-dihedral-SF-SD-PBE0}, but for $^{125}$Te in H$_2$Te$_2$.}
\label{fig:Te-dihedral-SF-SD-PBE0}
\end{figure*}

\begin{figure*}[ht!]
\centering
\begin{subfigure}{.48\textwidth}
  \centering
  \includegraphics[width=\linewidth]{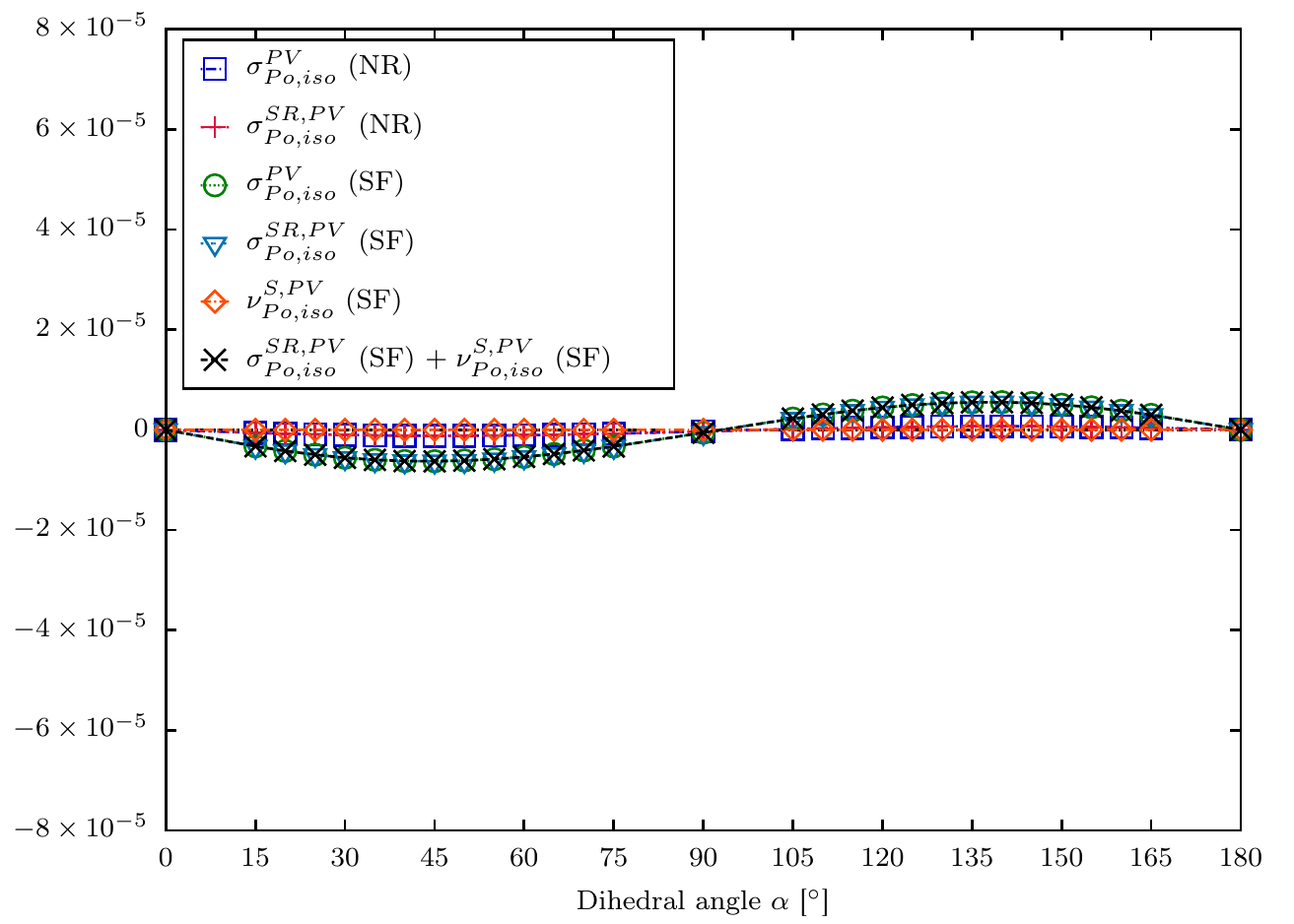}
  \caption{}
  \label{fig:Po-dihedral-SF-PBE0}
\end{subfigure}%
\begin{subfigure}{.48\textwidth}
  \centering
  \includegraphics[width=\linewidth]{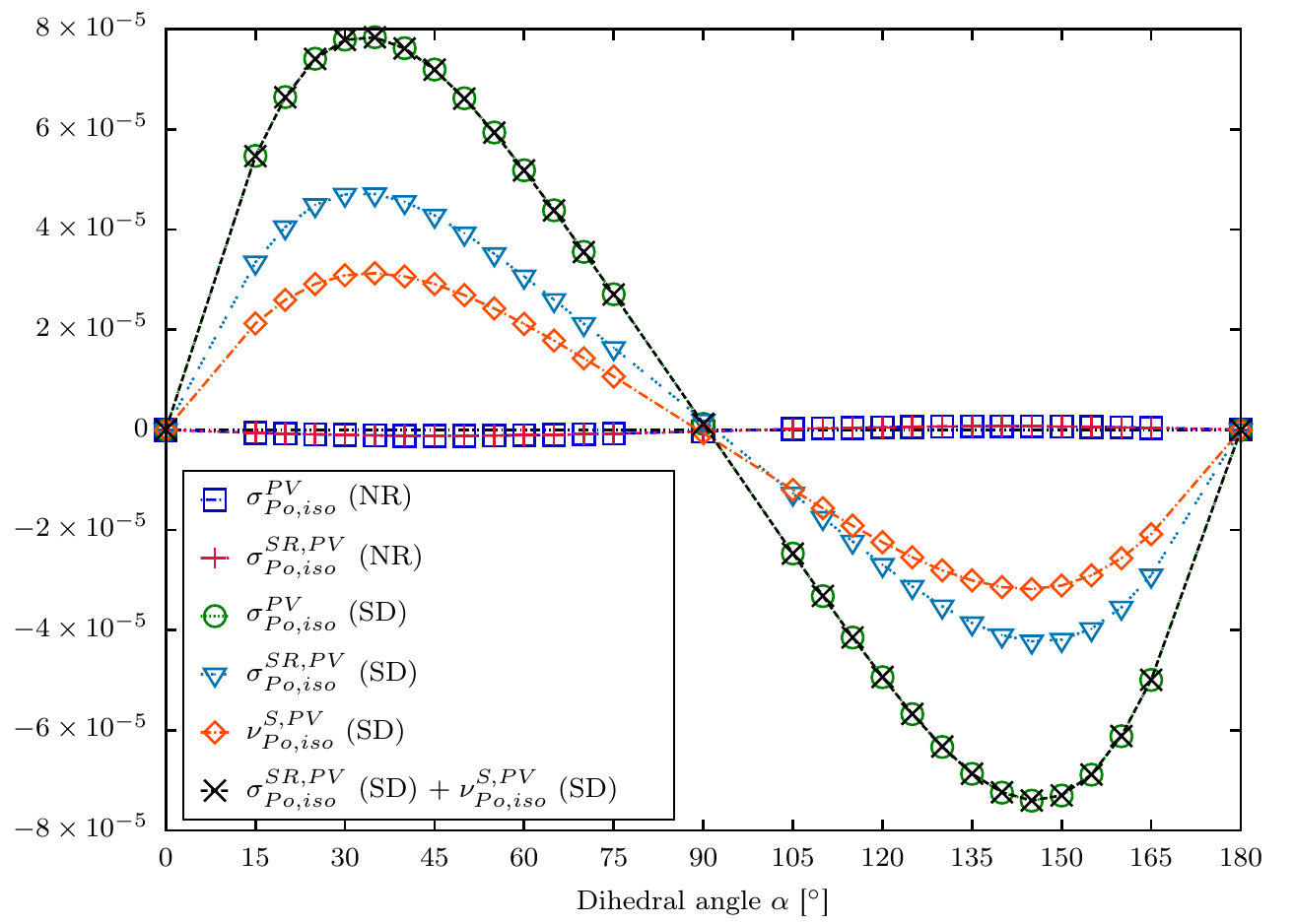}
  \caption{}
  \label{fig:Po-dihedral-SD-PBE0}
\end{subfigure}
\caption{Same as Fig.~\ref{fig:Se-dihedral-SF-SD-PBE0}, but for $^{209}$Po in H$_2$Po$_2$.}
\label{fig:Po-dihedral-SF-SD-PBE0}
\end{figure*}

\medskip

The main contribution of the present work is shown in Figs.~\ref{fig:Se-dihedral-RPA-PBE0} to \ref{fig:Po-dihedral-PBE0}, where it can be seen how the M-V model expressed in Eq.~\eqref{eq:M5-PV-final} ensures a good reproduction of the PV contributions to shieldings starting from PV-NSR values. The sum of $\sigma^{SR,PV}_{iso}$ and $\nu^{S,PV}_{iso}$ is in excellent agreement with the values of $\sigma^{PV}_{iso}$, meaning that the model being proposed here --which was originally developed in the context of PC properties\cite{Agus_Letter_2016,aucar-aucar2019}-- is well suited to the analysis of the PV-NMR shielding and PV-NSR constants.

\medskip

In Figs.~\ref{fig:Se-dihedral-SF-SD-PBE0}, \ref{fig:Te-dihedral-SF-SD-PBE0} and \ref{fig:Po-dihedral-SF-SD-PBE0} it can be seen that for selenium, tellurium and polonium the SD and SF contributions to $\sigma^{SR,PV}_{iso}$ and $\sigma^{PV}_{iso}$ vary with the dihedral angle as $\pm \textnormal{sin}(2\alpha)$, and the same occur for the lightest homologous oxygen and sulfur. In all these cases, SD contributions dominate the relativistic effects and even determine the qualitative behavior and sign of these properties. Their inclusion is, thus, mandatory. In the case of polonium, SD contributions are one order of magnitude larger than the SF ones, and the values of $\sigma^{SR,PV}_{iso}(^{209}\textnormal{Po})$ and $\sigma^{PV}_{iso}(^{209}\textnormal{Po})$ are almost completely due to these relativistic contributions. Therefore, they may be considered relativistic in nature, and their NR limits, negligible compared with their 4c counterparts.

We also note that the M-V model highlights the importance of the inclusion of SD contributions for the correct description of the PV-NMR-shielding tensor. The NR calculations of $\sigma^{PV}_{iso}$ and $\sigma^{SR,PV}_{iso}$, as well as the SF results of both properties, show that they agree with one another, meaning that the Ramsey-Flygare's relation holds for them\cite{Flygare74}. This is easily seen in all three Figs.~\ref{fig:Se-dihedral-SF-PBE0}, \ref{fig:Te-dihedral-SF-PBE0} and \ref{fig:Po-dihedral-SF-PBE0}. However, this is not the case for SD effects. In fact, Figs.~\ref{fig:Se-dihedral-SD-PBE0}, \ref{fig:Te-dihedral-SD-PBE0} and \ref{fig:Po-dihedral-SD-PBE0} show that, for a correct description of $\sigma^{PV}_{iso}$ starting from PV-NSR values, the inclusion of $\nu^{S,PV}_{iso}$ is mandatory. Calculated (or, eventually, measured) PV-NSR values can be used to obtain PV-NMR-shielding constants, in a completely similar way as for their PC homologous properties, i.e.~including the contributions given by $\nu^{S,PV}_{iso}$.

\section{\label{sec:conc}Conclusions}

We have analyzed the relation among the 4c relativistic nuclear-spin-dependent contributions of both, PV-NMR shieldings and PV-NSR constants using the H$_2X_2$ series of molecules (with $X=$ $^{17}$O, $^{33}$S, $^{77}$Se, $^{125}$Te and $^{209}$Po) as model systems.

The SF, SD and NR contributions to these properties show that the M-V model adequately reproduces the PV-NMR-shielding values starting from calculations of PV-NSR tensor elements. In particular, the (purely relativistic) $\nu^{S,PV}$ contribution to this model plays an important role for the SD contributions to the PV contributions to these properties.

Relativistic effects increase with the atomic number of the nucleus of interest, being not entirely negligible for oxygen and sulfur but becoming dominant for the heavier elements. In particular, the SD contributions are the main factors that determine the qualitative behavior of $\sigma^{PV}_{iso}$ and $\sigma^{SR,PV}_{iso}$ when we compare the NR values with its 4c values. They explain the change of signs. They also become the primary source of relativistic effects in both properties for tellurium and polonium.

In contrast, we found that the electronic correlation effects are of similar importance, but they are not much involved in the behavior of these properties as a function of the dihedral angle.

\begin{acknowledgments}
We acknowledge partial support from FONCYT by grant PICT 2016-2936.
IAA acknowledges partial support from FONCYT by grant PICT-2020-SerieA-00052 as well. We would also like to thank the Institute for Modeling and Innovation on Technologies (IMIT) of the National Scientific and Technical Research Council (CONICET) and the Northeastern University of Argentina (UNNE) for the support and for providing access to the IMIT high performance computing cluster.
\end{acknowledgments}

\section*{\label{sec:data}Data Availability}
The data that support the findings of this study are available from the corresponding author upon reasonable request.

\vspace{1cm}

\appendix

\section{Application of the LRESC model to $\hat{H}^{\bm{B}}$ and $\hat{H}^{\bm{J}}$}\label{sec:appendix}

The 4c perturbed operators due to the interaction of the electronic system with an external magnetic field and its coupling to the molecular rotation are given in Eqs.~\eqref{eq:H-B-SI} and \eqref{eq:H_J}.

Applying  Eqs.~\eqref{eq:matrixelementsH} and \eqref{eq:Onr2} to the magnetic perturbation of Eq.~\eqref{eq:H-B-SI}, we can write (in SI units)\cite{melo-lresc,Review-LRESC},
\begin{eqnarray}
 \hat{O}^{NR} (\hat{H}^{\bm{B}}) &=& \hat{H}^{OZ} + \hat{H}^{SZ} \nonumber \\
 \hat{O}^{(2)}(\hat{H}^{\bm{B}}) &=& \hat{H}^{OZ-K} + \hat{H}^{SZ-K} + \hat{H}^{B-SO},
\end{eqnarray}
\noindent where the NR perturbations are given by\cite{melo-lresc,Review-LRESC}
\begin{eqnarray}
 \hat{H}^{OZ}   &=& \frac{e}{2m} \bm{B}_0 \cdot \bm{L} = \frac{e}{2m} \bm{B}_0 \cdot \left[ \bm{r}_{GO} \times \bm{p} \right] \label{eq:H(OZ)} \nonumber \\
 \hat{H}^{SZ}   &=& \frac{e}{m} \bm{B}_0 \cdot \bm{S} = \frac{e\hslash}{2m} \bm{B}_0 \cdot \bm{\sigma}, \label{eq:H(SZ)}
\end{eqnarray}
\noindent with $\bm{L}$ and $\bm{S}$ being the $2\times2$ electronic orbital and spin angular momenta, respectively, and where the relativistic leading order perturbation operators are\cite{Review-LRESC}
\begin{eqnarray}
 \hat{H}^{OZ-K} &=& -\frac{1}{4m^2c^2} \left\{ p^2 , \hat{H}^{OZ} \right\} = -\frac{e}{8m^3c^2} \left\{ p^2 , \bm{B}_0 \cdot \bm{L} \right\} \label{eq:H(OZK)} \nonumber \\
 \hat{H}^{SZ-K} &=& -\frac{e}{4m^3c^2} \bm{B}_0 \cdot \left[ 3\bm{S}p^2 - (\bm{S}\cdot\bm{p}) \bm{p} \right] \label{eq:H(SZK)} \nonumber \\
 \hat{H}^{B-SO} &=& \frac{e}{2m^2c^2} \bm{B}_0 \cdot \left[ (\bm{r}\cdot\bm{\nabla})V_C \; \bm{S} - (\bm{r}\cdot\bm{S})\bm{\nabla}V_C) \right]. \label{eq:H(BSO)}
\end{eqnarray}

\bigskip

Similarly, for the operator $\hat{H}^{\bm{J}}$ we have\cite{Agus2012}
\begin{eqnarray}
 \hat{O}^{NR} (\hat{H}^{\bm{J}}) &=& \hat{H}^{BO-L} + \hat{H}^{BO-S} \nonumber \\
 \hat{O}^{(2)}(\hat{H}^{\bm{J}}) &=& 0,
\end{eqnarray}
\noindent with
\begin{eqnarray}
 \hat{H}^{BO-L}   &=& - \bm{\omega} \cdot \bm{L} = - \bm{\omega} \cdot \left[ \bm{r}_{CM} \times \bm{p} \right] \label{eq:H(BO-L)} \nonumber \\
 \hat{H}^{BO-S}   &=& - \bm{\omega} \cdot \bm{S} = - \frac{\hslash}{2} \bm{\omega} \cdot \bm{\sigma}.
\end{eqnarray}

\bibliography{PV-shi-LRESC}

\begin{thebibliography}{64}%
\makeatletter
\providecommand \@ifxundefined [1]{%
 \@ifx{#1\undefined}
}%
\providecommand \@ifnum [1]{%
 \ifnum #1\expandafter \@firstoftwo
 \else \expandafter \@secondoftwo
 \fi
}%
\providecommand \@ifx [1]{%
 \ifx #1\expandafter \@firstoftwo
 \else \expandafter \@secondoftwo
 \fi
}%
\providecommand \natexlab [1]{#1}%
\providecommand \enquote  [1]{``#1''}%
\providecommand \bibnamefont  [1]{#1}%
\providecommand \bibfnamefont [1]{#1}%
\providecommand \citenamefont [1]{#1}%
\providecommand \href@noop [0]{\@secondoftwo}%
\providecommand \href [0]{\begingroup \@sanitize@url \@href}%
\providecommand \@href[1]{\@@startlink{#1}\@@href}%
\providecommand \@@href[1]{\endgroup#1\@@endlink}%
\providecommand \@sanitize@url [0]{\catcode `\\12\catcode `\$12\catcode
  `\&12\catcode `\#12\catcode `\^12\catcode `\_12\catcode `\%12\relax}%
\providecommand \@@startlink[1]{}%
\providecommand \@@endlink[0]{}%
\providecommand \url  [0]{\begingroup\@sanitize@url \@url }%
\providecommand \@url [1]{\endgroup\@href {#1}{\urlprefix }}%
\providecommand \urlprefix  [0]{URL }%
\providecommand \Eprint [0]{\href }%
\providecommand \doibase [0]{https://doi.org/}%
\providecommand \selectlanguage [0]{\@gobble}%
\providecommand \bibinfo  [0]{\@secondoftwo}%
\providecommand \bibfield  [0]{\@secondoftwo}%
\providecommand \translation [1]{[#1]}%
\providecommand \BibitemOpen [0]{}%
\providecommand \bibitemStop [0]{}%
\providecommand \bibitemNoStop [0]{.\EOS\space}%
\providecommand \EOS [0]{\spacefactor3000\relax}%
\providecommand \BibitemShut  [1]{\csname bibitem#1\endcsname}%
\let\auto@bib@innerbib\@empty
%</preamble>
\bibitem [{\citenamefont {Lee}\ and\ \citenamefont {Yang}(1956)}]{LeeYang56}%
  \BibitemOpen
  \bibfield  {author} {\bibinfo {author} {\bibfnamefont {T.~D.}\ \bibnamefont
  {Lee}}\ and\ \bibinfo {author} {\bibfnamefont {C.~N.}\ \bibnamefont {Yang}},\
  }\bibfield  {title} {\bibinfo {title} {{Question of Parity Conservation in
  Weak Interactions}},\ }\href {https://doi.org/10.1103/PhysRev.104.254}
  {\bibfield  {journal} {\bibinfo  {journal} {Phys.~Rev.}\ }\textbf {\bibinfo
  {volume} {104}},\ \bibinfo {pages} {254} (\bibinfo {year}
  {1956})}\BibitemShut {NoStop}%
\bibitem [{\citenamefont {Wu}\ \emph {et~al.}(1957)\citenamefont {Wu},
  \citenamefont {Ambler}, \citenamefont {Hayward}, \citenamefont {Hoppes},\
  and\ \citenamefont {Hudson}}]{Wu57}%
  \BibitemOpen
  \bibfield  {author} {\bibinfo {author} {\bibfnamefont {C.~S.}\ \bibnamefont
  {Wu}}, \bibinfo {author} {\bibfnamefont {E.}~\bibnamefont {Ambler}}, \bibinfo
  {author} {\bibfnamefont {R.~W.}\ \bibnamefont {Hayward}}, \bibinfo {author}
  {\bibfnamefont {D.~D.}\ \bibnamefont {Hoppes}},\ and\ \bibinfo {author}
  {\bibfnamefont {R.~P.}\ \bibnamefont {Hudson}},\ }\bibfield  {title}
  {\bibinfo {title} {{Experimental Test of Parity Conservation in Beta
  Decay}},\ }\href {https://doi.org/10.1103/PhysRev.105.1413} {\bibfield
  {journal} {\bibinfo  {journal} {Phys.~Rev.}\ }\textbf {\bibinfo {volume}
  {105}},\ \bibinfo {pages} {1413} (\bibinfo {year} {1957})}\BibitemShut
  {NoStop}%
\bibitem [{\citenamefont {Bouchiat}(2012)}]{Bouchiat12}%
  \BibitemOpen
  \bibfield  {author} {\bibinfo {author} {\bibfnamefont {M.~A.}\ \bibnamefont
  {Bouchiat}},\ }\bibfield  {title} {\bibinfo {title} {{Atomic parity
  violation. Early days, present results, prospects}},\ }\href
  {https://doi.org/10.1393/ncc/i2012-11269-6} {\bibfield  {journal} {\bibinfo
  {journal} {Il Nuovo Cimento}\ }\textbf {\bibinfo {volume} {35}},\ \bibinfo
  {pages} {78} (\bibinfo {year} {2012})}\BibitemShut {NoStop}%
\bibitem [{\citenamefont {Zel’dovich}\ \emph {et~al.}(1977)\citenamefont
  {Zel’dovich}, \citenamefont {Saakyan},\ and\ \citenamefont
  {Sobel’man}}]{Zeldovich77}%
  \BibitemOpen
  \bibfield  {author} {\bibinfo {author} {\bibfnamefont {Y.~B.}\ \bibnamefont
  {Zel’dovich}}, \bibinfo {author} {\bibfnamefont {D.}~\bibnamefont
  {Saakyan}},\ and\ \bibinfo {author} {\bibfnamefont {I.}~\bibnamefont
  {Sobel’man}},\ }\bibfield  {title} {\bibinfo {title} {{Energy difference
  between right-hand and left-hand molecules, due to parity nonconservation in
  weak interactions of electrons with nuclei}},\ }\href
  {http://jetpletters.ru/ps/1388/article_21066.shtml} {\bibfield  {journal}
  {\bibinfo  {journal} {JETP Lett.}\ }\textbf {\bibinfo {volume} {25}},\
  \bibinfo {pages} {94} (\bibinfo {year} {1977})}\BibitemShut {NoStop}%
\bibitem [{\citenamefont {Quack}\ and\ \citenamefont
  {Stohner}(2005)}]{Quack05}%
  \BibitemOpen
  \bibfield  {author} {\bibinfo {author} {\bibfnamefont {M.}~\bibnamefont
  {Quack}}\ and\ \bibinfo {author} {\bibfnamefont {J.}~\bibnamefont
  {Stohner}},\ }\bibfield  {title} {\bibinfo {title} {{Parity Violation in
  Chiral Molecules}},\ }\href {https://doi.org/10.2533/000942905777676119}
  {\bibfield  {journal} {\bibinfo  {journal} {Chimia}\ }\textbf {\bibinfo
  {volume} {59}},\ \bibinfo {pages} {530} (\bibinfo {year} {2005})}\BibitemShut
  {NoStop}%
\bibitem [{\citenamefont {Jones}(2012)}]{Jones12}%
  \BibitemOpen
  \bibfield  {author} {\bibinfo {author} {\bibfnamefont {N.}~\bibnamefont
  {Jones}},\ }\bibfield  {title} {\bibinfo {title} {{Frontier experiments:
  Tough science}},\ }\href {https://doi.org/10.1038/481014a9} {\bibfield
  {journal} {\bibinfo  {journal} {Nature}\ }\textbf {\bibinfo {volume} {481}},\
  \bibinfo {pages} {14} (\bibinfo {year} {2012})}\BibitemShut {NoStop}%
\bibitem [{\citenamefont {Letokhov}(1975)}]{Letokhov75}%
  \BibitemOpen
  \bibfield  {author} {\bibinfo {author} {\bibfnamefont {V.~S.}\ \bibnamefont
  {Letokhov}},\ }\bibfield  {title} {\bibinfo {title} {{On difference of energy
  levels of left and right molecules due to weak interactions}},\ }\href
  {https://doi.org/10.1016/0375-9601(75)90064-X} {\bibfield  {journal}
  {\bibinfo  {journal} {Phys.~Lett.~A}\ }\textbf {\bibinfo {volume} {53}},\
  \bibinfo {pages} {275} (\bibinfo {year} {1975})}\BibitemShut {NoStop}%
\bibitem [{\citenamefont {Berger}(2004)}]{Berger04-bookchapter}%
  \BibitemOpen
  \bibfield  {author} {\bibinfo {author} {\bibfnamefont {R.}~\bibnamefont
  {Berger}},\ }\bibfield  {title} {\bibinfo {title} {{Parity-Violation Effects
  in Molecules}},\ }in\ \href {https://doi.org/10.1016/S1380-7323(04)80031-1}
  {\emph {\bibinfo {booktitle} {Relativistic Electronic Structure Theory}}},\
  Vol.~\bibinfo {volume} {14},\ \bibinfo {editor} {edited by\ \bibinfo {editor}
  {\bibfnamefont {P.}~\bibnamefont {Schwerdtfeger}}}\ (\bibinfo  {publisher}
  {Elsevier},\ \bibinfo {year} {2004})\ pp.\ \bibinfo {pages}
  {188--288}\BibitemShut {NoStop}%
\bibitem [{\citenamefont {Darquié}\ \emph {et~al.}(2010)\citenamefont
  {Darquié}, \citenamefont {Stoeffler}, \citenamefont {Shelkovnikov},
  \citenamefont {Daussy}, \citenamefont {Amy-Klein}, \citenamefont
  {Chardonnet}, \citenamefont {Zrig}, \citenamefont {Guy}, \citenamefont
  {Crassous}, \citenamefont {Soulard} \emph {et~al.}}]{DarStoShe10}%
  \BibitemOpen
  \bibfield  {author} {\bibinfo {author} {\bibfnamefont {B.}~\bibnamefont
  {Darquié}}, \bibinfo {author} {\bibfnamefont {C.}~\bibnamefont {Stoeffler}},
  \bibinfo {author} {\bibfnamefont {A.}~\bibnamefont {Shelkovnikov}}, \bibinfo
  {author} {\bibfnamefont {C.}~\bibnamefont {Daussy}}, \bibinfo {author}
  {\bibfnamefont {A.}~\bibnamefont {Amy-Klein}}, \bibinfo {author}
  {\bibfnamefont {C.}~\bibnamefont {Chardonnet}}, \bibinfo {author}
  {\bibfnamefont {S.}~\bibnamefont {Zrig}}, \bibinfo {author} {\bibfnamefont
  {L.}~\bibnamefont {Guy}}, \bibinfo {author} {\bibfnamefont {J.}~\bibnamefont
  {Crassous}}, \bibinfo {author} {\bibfnamefont {P.}~\bibnamefont {Soulard}},
  \emph {et~al.},\ }\bibfield  {title} {\bibinfo {title} {{Progress toward the
  first observation of parity violation in chiral molecules by high-resolution
  laser spectroscopy}},\ }\href {https://doi.org/10.1002/chir.20911} {\bibfield
   {journal} {\bibinfo  {journal} {Chirality}\ }\textbf {\bibinfo {volume}
  {22}},\ \bibinfo {pages} {870} (\bibinfo {year} {2010})}\BibitemShut
  {NoStop}%
\bibitem [{\citenamefont {Hobi}\ \emph {et~al.}(2013)\citenamefont {Hobi},
  \citenamefont {Berger},\ and\ \citenamefont {Stohner}}]{Hobi2013}%
  \BibitemOpen
  \bibfield  {author} {\bibinfo {author} {\bibfnamefont {F.}~\bibnamefont
  {Hobi}}, \bibinfo {author} {\bibfnamefont {R.}~\bibnamefont {Berger}},\ and\
  \bibinfo {author} {\bibfnamefont {J.}~\bibnamefont {Stohner}},\ }\bibfield
  {title} {\bibinfo {title} {{Investigation of parity violation in nuclear
  spin-rotation interaction of fluorooxirane}},\ }\href
  {https://doi.org/10.1080/00268976.2013.816444} {\bibfield  {journal}
  {\bibinfo  {journal} {Mol.~Phys.}\ }\textbf {\bibinfo {volume} {111}},\
  \bibinfo {pages} {2345} (\bibinfo {year} {2013})}\BibitemShut {NoStop}%
\bibitem [{\citenamefont {Cournol}\ \emph {et~al.}(2019)\citenamefont
  {Cournol}, \citenamefont {Manceau}, \citenamefont {Pierens}, \citenamefont
  {Lecordier}, \citenamefont {Tran}, \citenamefont {Santagata}, \citenamefont
  {Argence}, \citenamefont {Goncharov}, \citenamefont {Lopez}, \citenamefont
  {Abgrall} \emph {et~al.}}]{CouManPie19}%
  \BibitemOpen
  \bibfield  {author} {\bibinfo {author} {\bibfnamefont {A.}~\bibnamefont
  {Cournol}}, \bibinfo {author} {\bibfnamefont {M.}~\bibnamefont {Manceau}},
  \bibinfo {author} {\bibfnamefont {M.}~\bibnamefont {Pierens}}, \bibinfo
  {author} {\bibfnamefont {L.}~\bibnamefont {Lecordier}}, \bibinfo {author}
  {\bibfnamefont {D.~B.~A.}\ \bibnamefont {Tran}}, \bibinfo {author}
  {\bibfnamefont {R.}~\bibnamefont {Santagata}}, \bibinfo {author}
  {\bibfnamefont {B.}~\bibnamefont {Argence}}, \bibinfo {author} {\bibfnamefont
  {A.}~\bibnamefont {Goncharov}}, \bibinfo {author} {\bibfnamefont
  {O.}~\bibnamefont {Lopez}}, \bibinfo {author} {\bibfnamefont
  {M.}~\bibnamefont {Abgrall}}, \emph {et~al.},\ }\bibfield  {title} {\bibinfo
  {title} {{A new experiment to test parity symmetry in cold chiral molecules
  using vibrational spectroscopy}},\ }\href {https://doi.org/10.1070/qel16880}
  {\bibfield  {journal} {\bibinfo  {journal} {Quantum Electron.}\ }\textbf
  {\bibinfo {volume} {49}},\ \bibinfo {pages} {288} (\bibinfo {year}
  {2019})}\BibitemShut {NoStop}%
\bibitem [{\citenamefont {Gorshkov}\ \emph {et~al.}(1982)\citenamefont
  {Gorshkov}, \citenamefont {Kozlov},\ and\ \citenamefont
  {Labzovskii}}]{Gorshkov82}%
  \BibitemOpen
  \bibfield  {author} {\bibinfo {author} {\bibfnamefont {V.~G.}\ \bibnamefont
  {Gorshkov}}, \bibinfo {author} {\bibfnamefont {M.~G.}\ \bibnamefont
  {Kozlov}},\ and\ \bibinfo {author} {\bibfnamefont {L.~N.}\ \bibnamefont
  {Labzovskii}},\ }\bibfield  {title} {\bibinfo {title} {{P-odd effects in
  polyatomic molecules}},\ }\href@noop {} {\bibfield  {journal} {\bibinfo
  {journal} {Sov.~Phys.~JETP}\ }\textbf {\bibinfo {volume} {55}},\ \bibinfo
  {pages} {1042} (\bibinfo {year} {1982})}\BibitemShut {NoStop}%
\bibitem [{\citenamefont {Barra}\ \emph {et~al.}(1986)\citenamefont {Barra},
  \citenamefont {Robert},\ and\ \citenamefont {Wiesenfeld}}]{Barra1986}%
  \BibitemOpen
  \bibfield  {author} {\bibinfo {author} {\bibfnamefont {A.~L.}\ \bibnamefont
  {Barra}}, \bibinfo {author} {\bibfnamefont {J.~B.}\ \bibnamefont {Robert}},\
  and\ \bibinfo {author} {\bibfnamefont {L.}~\bibnamefont {Wiesenfeld}},\
  }\bibfield  {title} {\bibinfo {title} {{Parity non-conservation and NMR
  observables. calculation of Tl resonance frequency differences in
  enantiomers}},\ }\href {https://doi.org/10.1016/0375-9601(86)90072-1}
  {\bibfield  {journal} {\bibinfo  {journal} {Phys.~Lett.~A}\ }\textbf
  {\bibinfo {volume} {115}},\ \bibinfo {pages} {443} (\bibinfo {year}
  {1986})}\BibitemShut {NoStop}%
\bibitem [{\citenamefont {Barra}\ \emph {et~al.}(1988)\citenamefont {Barra},
  \citenamefont {Robert},\ and\ \citenamefont {Wiesenfeld}}]{Barra1988}%
  \BibitemOpen
  \bibfield  {author} {\bibinfo {author} {\bibfnamefont {A.~L.}\ \bibnamefont
  {Barra}}, \bibinfo {author} {\bibfnamefont {J.~B.}\ \bibnamefont {Robert}},\
  and\ \bibinfo {author} {\bibfnamefont {L.}~\bibnamefont {Wiesenfeld}},\
  }\bibfield  {title} {\bibinfo {title} {{Possible Observation of Parity
  Nonconservation by High-Resolution {NMR}}},\ }\href
  {https://doi.org/10.1209/0295-5075/5/3/006} {\bibfield  {journal} {\bibinfo
  {journal} {EPL}\ }\textbf {\bibinfo {volume} {5}},\ \bibinfo {pages} {217}
  (\bibinfo {year} {1988})}\BibitemShut {NoStop}%
\bibitem [{\citenamefont {Barra}\ and\ \citenamefont
  {Robert}(1996)}]{Barra1996}%
  \BibitemOpen
  \bibfield  {author} {\bibinfo {author} {\bibfnamefont {A.~L.}\ \bibnamefont
  {Barra}}\ and\ \bibinfo {author} {\bibfnamefont {J.~B.}\ \bibnamefont
  {Robert}},\ }\bibfield  {title} {\bibinfo {title} {{Parity non-conservation
  and NMR parameters}},\ }\href {https://doi.org/10.1080/00268979609484479}
  {\bibfield  {journal} {\bibinfo  {journal} {Mol.~Phys.}\ }\textbf {\bibinfo
  {volume} {88}},\ \bibinfo {pages} {875} (\bibinfo {year} {1996})}\BibitemShut
  {NoStop}%
\bibitem [{\citenamefont {Robert}\ and\ \citenamefont
  {Barra}(2001)}]{Robert2001}%
  \BibitemOpen
  \bibfield  {author} {\bibinfo {author} {\bibfnamefont {J.~B.}\ \bibnamefont
  {Robert}}\ and\ \bibinfo {author} {\bibfnamefont {A.~L.}\ \bibnamefont
  {Barra}},\ }\bibfield  {title} {\bibinfo {title} {{NMR and parity
  nonconservation. Experimental requirements to observe a difference between
  enantiomer signals}},\ }\href {https://doi.org/10.1002/chir.10003} {\bibfield
   {journal} {\bibinfo  {journal} {Chirality}\ }\textbf {\bibinfo {volume}
  {13}},\ \bibinfo {pages} {699} (\bibinfo {year} {2001})}\BibitemShut
  {NoStop}%
\bibitem [{\citenamefont {Soncini}\ \emph {et~al.}(2003)\citenamefont
  {Soncini}, \citenamefont {Faglioni},\ and\ \citenamefont
  {Lazzeretti}}]{Soncini2003}%
  \BibitemOpen
  \bibfield  {author} {\bibinfo {author} {\bibfnamefont {A.}~\bibnamefont
  {Soncini}}, \bibinfo {author} {\bibfnamefont {F.}~\bibnamefont {Faglioni}},\
  and\ \bibinfo {author} {\bibfnamefont {P.}~\bibnamefont {Lazzeretti}},\
  }\bibfield  {title} {\bibinfo {title} {{Parity-violating contributions to
  nuclear magnetic shielding}},\ }\href
  {https://doi.org/10.1103/PhysRevA.68.033402} {\bibfield  {journal} {\bibinfo
  {journal} {Phys.~Rev.~A}\ }\textbf {\bibinfo {volume} {68}},\ \bibinfo
  {pages} {033402} (\bibinfo {year} {2003})}\BibitemShut {NoStop}%
\bibitem [{\citenamefont {Laubender}\ and\ \citenamefont
  {Berger}(2003)}]{Laubender2003}%
  \BibitemOpen
  \bibfield  {author} {\bibinfo {author} {\bibfnamefont {G.}~\bibnamefont
  {Laubender}}\ and\ \bibinfo {author} {\bibfnamefont {R.}~\bibnamefont
  {Berger}},\ }\bibfield  {title} {\bibinfo {title} {{Ab Initio Calculation of
  Parity-Violating Chemical Shifts in NMR Spectra of Chiral Molecules}},\
  }\href {https://doi.org/10.1002/cphc.200390070} {\bibfield  {journal}
  {\bibinfo  {journal} {ChemPhysChem}\ }\textbf {\bibinfo {volume} {4}},\
  \bibinfo {pages} {395} (\bibinfo {year} {2003})}\BibitemShut {NoStop}%
\bibitem [{\citenamefont {Weijo}\ \emph {et~al.}(2005)\citenamefont {Weijo},
  \citenamefont {Manninen},\ and\ \citenamefont {Vaara}}]{Weijo2005}%
  \BibitemOpen
  \bibfield  {author} {\bibinfo {author} {\bibfnamefont {V.}~\bibnamefont
  {Weijo}}, \bibinfo {author} {\bibfnamefont {P.}~\bibnamefont {Manninen}},\
  and\ \bibinfo {author} {\bibfnamefont {J.}~\bibnamefont {Vaara}},\ }\bibfield
   {title} {\bibinfo {title} {{Perturbational calculations of parity-violating
  effects in nuclear-magnetic-resonance parameters}},\ }\href
  {https://doi.org/10.1063/1.1961321} {\bibfield  {journal} {\bibinfo
  {journal} {J.~Chem.~Phys.}\ }\textbf {\bibinfo {volume} {123}},\ \bibinfo
  {pages} {054501} (\bibinfo {year} {2005})}\BibitemShut {NoStop}%
\bibitem [{\citenamefont {Laubender}\ and\ \citenamefont
  {Berger}(2006)}]{LauBer06}%
  \BibitemOpen
  \bibfield  {author} {\bibinfo {author} {\bibfnamefont {G.}~\bibnamefont
  {Laubender}}\ and\ \bibinfo {author} {\bibfnamefont {R.}~\bibnamefont
  {Berger}},\ }\bibfield  {title} {\bibinfo {title} {{Electroweak quantum
  chemistry for nuclear-magnetic-resonance-shielding constants: Impact of
  electron correlation}},\ }\href {https://doi.org/10.1103/PhysRevA.74.032105}
  {\bibfield  {journal} {\bibinfo  {journal} {Phys.~Rev.~A}\ }\textbf {\bibinfo
  {volume} {74}},\ \bibinfo {pages} {032105} (\bibinfo {year}
  {2006})}\BibitemShut {NoStop}%
\bibitem [{\citenamefont {Bast}\ \emph {et~al.}(2006)\citenamefont {Bast},
  \citenamefont {Schwerdtfeger},\ and\ \citenamefont {Saue}}]{Bast2006}%
  \BibitemOpen
  \bibfield  {author} {\bibinfo {author} {\bibfnamefont {R.}~\bibnamefont
  {Bast}}, \bibinfo {author} {\bibfnamefont {P.}~\bibnamefont
  {Schwerdtfeger}},\ and\ \bibinfo {author} {\bibfnamefont {T.}~\bibnamefont
  {Saue}},\ }\bibfield  {title} {\bibinfo {title} {{Parity nonconservation
  contribution to the nuclear magnetic resonance shielding constants of chiral
  molecules: A four-component relativistic study}},\ }\href
  {https://doi.org/10.1063/1.2218333} {\bibfield  {journal} {\bibinfo
  {journal} {J.~Chem.~Phys.}\ }\textbf {\bibinfo {volume} {125}},\ \bibinfo
  {pages} {064504} (\bibinfo {year} {2006})}\BibitemShut {NoStop}%
\bibitem [{\citenamefont {Nahrwold}\ and\ \citenamefont
  {Berger}(2009)}]{Nahrwold2009}%
  \BibitemOpen
  \bibfield  {author} {\bibinfo {author} {\bibfnamefont {S.}~\bibnamefont
  {Nahrwold}}\ and\ \bibinfo {author} {\bibfnamefont {R.}~\bibnamefont
  {Berger}},\ }\bibfield  {title} {\bibinfo {title} {{Zeroth order regular
  approximation approach to parity violating nuclear magnetic resonance
  shielding tensors}},\ }\href {https://doi.org/10.1063/1.3103643} {\bibfield
  {journal} {\bibinfo  {journal} {J.~Chem.~Phys.}\ }\textbf {\bibinfo {volume}
  {130}},\ \bibinfo {pages} {214101} (\bibinfo {year} {2009})}\BibitemShut
  {NoStop}%
\bibitem [{\citenamefont {Eills}\ \emph {et~al.}(2017)\citenamefont {Eills},
  \citenamefont {Blanchard}, \citenamefont {Bougas}, \citenamefont {Kozlov},
  \citenamefont {Pines},\ and\ \citenamefont {Budker}}]{Eills2017}%
  \BibitemOpen
  \bibfield  {author} {\bibinfo {author} {\bibfnamefont {J.}~\bibnamefont
  {Eills}}, \bibinfo {author} {\bibfnamefont {J.~W.}\ \bibnamefont
  {Blanchard}}, \bibinfo {author} {\bibfnamefont {L.}~\bibnamefont {Bougas}},
  \bibinfo {author} {\bibfnamefont {M.~G.}\ \bibnamefont {Kozlov}}, \bibinfo
  {author} {\bibfnamefont {A.}~\bibnamefont {Pines}},\ and\ \bibinfo {author}
  {\bibfnamefont {D.}~\bibnamefont {Budker}},\ }\bibfield  {title} {\bibinfo
  {title} {{Measuring molecular parity nonconservation using
  nuclear-magnetic-resonance spectroscopy}},\ }\href
  {https://doi.org/10.1103/PhysRevA.96.042119} {\bibfield  {journal} {\bibinfo
  {journal} {Phys.~Rev.~A}\ }\textbf {\bibinfo {volume} {96}},\ \bibinfo
  {pages} {042119} (\bibinfo {year} {2017})}\BibitemShut {NoStop}%
\bibitem [{\citenamefont {Bouchiat}\ and\ \citenamefont
  {Bouchiat}(1974)}]{Bouchiat74}%
  \BibitemOpen
  \bibfield  {author} {\bibinfo {author} {\bibfnamefont {M.~A.}\ \bibnamefont
  {Bouchiat}}\ and\ \bibinfo {author} {\bibfnamefont {C.}~\bibnamefont
  {Bouchiat}},\ }\bibfield  {title} {\bibinfo {title} {{I. Parity violation
  induced by weak neutral currents in atomic physics}},\ }\href
  {https://doi.org/10.1051/jphys:019740035012089900} {\bibfield  {journal}
  {\bibinfo  {journal} {J.~Phys.~France}\ }\textbf {\bibinfo {volume} {35}},\
  \bibinfo {pages} {899} (\bibinfo {year} {1974})}\BibitemShut {NoStop}%
\bibitem [{\citenamefont {Bouchiat}\ and\ \citenamefont
  {Bouchiat}(1997)}]{Bouchiat97}%
  \BibitemOpen
  \bibfield  {author} {\bibinfo {author} {\bibfnamefont {M.~A.}\ \bibnamefont
  {Bouchiat}}\ and\ \bibinfo {author} {\bibfnamefont {C.}~\bibnamefont
  {Bouchiat}},\ }\bibfield  {title} {\bibinfo {title} {{Parity violation in
  atoms}},\ }\href {https://doi.org/10.1088/0034-4885/60/11/004} {\bibfield
  {journal} {\bibinfo  {journal} {Rep.~Prog.~Phys.}\ }\textbf {\bibinfo
  {volume} {60}},\ \bibinfo {pages} {1351} (\bibinfo {year}
  {1997})}\BibitemShut {NoStop}%
\bibitem [{\citenamefont {Aucar}\ and\ \citenamefont
  {Borschevsky}(2021)}]{Aucar-PVSR2021}%
  \BibitemOpen
  \bibfield  {author} {\bibinfo {author} {\bibfnamefont {I.~A.}\ \bibnamefont
  {Aucar}}\ and\ \bibinfo {author} {\bibfnamefont {A.}~\bibnamefont
  {Borschevsky}},\ }\bibfield  {title} {\bibinfo {title} {{Relativistic study
  of parity-violating nuclear spin-rotation tensors}},\ }\href
  {https://doi.org/10.1063/5.0065487} {\bibfield  {journal} {\bibinfo
  {journal} {J.~Chem.~Phys.}\ }\textbf {\bibinfo {volume} {155}},\ \bibinfo
  {pages} {134307} (\bibinfo {year} {2021})}\BibitemShut {NoStop}%
\bibitem [{\citenamefont {Oddershede}(1978)}]{Oddershede1978}%
  \BibitemOpen
  \bibfield  {author} {\bibinfo {author} {\bibfnamefont {J.}~\bibnamefont
  {Oddershede}},\ }\bibfield  {title} {\bibinfo {title} {{Polarization
  Propagator Calculations}},\ }in\ \href
  {https://doi.org/10.1016/S0065-3276(08)60240-3} {\emph {\bibinfo {booktitle}
  {Advances in Quantum Chemistry}}},\ Vol.~\bibinfo {volume} {11},\ \bibinfo
  {editor} {edited by\ \bibinfo {editor} {\bibfnamefont {P.-O.}\ \bibnamefont
  {Löwdin}}}\ (\bibinfo  {publisher} {Academic},\ \bibinfo {address} {San
  Diego},\ \bibinfo {year} {1978})\ pp.\ \bibinfo {pages}
  {275--352}\BibitemShut {NoStop}%
\bibitem [{\citenamefont {Helgaker}\ \emph {et~al.}(2012)\citenamefont
  {Helgaker}, \citenamefont {Coriani}, \citenamefont {J{\o}rgensen},
  \citenamefont {Kristensen}, \citenamefont {Olsen},\ and\ \citenamefont
  {Ruud}}]{Helgaker2012}%
  \BibitemOpen
  \bibfield  {author} {\bibinfo {author} {\bibfnamefont {T.}~\bibnamefont
  {Helgaker}}, \bibinfo {author} {\bibfnamefont {S.}~\bibnamefont {Coriani}},
  \bibinfo {author} {\bibfnamefont {P.}~\bibnamefont {J{\o}rgensen}}, \bibinfo
  {author} {\bibfnamefont {K.}~\bibnamefont {Kristensen}}, \bibinfo {author}
  {\bibfnamefont {J.}~\bibnamefont {Olsen}},\ and\ \bibinfo {author}
  {\bibfnamefont {K.}~\bibnamefont {Ruud}},\ }\bibfield  {title} {\bibinfo
  {title} {{Recent Advances in Wave Function-Based Methods of
  Molecular-Property Calculations}},\ }\href
  {https://doi.org/10.1021/cr2002239} {\bibfield  {journal} {\bibinfo
  {journal} {Chem.~Rev.}\ }\textbf {\bibinfo {volume} {112}},\ \bibinfo {pages}
  {543} (\bibinfo {year} {2012})},\ \bibinfo {note} {pMID:
  22236047}\BibitemShut {NoStop}%
\bibitem [{\citenamefont {Rusakov}\ and\ \citenamefont
  {Krivdin}(2013)}]{Rusakov2013}%
  \BibitemOpen
  \bibfield  {author} {\bibinfo {author} {\bibfnamefont {Y.}~\bibnamefont
  {Rusakov}}\ and\ \bibinfo {author} {\bibfnamefont {L.~B.}\ \bibnamefont
  {Krivdin}},\ }\bibfield  {title} {\bibinfo {title} {{Modern quantum chemical
  methods for calculating spin{\textendash}spin coupling constants: theoretical
  basis and structural applications in chemistry}},\ }\href
  {https://doi.org/10.1070/rc2013v082n02abeh004350} {\bibfield  {journal}
  {\bibinfo  {journal} {Russ.~Chem.~Rev.}\ }\textbf {\bibinfo {volume} {82}},\
  \bibinfo {pages} {99} (\bibinfo {year} {2013})}\BibitemShut {NoStop}%
\bibitem [{\citenamefont {Aucar}(2014)}]{Aucar2014}%
  \BibitemOpen
  \bibfield  {author} {\bibinfo {author} {\bibfnamefont {G.~A.}\ \bibnamefont
  {Aucar}},\ }\bibfield  {title} {\bibinfo {title} {{Toward a QFT-based theory
  of atomic and molecular properties}},\ }\href
  {https://doi.org/10.1039/C3CP52685B} {\bibfield  {journal} {\bibinfo
  {journal} {Phys.~Chem.~Chem.~Phys.}\ }\textbf {\bibinfo {volume} {16}},\
  \bibinfo {pages} {4420} (\bibinfo {year} {2014})}\BibitemShut {NoStop}%
\bibitem [{\citenamefont {Aucar}\ \emph {et~al.}(2010)\citenamefont {Aucar},
  \citenamefont {Romero},\ and\ \citenamefont {Maldonado}}]{Aucar2010}%
  \BibitemOpen
  \bibfield  {author} {\bibinfo {author} {\bibfnamefont {G.~A.}\ \bibnamefont
  {Aucar}}, \bibinfo {author} {\bibfnamefont {R.~H.}\ \bibnamefont {Romero}},\
  and\ \bibinfo {author} {\bibfnamefont {A.~F.}\ \bibnamefont {Maldonado}},\
  }\bibfield  {title} {\bibinfo {title} {{Polarization propagators: A powerful
  theoretical tool for a deeper understanding of NMR spectroscopic
  parameters}},\ }\href {https://doi.org/10.1080/01442350903432865} {\bibfield
  {journal} {\bibinfo  {journal} {Int.~Rev.~Phys.~Chem.}\ }\textbf {\bibinfo
  {volume} {29}},\ \bibinfo {pages} {1} (\bibinfo {year} {2010})}\BibitemShut
  {NoStop}%
\bibitem [{\citenamefont {Saue}\ and\ \citenamefont {Jensen}(2003)}]{Saue2003}%
  \BibitemOpen
  \bibfield  {author} {\bibinfo {author} {\bibfnamefont {T.}~\bibnamefont
  {Saue}}\ and\ \bibinfo {author} {\bibfnamefont {H.~J.~{\relax Aa}.}\
  \bibnamefont {Jensen}},\ }\bibfield  {title} {\bibinfo {title} {{Linear
  response at the 4-component relativistic level: Application to the
  frequency-dependent dipole polarizabilities of the coinage metal dimers}},\
  }\href {https://doi.org/10.1063/1.1522407} {\bibfield  {journal} {\bibinfo
  {journal} {J.~Chem.~Phys.}\ }\textbf {\bibinfo {volume} {118}},\ \bibinfo
  {pages} {522} (\bibinfo {year} {2003})}\BibitemShut {NoStop}%
\bibitem [{\citenamefont {Aucar}\ \emph {et~al.}(2012)\citenamefont {Aucar},
  \citenamefont {Gómez}, \citenamefont {{Ruiz de Azúa}},\ and\ \citenamefont
  {Giribet}}]{Agus2012}%
  \BibitemOpen
  \bibfield  {author} {\bibinfo {author} {\bibfnamefont {I.~A.}\ \bibnamefont
  {Aucar}}, \bibinfo {author} {\bibfnamefont {S.~S.}\ \bibnamefont {Gómez}},
  \bibinfo {author} {\bibfnamefont {M.~C.}\ \bibnamefont {{Ruiz de Azúa}}},\
  and\ \bibinfo {author} {\bibfnamefont {C.~G.}\ \bibnamefont {Giribet}},\
  }\bibfield  {title} {\bibinfo {title} {{Theoretical study of the nuclear
  spin-molecular rotation coupling for relativistic electrons and
  non-relativistic nuclei}},\ }\href {https://doi.org/10.1063/1.4721627}
  {\bibfield  {journal} {\bibinfo  {journal} {J.~Chem.~Phys.}\ }\textbf
  {\bibinfo {volume} {136}},\ \bibinfo {pages} {204119} (\bibinfo {year}
  {2012})}\BibitemShut {NoStop}%
\bibitem [{\citenamefont {Aucar}\ \emph
  {et~al.}(2013{\natexlab{a}})\citenamefont {Aucar}, \citenamefont {Gomez},
  \citenamefont {Melo}, \citenamefont {Giribet},\ and\ \citenamefont {{Ruiz de
  Az{\'u}a}}}]{Agus2013-1}%
  \BibitemOpen
  \bibfield  {author} {\bibinfo {author} {\bibfnamefont {I.~A.}\ \bibnamefont
  {Aucar}}, \bibinfo {author} {\bibfnamefont {S.~S.}\ \bibnamefont {Gomez}},
  \bibinfo {author} {\bibfnamefont {J.~I.}\ \bibnamefont {Melo}}, \bibinfo
  {author} {\bibfnamefont {C.~G.}\ \bibnamefont {Giribet}},\ and\ \bibinfo
  {author} {\bibfnamefont {M.~C.}\ \bibnamefont {{Ruiz de Az{\'u}a}}},\
  }\bibfield  {title} {\bibinfo {title} {{Theoretical Study of the Nuclear
  Spin-Molecular Rotation Coupling for Relativistic Electrons and
  Non-Relativistic Nuclei. II. Quantitative Results in ${\mathrm{HX}}$
  (${\mathrm{X=H,F,Cl,Br,I}}$) Compounds}},\ }\href
  {https://doi.org/10.1063/1.4796461} {\bibfield  {journal} {\bibinfo
  {journal} {J.~Chem.~Phys.}\ }\textbf {\bibinfo {volume} {138}},\ \bibinfo
  {pages} {134107} (\bibinfo {year} {2013}{\natexlab{a}})}\BibitemShut
  {NoStop}%
\bibitem [{\citenamefont {Aucar}\ and\ \citenamefont
  {Aucar}(2019)}]{aucar-aucar2019}%
  \BibitemOpen
  \bibfield  {author} {\bibinfo {author} {\bibfnamefont {G.~A.}\ \bibnamefont
  {Aucar}}\ and\ \bibinfo {author} {\bibfnamefont {I.~A.}\ \bibnamefont
  {Aucar}},\ }\bibfield  {title} {\bibinfo {title} {{Recent Developments in
  Absolute Shielding Scales for NMR Spectroscopy}},\ }in\ \href
  {https://doi.org/10.1016/bs.arnmr.2018.08.001} {\emph {\bibinfo {booktitle}
  {Annual Reports on NMR Spectroscopy}}},\ Vol.~\bibinfo {volume} {96},\
  \bibinfo {editor} {edited by\ \bibinfo {editor} {\bibfnamefont {G.~A.}\
  \bibnamefont {Webb}}}\ (\bibinfo  {publisher} {Academic},\ \bibinfo {address}
  {San Diego},\ \bibinfo {year} {2019})\ pp.\ \bibinfo {pages}
  {77--141}\BibitemShut {NoStop}%
\bibitem [{\citenamefont {Aucar}\ \emph
  {et~al.}(2013{\natexlab{b}})\citenamefont {Aucar}, \citenamefont {Gómez},
  \citenamefont {Giribet},\ and\ \citenamefont {{Ruiz de Azúa}}}]{Agus2013-2}%
  \BibitemOpen
  \bibfield  {author} {\bibinfo {author} {\bibfnamefont {I.~A.}\ \bibnamefont
  {Aucar}}, \bibinfo {author} {\bibfnamefont {S.~S.}\ \bibnamefont {Gómez}},
  \bibinfo {author} {\bibfnamefont {C.~G.}\ \bibnamefont {Giribet}},\ and\
  \bibinfo {author} {\bibfnamefont {M.~C.}\ \bibnamefont {{Ruiz de Azúa}}},\
  }\bibfield  {title} {\bibinfo {title} {{Breit interaction effects in
  relativistic theory of the nuclear spin-rotation tensor}},\ }\href
  {https://doi.org/10.1063/1.4819958} {\bibfield  {journal} {\bibinfo
  {journal} {J.~Chem.~Phys.}\ }\textbf {\bibinfo {volume} {139}},\ \bibinfo
  {pages} {094112} (\bibinfo {year} {2013}{\natexlab{b}})}\BibitemShut
  {NoStop}%
\bibitem [{\citenamefont {Flambaum}\ and\ \citenamefont
  {Khriplovich}(1980)}]{Flambaum1980}%
  \BibitemOpen
  \bibfield  {author} {\bibinfo {author} {\bibfnamefont {V.}~\bibnamefont
  {Flambaum}}\ and\ \bibinfo {author} {\bibfnamefont {I.}~\bibnamefont
  {Khriplovich}},\ }\bibfield  {title} {\bibinfo {title} {{P-odd nuclear forces
  -- a source of parity violation in atoms}},\ }\href
  {http://www.jetp.ac.ru/cgi-bin/dn/e_052_05_0835.pdf} {\bibfield  {journal}
  {\bibinfo  {journal} {Sov.~Phys.~JETP}\ }\textbf {\bibinfo {volume} {52}},\
  \bibinfo {pages} {835} (\bibinfo {year} {1980})}\BibitemShut {NoStop}%
\bibitem [{\citenamefont {Blundell}\ \emph {et~al.}(1992)\citenamefont
  {Blundell}, \citenamefont {Sapirstein},\ and\ \citenamefont
  {Johnson}}]{Blundell1992}%
  \BibitemOpen
  \bibfield  {author} {\bibinfo {author} {\bibfnamefont {S.~A.}\ \bibnamefont
  {Blundell}}, \bibinfo {author} {\bibfnamefont {J.}~\bibnamefont
  {Sapirstein}},\ and\ \bibinfo {author} {\bibfnamefont {W.~R.}\ \bibnamefont
  {Johnson}},\ }\bibfield  {title} {\bibinfo {title} {{High-accuracy
  calculation of parity nonconservation in cesium and implications for particle
  physics}},\ }\href {https://doi.org/10.1103/PhysRevD.45.1602} {\bibfield
  {journal} {\bibinfo  {journal} {Phys.~Rev.~D}\ }\textbf {\bibinfo {volume}
  {45}},\ \bibinfo {pages} {1602} (\bibinfo {year} {1992})}\BibitemShut
  {NoStop}%
\bibitem [{\citenamefont {{Tiesinga, E. and Mohr, P.~J. and Newell, D.~B. and
  Taylor, B.~N.}}()}]{codata2018}%
  \BibitemOpen
  \bibfield  {author} {\bibinfo {author} {\bibnamefont {{Tiesinga, E. and Mohr,
  P.~J. and Newell, D.~B. and Taylor, B.~N.}}},\ }\href@noop {} {\bibinfo
  {title} {{The 2018 CODATA Recommended Values of the Fundamental Physical
  Constants (Web Version 8.1)}}},\ \bibinfo {howpublished}
  {\url{http://physics.nist.gov/constants}},\ \bibinfo {note} {{D}atabase
  developed by J. Baker, M. Douma, and S. Kotochigova. National Institute of
  Standards and Technology, Gaithersburg, MD 20899}\BibitemShut {NoStop}%
\bibitem [{\citenamefont {Zyla}\ \emph {et~al.}(2020)\citenamefont {Zyla} \emph
  {et~al.}}]{Zyla2020}%
  \BibitemOpen
  \bibfield  {author} {\bibinfo {author} {\bibfnamefont {P.~A.}\ \bibnamefont
  {Zyla}} \emph {et~al.} (\bibinfo {collaboration} {Particle Data Group}),\
  }\bibfield  {title} {\bibinfo {title} {{Review of Particle Physics}},\ }\href
  {https://doi.org/10.1093/ptep/ptaa104} {\bibfield  {journal} {\bibinfo
  {journal} {PTEP}\ }\textbf {\bibinfo {volume} {2020}},\ \bibinfo {pages}
  {083C01} (\bibinfo {year} {2020})}\BibitemShut {NoStop}%
\bibitem [{\citenamefont {Montanet}\ \emph {et~al.}(1994)\citenamefont
  {Montanet} \emph {et~al.}}]{sintheta}%
  \BibitemOpen
  \bibfield  {author} {\bibinfo {author} {\bibfnamefont {L.}~\bibnamefont
  {Montanet}} \emph {et~al.} (\bibinfo {collaboration} {Particle Data Group}),\
  }\bibfield  {title} {\bibinfo {title} {{Review of Particle Properties}},\
  }\href {https://doi.org/10.1103/PhysRevD.50.1173} {\bibfield  {journal}
  {\bibinfo  {journal} {Phys.~Rev.~D}\ }\textbf {\bibinfo {volume} {50}},\
  \bibinfo {pages} {1173} (\bibinfo {year} {1994})}\BibitemShut {NoStop}%
\bibitem [{\citenamefont {Aucar}\ \emph {et~al.}(2022)\citenamefont {Aucar},
  \citenamefont {{Chamorro Mena}},\ and\ \citenamefont
  {Borschevsky}}]{Aucar-PVSR2022}%
  \BibitemOpen
  \bibfield  {author} {\bibinfo {author} {\bibfnamefont {I.~A.}\ \bibnamefont
  {Aucar}}, \bibinfo {author} {\bibfnamefont {Y.~A.}\ \bibnamefont {{Chamorro
  Mena}}},\ and\ \bibinfo {author} {\bibfnamefont {A.}~\bibnamefont
  {Borschevsky}},\ }\bibfield  {title} {\bibinfo {title} {{Parity-violating
  nuclear spin-rotation and NMR shielding tensors in tetrahedral molecules}},\
  }\href@noop {} {\  (\bibinfo {year} {2022})},\ \bibinfo {note} {in
  preparation}\BibitemShut {NoStop}%
\bibitem [{\citenamefont {Melo}\ \emph {et~al.}(2003)\citenamefont {Melo},
  \citenamefont {{Ruiz de Az{\'u}a}}, \citenamefont {Giribet}, \citenamefont
  {Aucar},\ and\ \citenamefont {Romero}}]{melo-lresc}%
  \BibitemOpen
  \bibfield  {author} {\bibinfo {author} {\bibfnamefont {J.~I.}\ \bibnamefont
  {Melo}}, \bibinfo {author} {\bibfnamefont {M.~C.}\ \bibnamefont {{Ruiz de
  Az{\'u}a}}}, \bibinfo {author} {\bibfnamefont {C.~G.}\ \bibnamefont
  {Giribet}}, \bibinfo {author} {\bibfnamefont {G.~A.}\ \bibnamefont {Aucar}},\
  and\ \bibinfo {author} {\bibfnamefont {R.~H.}\ \bibnamefont {Romero}},\
  }\bibfield  {title} {\bibinfo {title} {{Relativistic effects on the nuclear
  magnetic shielding tensor}},\ }\href {https://doi.org/10.1063/1.1525808}
  {\bibfield  {journal} {\bibinfo  {journal} {J.~Chem.~Phys.}\ }\textbf
  {\bibinfo {volume} {118}},\ \bibinfo {pages} {471} (\bibinfo {year}
  {2003})}\BibitemShut {NoStop}%
\bibitem [{\citenamefont {Aucar}\ \emph
  {et~al.}(2018{\natexlab{a}})\citenamefont {Aucar}, \citenamefont {Melo},
  \citenamefont {Aucar},\ and\ \citenamefont {Maldonado}}]{Review-LRESC}%
  \BibitemOpen
  \bibfield  {author} {\bibinfo {author} {\bibfnamefont {G.~A.}\ \bibnamefont
  {Aucar}}, \bibinfo {author} {\bibfnamefont {J.~I.}\ \bibnamefont {Melo}},
  \bibinfo {author} {\bibfnamefont {I.~A.}\ \bibnamefont {Aucar}},\ and\
  \bibinfo {author} {\bibfnamefont {A.~F.}\ \bibnamefont {Maldonado}},\
  }\bibfield  {title} {\bibinfo {title} {{Foundations of the LRESC Model for
  Response Properties and Some Applications}},\ }\href
  {https://doi.org/10.1002/qua.25487} {\bibfield  {journal} {\bibinfo
  {journal} {Int.~J.~Quantum Chem.}\ }\textbf {\bibinfo {volume} {118}},\
  \bibinfo {pages} {e25487} (\bibinfo {year} {2018}{\natexlab{a}})}\BibitemShut
  {NoStop}%
\bibitem [{\citenamefont {Aucar}\ \emph {et~al.}(2014)\citenamefont {Aucar},
  \citenamefont {Gomez}, \citenamefont {Giribet},\ and\ \citenamefont {{Ruiz de
  Az{\'u}a}}}]{Agus_g_2014}%
  \BibitemOpen
  \bibfield  {author} {\bibinfo {author} {\bibfnamefont {I.~A.}\ \bibnamefont
  {Aucar}}, \bibinfo {author} {\bibfnamefont {S.~S.}\ \bibnamefont {Gomez}},
  \bibinfo {author} {\bibfnamefont {C.~G.}\ \bibnamefont {Giribet}},\ and\
  \bibinfo {author} {\bibfnamefont {M.~C.}\ \bibnamefont {{Ruiz de
  Az{\'u}a}}},\ }\bibfield  {title} {\bibinfo {title} {{Theoretical study of
  the relativistic molecular rotational g-tensor}},\ }\href
  {https://doi.org/10.1063/1.4721627} {\bibfield  {journal} {\bibinfo
  {journal} {J.~Chem.~Phys.}\ }\textbf {\bibinfo {volume} {141}},\ \bibinfo
  {pages} {194103} (\bibinfo {year} {2014})}\BibitemShut {NoStop}%
\bibitem [{\citenamefont {Aucar}\ \emph {et~al.}(1999)\citenamefont {Aucar},
  \citenamefont {Saue}, \citenamefont {Visscher},\ and\ \citenamefont
  {Jensen}}]{Aucar1999}%
  \BibitemOpen
  \bibfield  {author} {\bibinfo {author} {\bibfnamefont {G.~A.}\ \bibnamefont
  {Aucar}}, \bibinfo {author} {\bibfnamefont {T.}~\bibnamefont {Saue}},
  \bibinfo {author} {\bibfnamefont {L.}~\bibnamefont {Visscher}},\ and\
  \bibinfo {author} {\bibfnamefont {H.~J.~{\relax Aa}.}\ \bibnamefont
  {Jensen}},\ }\bibfield  {title} {\bibinfo {title} {{On the origin and
  contribution of the diamagnetic term in four-component relativistic
  calculations of magnetic properties}},\ }\href
  {https://doi.org/10.1063/1.479181} {\bibfield  {journal} {\bibinfo  {journal}
  {J.~Chem.~Phys.}\ }\textbf {\bibinfo {volume} {110}},\ \bibinfo {pages}
  {6208} (\bibinfo {year} {1999})}\BibitemShut {NoStop}%
\bibitem [{\citenamefont {Aucar}\ \emph
  {et~al.}(2016{\natexlab{a}})\citenamefont {Aucar}, \citenamefont {Gomez},
  \citenamefont {Giribet},\ and\ \citenamefont {Aucar}}]{Agus_PCCP_2016}%
  \BibitemOpen
  \bibfield  {author} {\bibinfo {author} {\bibfnamefont {I.~A.}\ \bibnamefont
  {Aucar}}, \bibinfo {author} {\bibfnamefont {S.~S.}\ \bibnamefont {Gomez}},
  \bibinfo {author} {\bibfnamefont {C.~G.}\ \bibnamefont {Giribet}},\ and\
  \bibinfo {author} {\bibfnamefont {G.~A.}\ \bibnamefont {Aucar}},\ }\bibfield
  {title} {\bibinfo {title} {{Toward an Absolute NMR Shielding Scale Using the
  Spin-Rotation Tensor Within a Relativistic Framework}},\ }\href
  {https://doi.org/10.1039/c6cp03355e} {\bibfield  {journal} {\bibinfo
  {journal} {Phys.~Chem.~Chem.~Phys.}\ }\textbf {\bibinfo {volume} {18}},\
  \bibinfo {pages} {23572} (\bibinfo {year} {2016}{\natexlab{a}})}\BibitemShut
  {NoStop}%
\bibitem [{\citenamefont {Flygare}(1964)}]{Flygare64}%
  \BibitemOpen
  \bibfield  {author} {\bibinfo {author} {\bibfnamefont {W.~H.}\ \bibnamefont
  {Flygare}},\ }\bibfield  {title} {\bibinfo {title} {{Spin-Rotation
  Interaction and Magnetic Shielding in Molecules}},\ }\href
  {https://doi.org/10.1063/1.1725962} {\bibfield  {journal} {\bibinfo
  {journal} {J.~Chem.~Phys.}\ }\textbf {\bibinfo {volume} {41}},\ \bibinfo
  {pages} {793} (\bibinfo {year} {1964})}\BibitemShut {NoStop}%
\bibitem [{\citenamefont {Flygare}(1974)}]{Flygare74}%
  \BibitemOpen
  \bibfield  {author} {\bibinfo {author} {\bibfnamefont {W.~H.}\ \bibnamefont
  {Flygare}},\ }\bibfield  {title} {\bibinfo {title} {{Magnetic Interactions in
  Molecules and an Analysis of Molecular Electronic Charge Distribution from
  Magnetic Parameters}},\ }\href {https://doi.org/10.1021/cr60292a003}
  {\bibfield  {journal} {\bibinfo  {journal} {Chem.~Rev.}\ }\textbf {\bibinfo
  {volume} {74}},\ \bibinfo {pages} {653} (\bibinfo {year} {1974})}\BibitemShut
  {NoStop}%
\bibitem [{\citenamefont {Aucar}\ \emph
  {et~al.}(2016{\natexlab{b}})\citenamefont {Aucar}, \citenamefont {Gomez},
  \citenamefont {Giribet},\ and\ \citenamefont {Aucar}}]{Agus_Letter_2016}%
  \BibitemOpen
  \bibfield  {author} {\bibinfo {author} {\bibfnamefont {I.~A.}\ \bibnamefont
  {Aucar}}, \bibinfo {author} {\bibfnamefont {S.~S.}\ \bibnamefont {Gomez}},
  \bibinfo {author} {\bibfnamefont {C.~G.}\ \bibnamefont {Giribet}},\ and\
  \bibinfo {author} {\bibfnamefont {G.~A.}\ \bibnamefont {Aucar}},\ }\bibfield
  {title} {\bibinfo {title} {{Role of Spin Dependent Terms on the Relationship
  Among Nuclear Spin-Rotation and NMR Magnetic Shielding Tensors}},\ }\href
  {https://doi.org/10.1021/acs.jpclett.6b02361} {\bibfield  {journal} {\bibinfo
   {journal} {J.~Phys.~Chem.~Lett.}\ }\textbf {\bibinfo {volume} {7}},\
  \bibinfo {pages} {5188} (\bibinfo {year} {2016}{\natexlab{b}})}\BibitemShut
  {NoStop}%
\bibitem [{DIR()}]{DIRAC21}%
  \BibitemOpen
  \href@noop {} {}\bibinfo {note} {\textsc{dirac}, a relativistic \textit{ab
  initio} electronic structure program, Release \textsc{dirac}21 (2021),
  written by R.~Bast, A.~S.~P.~Gomes, T.~Saue, L.~Visscher, and H.~J.~{\relax
  Aa}.~Jensen, with contributions from I.~A.~Aucar, V.~Bakken, K.~G.~Dyall,
  S.~Dubillard, U.~Ekstr{\"o}m, et~al., available at
  \url{http://dx.doi.org/10.5281/zenodo.3572669}, see also
  \url{http://www.diracprogram.org})}\BibitemShut {NoStop}%
\bibitem [{\citenamefont {Saue}\ \emph {et~al.}(2020)\citenamefont {Saue},
  \citenamefont {Bast}, \citenamefont {Gomes}, \citenamefont {Jensen},
  \citenamefont {Visscher}, \citenamefont {Aucar}, \citenamefont {{Di
  Remigio}}, \citenamefont {Dyall}, \citenamefont {Eliav}, \citenamefont
  {Fa{\ss}hauer} \emph {et~al.}}]{dirac-paper}%
  \BibitemOpen
  \bibfield  {author} {\bibinfo {author} {\bibfnamefont {T.}~\bibnamefont
  {Saue}}, \bibinfo {author} {\bibfnamefont {R.}~\bibnamefont {Bast}}, \bibinfo
  {author} {\bibfnamefont {A.~S.~P.}\ \bibnamefont {Gomes}}, \bibinfo {author}
  {\bibfnamefont {H.~J.~{\relax Aa}.}\ \bibnamefont {Jensen}}, \bibinfo
  {author} {\bibfnamefont {L.}~\bibnamefont {Visscher}}, \bibinfo {author}
  {\bibfnamefont {I.~A.}\ \bibnamefont {Aucar}}, \bibinfo {author}
  {\bibfnamefont {R.}~\bibnamefont {{Di Remigio}}}, \bibinfo {author}
  {\bibfnamefont {K.~G.}\ \bibnamefont {Dyall}}, \bibinfo {author}
  {\bibfnamefont {E.}~\bibnamefont {Eliav}}, \bibinfo {author} {\bibfnamefont
  {E.}~\bibnamefont {Fa{\ss}hauer}}, \emph {et~al.},\ }\bibfield  {title}
  {\bibinfo {title} {The {DIRAC} code for relativistic molecular
  calculations},\ }\href {https://dx.doi.org/10.1063/5.0004844} {\bibfield
  {journal} {\bibinfo  {journal} {J.~Chem.~Phys.}\ }\textbf {\bibinfo {volume}
  {152}},\ \bibinfo {pages} {204104} (\bibinfo {year} {2020})}\BibitemShut
  {NoStop}%
\bibitem [{\citenamefont {Laerdahl}\ and\ \citenamefont
  {Schwerdtfeger}(1999)}]{Laerdahl1999}%
  \BibitemOpen
  \bibfield  {author} {\bibinfo {author} {\bibfnamefont {J.~K.}\ \bibnamefont
  {Laerdahl}}\ and\ \bibinfo {author} {\bibfnamefont {P.}~\bibnamefont
  {Schwerdtfeger}},\ }\bibfield  {title} {\bibinfo {title} {{Fully relativistic
  ab initio calculations of the energies of chiral molecules including
  parity-violating weak interactions}},\ }\href
  {https://doi.org/10.1103/PhysRevA.60.4439} {\bibfield  {journal} {\bibinfo
  {journal} {Phys.~Rev.~A}\ }\textbf {\bibinfo {volume} {60}},\ \bibinfo
  {pages} {4439} (\bibinfo {year} {1999})}\BibitemShut {NoStop}%
\bibitem [{\citenamefont {Raghavan}(1989)}]{Raghavan89}%
  \BibitemOpen
  \bibfield  {author} {\bibinfo {author} {\bibfnamefont {P.}~\bibnamefont
  {Raghavan}},\ }\bibfield  {title} {\bibinfo {title} {{Table of Nuclear
  Moments}},\ }\href {https://doi.org/10.1016/0092-640X(89)90008-9} {\bibfield
  {journal} {\bibinfo  {journal} {At.~Data Nucl.~Data Tables}\ }\textbf
  {\bibinfo {volume} {42}},\ \bibinfo {pages} {189} (\bibinfo {year}
  {1989})}\BibitemShut {NoStop}%
\bibitem [{\citenamefont {Dyall}(2006)}]{KD06}%
  \BibitemOpen
  \bibfield  {author} {\bibinfo {author} {\bibfnamefont {K.~G.}\ \bibnamefont
  {Dyall}},\ }\bibfield  {title} {\bibinfo {title} {{Relativistic
  Quadruple-Zeta and Revised Triple-Zeta and Double-Zeta Basis Sets for the 4p,
  5p, and 6p Elements}},\ }\href {http://dx.doi.org/10.1007/s00214-006-0126-0}
  {\bibfield  {journal} {\bibinfo  {journal} {Theor.~Chem.~Acc.}\ }\textbf
  {\bibinfo {volume} {115}},\ \bibinfo {pages} {441} (\bibinfo {year}
  {2006})}\BibitemShut {NoStop}%
\bibitem [{\citenamefont {Dyall}(2012)}]{KD12}%
  \BibitemOpen
  \bibfield  {author} {\bibinfo {author} {\bibfnamefont {K.~G.}\ \bibnamefont
  {Dyall}},\ }\bibfield  {title} {\bibinfo {title} {{Core correlating basis
  functions for elements 31–118}},\ }\href
  {https://doi.org/10.1007/s00214-012-1217-8} {\bibfield  {journal} {\bibinfo
  {journal} {Theor.~Chem.~Acc.}\ }\textbf {\bibinfo {volume} {131}},\ \bibinfo
  {pages} {1217} (\bibinfo {year} {2012})}\BibitemShut {NoStop}%
\bibitem [{\citenamefont {Dyall}(2016)}]{KD16}%
  \BibitemOpen
  \bibfield  {author} {\bibinfo {author} {\bibfnamefont {K.~G.}\ \bibnamefont
  {Dyall}},\ }\bibfield  {title} {\bibinfo {title} {{Relativistic double-zeta,
  triple-zeta, and quadruple-zeta basis sets for the light elements H--Ar}},\
  }\href {https://doi.org/10.1007/s00214-016-1884-y} {\bibfield  {journal}
  {\bibinfo  {journal} {Theor.~Chem.~Acc.}\ }\textbf {\bibinfo {volume}
  {135}},\ \bibinfo {pages} {128} (\bibinfo {year} {2016})}\BibitemShut
  {NoStop}%
\bibitem [{\citenamefont {Saue}(2005)}]{Saue2005}%
  \BibitemOpen
  \bibfield  {author} {\bibinfo {author} {\bibfnamefont {T.}~\bibnamefont
  {Saue}},\ }\bibfield  {title} {\bibinfo {title} {{Spin-Interactions and the
  Non-relativistic Limit of Electrodynamics}},\ }in\ \href
  {https://doi.org/10.1016/S0065-3276(05)48020-X} {\emph {\bibinfo {booktitle}
  {Advances in Quantum Chemistry}}},\ Vol.~\bibinfo {volume} {48},\ \bibinfo
  {editor} {edited by\ \bibinfo {editor} {\bibfnamefont {J.~R.}\ \bibnamefont
  {Sabin}}}\ (\bibinfo  {publisher} {Academic},\ \bibinfo {address} {San
  Diego},\ \bibinfo {year} {2005})\ pp.\ \bibinfo {pages}
  {383--405}\BibitemShut {NoStop}%
\bibitem [{\citenamefont {Visscher}\ and\ \citenamefont
  {Saue}(2000)}]{Visscher2000}%
  \BibitemOpen
  \bibfield  {author} {\bibinfo {author} {\bibfnamefont {L.}~\bibnamefont
  {Visscher}}\ and\ \bibinfo {author} {\bibfnamefont {T.}~\bibnamefont
  {Saue}},\ }\bibfield  {title} {\bibinfo {title} {{Approximate relativistic
  electronic structure methods based on the quaternion modified Dirac
  equation}},\ }\href {https://doi.org/10.1063/1.1288371} {\bibfield  {journal}
  {\bibinfo  {journal} {J.~Chem.~Phys.}\ }\textbf {\bibinfo {volume} {113}},\
  \bibinfo {pages} {3996} (\bibinfo {year} {2000})}\BibitemShut {NoStop}%
\bibitem [{\citenamefont {Visscher}(1997)}]{Visscher1997-SSSS}%
  \BibitemOpen
  \bibfield  {author} {\bibinfo {author} {\bibfnamefont {L.}~\bibnamefont
  {Visscher}},\ }\bibfield  {title} {\bibinfo {title} {{Approximate molecular
  relativistic Dirac--Coulomb calculations using a simple Coulombic
  correction}},\ }\href {https://doi.org/10.1007/s002140050280} {\bibfield
  {journal} {\bibinfo  {journal} {Theor.~Chem.~Acc.}\ }\textbf {\bibinfo
  {volume} {98}},\ \bibinfo {pages} {68} (\bibinfo {year} {1997})}\BibitemShut
  {NoStop}%
\bibitem [{\citenamefont {Visscher}\ and\ \citenamefont
  {Dyall}(1997)}]{Visscher1997}%
  \BibitemOpen
  \bibfield  {author} {\bibinfo {author} {\bibfnamefont {L.}~\bibnamefont
  {Visscher}}\ and\ \bibinfo {author} {\bibfnamefont {K.~G.}\ \bibnamefont
  {Dyall}},\ }\bibfield  {title} {\bibinfo {title} {{Dirac--Fock Atomic
  Electronic Structure Calculations Using Different Nuclear Charge
  Distributions}},\ }\href {https://doi.org/10.1006/adnd.1997.0751} {\bibfield
  {journal} {\bibinfo  {journal} {At.~Data Nucl.~Data Tables}\ }\textbf
  {\bibinfo {volume} {67}},\ \bibinfo {pages} {207} (\bibinfo {year}
  {1997})}\BibitemShut {NoStop}%
\bibitem [{\citenamefont {Aucar}\ \emph
  {et~al.}(2018{\natexlab{b}})\citenamefont {Aucar}, \citenamefont
  {Gim{\'e}nez},\ and\ \citenamefont {Aucar}}]{Agus_NChDE_RSC2018}%
  \BibitemOpen
  \bibfield  {author} {\bibinfo {author} {\bibfnamefont {I.~A.}\ \bibnamefont
  {Aucar}}, \bibinfo {author} {\bibfnamefont {C.~A.}\ \bibnamefont
  {Gim{\'e}nez}},\ and\ \bibinfo {author} {\bibfnamefont {G.~A.}\ \bibnamefont
  {Aucar}},\ }\bibfield  {title} {\bibinfo {title} {{Influence of the nuclear
  charge distribution and electron correlation effects on magnetic shieldings
  and spin-rotation tensors of linear molecules}},\ }\href
  {https://doi.org/10.1039/C8RA03948H} {\bibfield  {journal} {\bibinfo
  {journal} {RSC Adv.}\ }\textbf {\bibinfo {volume} {8}},\ \bibinfo {pages}
  {20234} (\bibinfo {year} {2018}{\natexlab{b}})}\BibitemShut {NoStop}%
\bibitem [{\citenamefont {Adamo}\ and\ \citenamefont {Barone}(1999)}]{PBE0}%
  \BibitemOpen
  \bibfield  {author} {\bibinfo {author} {\bibfnamefont {C.}~\bibnamefont
  {Adamo}}\ and\ \bibinfo {author} {\bibfnamefont {V.}~\bibnamefont {Barone}},\
  }\bibfield  {title} {\bibinfo {title} {{Toward Reliable Density Functional
  Methods without Adjustable Parameters: The PBE0 Model}},\ }\href
  {http://dx.doi.org/10.1063/1.478522} {\bibfield  {journal} {\bibinfo
  {journal} {J.~Chem.~Phys.}\ }\textbf {\bibinfo {volume} {110}},\ \bibinfo
  {pages} {6158} (\bibinfo {year} {1999})}\BibitemShut {NoStop}%
\bibitem [{\citenamefont {Bajac}\ \emph {et~al.}(2021)\citenamefont {Bajac},
  \citenamefont {Aucar},\ and\ \citenamefont {Aucar}}]{Aucar_CH3X}%
  \BibitemOpen
  \bibfield  {author} {\bibinfo {author} {\bibfnamefont {D.~F.~E.}\
  \bibnamefont {Bajac}}, \bibinfo {author} {\bibfnamefont {I.~A.}\ \bibnamefont
  {Aucar}},\ and\ \bibinfo {author} {\bibfnamefont {G.~A.}\ \bibnamefont
  {Aucar}},\ }\bibfield  {title} {\bibinfo {title} {{Absolute NMR shielding
  scales in methyl halides obtained from experimental and calculated nuclear
  spin-rotation constants}},\ }\href
  {https://doi.org/10.1103/PhysRevA.104.012805} {\bibfield  {journal} {\bibinfo
   {journal} {Phys.~Rev.~A}\ }\textbf {\bibinfo {volume} {104}},\ \bibinfo
  {pages} {012805} (\bibinfo {year} {2021})}\BibitemShut {NoStop}%
\end{thebibliography}%

\end{document}